\documentclass[superscriptaddress,twocolumn,secnumarabic,amssymb,amsmath,nobibnotes,aps,
prd,showpacs,nofootinbib]{revtex4}
\usepackage{epsf}
\usepackage{bm}
\usepackage{amsmath}
\usepackage{amsfonts}
\usepackage{amssymb}
\usepackage{epstopdf}%
\usepackage{epsfig}
\usepackage{subfigure,graphicx, latexsym, amssymb, amscd, psfrag}

\begin{document}

\title{Evolution of spherical domain walls in solitonic symmetron models}

\author{Marzieh Peyravi}
\email{marziyeh.peyravi@stu-mail.um.ac.ir}
\affiliation{Department of Physics, School of Sciences, Ferdowsi
University of Mashhad, Mashhad 91775-1436, Iran}
\affiliation{Instituto de Astrof\'{\i}sica e Ci\^{e}ncias do Espa\c{c}o, Faculdade de
Ci\^encias da Universidade de Lisboa, Edif\'{\i}cio C8, Campo Grande,
P-1749-016 Lisbon, Portugal.}

\author{Nematollah Riazi}
\email{n_riazi@sbu.ac.ir}
\affiliation{Physics Department, Shahid Beheshti University, Evin, Tehran
19839, Iran.}

\author{Francisco S. N. Lobo}
\email{fslobo@fc.ul.pt}
\affiliation{Instituto de Astrof\'{\i}sica e Ci\^{e}ncias do Espa\c{c}o, Faculdade de
Ci\^encias da Universidade de Lisboa, Edif\'{\i}cio C8, Campo Grande,
P-1749-016 Lisbon, Portugal.}


\date{\today}

\begin{abstract}
In this work, inspired by the symmetron model, we analyse the evolution of spherical domain walls by considering specific potentials that ensure symmetry breaking and the occurrence of degenerate vacua that are necessary for the formation of domain walls. By considering a simple analytical model of spherical domain wall collapse in vacuum, it is shown that this model fits the more accurate numerical results very well until full collapse, after which oscillations and scalar radiation take place. Furthermore, we explore the effect of a central non-relativistic matter lump on the evolution of a spherical domain wall and show that the central lump can prevent the full collapse and annihilation of the domain wall bubble, due to the repulsion between the domain wall and matter over-density within the adopted symmetron inspired model.
\end{abstract}

\maketitle

\section{Introduction}

Substantial observational evidence, such as the late-time accelerated expansion of the Universe \cite{Perlmutter:1998np,Riess:1998cb}, supports the idea of an exotic cosmic fluid denoted as {\it dark energy} \cite{W9, AT10, LRM}. However, the nature of dark energy is not yet understood and there are two different points of view: first, dark energy is a kind of unknown matter/energy with highly negative pressure and second, that General Relativity (GR) needs to be modified. Moreover, the requirement of the first approach consists of finding a new type of matter with an equation of state of the form $w\equiv p/\rho\approx-1$, and its detection will be a milestone for particle physics. Nevertheless, according to the second paradigm, it is conceivable that GR is a first order approximation to a more fundamental theory. Such an idea is known as modified gravity \cite{Clifton:2011jh}. Accordingly, numerous theories of high energy physics, such as string theory and supergravity, predict light, gravitationally coupled scalar fields \cite{ann, B6, L8}. In fact, in all of these theories a scalar field can play the role of dark energy \cite{ann,MG}. In addition, a fifth force emerges from this scalar field.

In fact, one of the best motivated modifications of GR are scalar-tensor theories \cite{MG}. These can be interpreted as a generalized form of quintessence models, which contain a scalar field coupled to matter. It seems natural to assume that the order of this coupling constant is unity and one may interpret it as a source of the fifth force \cite{MG}. It is necessary to emphasize that the detection of this force not only depends on the value of the coupling constant, but is also associated with the average matter density of the environment. This idea can be formulated by a ``screening mechanism'' that leads to the suppression of this additional force in a medium with high average matter density such as the solar system \cite{ann, MG, CL}.
Recently, two screening mechanisms have been introduced, namely, the chameleon mechanism and the symmetron model. Briefly, they work through different mechanisms, although, they are similar in some respects \cite{MG}. While in the chameleon mechanism \cite{MG,BRA,Khoa, Khob}, the effective mass of the field depends on the local matter density, in the symmetron model \cite{MG,KJAA,KJ,OP}, the vacuum-expectation value (VEV) of the scalar field and the symmetry of the potential are dependent on the local matter density.

In this paper, we focus on the symmetron model and explore alternative potentials to those proposed in the symmetron literature. As mentioned above, the symmetron model has been formulated based on a scalar field and matter interaction, undergoing symmetry breaking. An interesting consequence of symmetry breaking that appears in many different physical theories is that of {\it domain walls} \cite{nor, man, pp, 2, 3, 4, 6, Pe, the}. Topologically, domain walls can form if the field potential has disconnected vacua \cite{nor}. Furthermore, in cosmology and during the early universe, it is assumed that the cosmic medium cools as it expands, so that  cosmological phase transitions could occur due to the breaking of fundamental symmetries \cite{Mukh}. Formerly, cosmological domain walls were expected to form during a symmetry breaking in the early universe via a second-order phase transition, through a process known as the Kibble mechanism \cite{Va,Wag,Kib} and it was speculated that they lead to the formation of large-scale structures.
However, since no observational evidence has yet been found in favour of such objects in the cosmic microwave background radiation (CMB), such a scenario is nowadays usually discarded.

As mentioned before, scalar fields with a strong coupling to matter can be present in the universe while invisible to local observations. This may be so if these fields are subject to a screening mechanism, such as in the symmetron model. Note that in this model, regions of high density shield the fifth force resulting from the scalar field. In fact, in \cite{ann}, a structure formation analysis in the symmetron model via $N-$body simulations confirms the suppression of the scalar fifth force in high-density regions. Moreover, the properties of domain walls in the symmetron model have been studied in \cite{CL}, where numerical simulations of representative interactions between domain walls and matter over-densities have been investigated.

In the present paper, motivated by the symmetron model, we analyse the dynamics (collapse/expansion) of spherical domain walls which are governed by popular scalar field theory potentials capable of producing domain walls. We will present several ways of doing analytical and numerical calculations showing the collapse (or expansion) of the spherical domain wall, with/without the gravitational interaction and with/without direct interaction between the symmetron field $\varphi$ and the matter.
The outline of this paper is as follows: In Section \ref{Sym}, we present a short review of the symmetron model and consider alternative potentials that will be analysed throughout the paper. In Section \ref{wom}, we calculate the collapse of a spherical domain wall for each model, followed by a simple analytical model which closely agrees with the numerical results. In Section \ref{wm}, we discuss the evolution of spherical domain walls in the presence of central matter density. Section \ref{gravity} involves the gravitational effects of the central mass as well as the self gravity of the domain wall through a collective coordinate approximation. We present our conclusions in Section \ref{Conclusion}.

\section{Symmetron model and alternative potentials}\label{Sym}

\subsection{Symmetron model: General formalism}

The action of the symmetron model in the Einstein
frame is given by \cite{ann,MG,CL,KJ}, with metric signature $(-,+,+,+)$,
\begin{eqnarray}
S=\int d^{4}x \Bigg[\sqrt{-g}\left(\frac{{\cal R}}{2}M_{pl}^{2}-\frac{1}{2}g^{\mu\nu}\partial_{\mu}\varphi\partial_{\nu}\varphi-V(\varphi)\right)
   \nonumber  \\
+\sqrt{-\tilde{g}}{\cal L} _{m}(\psi,\tilde{g}_{\mu\nu})\Big],
\end{eqnarray}
where $M_{pl}\equiv 1/\sqrt{8\pi G}$ with $G$ as a Newton's constant\footnote{We will use units in which $G=1$ and $M_{pl}=1/\sqrt{8\pi}$.}, $\psi$ is representative of the matter fields and $\varphi$ is the scalar (symmetron) field which is coupled to the Jordan frame metric via a conformal rescaling, given by $\tilde{g}_{\mu\nu}\equiv A^{2}(\varphi)g_{\mu\nu}$ \cite{ann,KJAA,MG,CL}.
The coupling function $A(\varphi)$ is usually chosen to be an even polynomial with respect to $\varphi$, in order to be compatible with the transformation $\varphi \rightarrow -\varphi$, as shown below.

As one can obtain from the action, the scalar field equation of motion is given by
\cite{KJAA,MG}
\begin{equation}\label{a}
\Box \varphi-V_{,\varphi}+A^{3}(\varphi)A_{,\varphi}(\varphi)\tilde{T}=0,
\end{equation}
where $\tilde{T}=\tilde{g}^{\mu\nu}\tilde{T}_{\mu\nu}$ is the trace of the Jordan frame matter energy-momentum tensor \cite{KJAA,MG,CL}. The latter is defined as $\tilde{T}_{\mu\nu}=-\frac{2}{\sqrt{-\tilde{g}}}\frac{\delta{\cal L}_{m}}{\delta\tilde{g}^{\mu\nu}}$.
Considering non-relativistic matter ($\tilde{T}\approx-\tilde{\rho}$) and $\rho=A^{3}(\varphi)\tilde{\rho}$, then Eq. (\ref{a}) takes the form
\begin{equation}\label{5}
\Box \varphi-V_{,\varphi}-A_{,\varphi}(\varphi)\rho=0.
\end{equation}
Furthermore, by interpreting $V(\varphi)+\rho A(\varphi)$, as an effective potential, the field equation reduces to \cite{ann,MG}
\begin{equation}\label{6}
\Box \varphi=V_{{\rm eff},\varphi}.
\end{equation}
The form of the functions $A(\varphi)$ and $V(\varphi)$ is fundamental for the symmetron model. In general, it is assumed that these functions are symmetric under the transformation $\varphi\longrightarrow-\varphi$ and work in such a way that they could control the vacuum expectation value (VEV) of the effective potential during symmetry breaking \cite{KJAA,MG}.

The simplest symmetron models use the following functions \cite{ann,KJAA,MG,CL}:
\begin{equation}\label{8}
A(\varphi)=1+\frac{\varphi^{2}}{2M^{2}}+\mathcal{O}\left(\frac{\varphi^{4}}{M^{4}}\right),
\end{equation}
where $\varphi/M \ll 1$ is considered, and the potential
\begin{equation}
V(\varphi)=V_{0}-\frac{1}{2}\mu^{2}\varphi^{2}+\frac{1}{4}\lambda\varphi^{4},
 \label{symmpotential}
\end{equation}
is chosen to be of the symmetry breaking form.

The constants $\mu$ and $M$ have mass dimensions and $\lambda$ is a positive dimensionless coupling \cite{ann,MG,CL}.
However, it is more appropriate to work with physically intuitive quantities such as $L$, $\chi$ and $\rho_{\rm SSB}$, where $L$ is the cosmological range of the fifth force in $Mpc/h$ [here $h=H_{0}/(100{\rm kms}^{-1}{\rm Mpc}^{-1})$], $\chi$ is the strength of the fifth force relative to gravity and $\rho_{\rm SSB}$ is related to the density at which the spontaneous symmetry breaking (SSB) takes place in the cosmological background \cite{ann,CL}.

In the symmetron model, SSB is governed by the coupling
between matter and the scalar field which results in the following effective potential
\cite{ann,KJAA,MG,CL}:
\begin{equation}
V_{\rm eff}\equiv\frac{1}{2}\left(\frac{\rho_m}{M^2}-\mu^2\right)\varphi^2
+\frac{1}{4}\lambda\varphi^{4}+V_{0},
 \label{symmetronVeff}
\end{equation}
where $M$ is the mass scale for SSB, and $\rho_{\rm SSB}=\mu^{2}M^{2}$. Note that $\rho_{m}>\mu^{2}M^{2}$ corresponds to the symmetric phase, while $\rho_{m}<\mu^{2}M^{2}$
leads to SSB.

The introduction of cosmic scalar fields are severely constrained by
observations of the behavior of local gravitational fields. Such a
field -- if in existence -- should be coupled to the matter field in
such a way that its physical effects (the so-called fifth force)
is screened at solar system scales. The range of the symmetron
field depends on the effective mass of the field near the
minimum of the effective potential. Note that the effective mass of the scalar field is defined as $m_{\varphi}\equiv \left(\partial^{2}V_{\rm eff}/\partial
\varphi^2\right)^{1/2}\big|_{\rm vac}$, and using Eq. (\ref{symmetronVeff}), takes the form
\begin{equation}
m_{\varphi}=\left(\frac{\rho_m}{\rho_{\rm SSB}}-1\right)\mu^{2}+3\lambda
\varphi_{\rm min}^2.
\end{equation}
where $\varphi_{\rm min}=\pm\varphi_{0}\sqrt{1-\rho_{m}/\rho_{\rm SSB}}$,
and $\varphi_{0}\equiv \mu/\sqrt{\lambda}$ is the symmetry breaking VEV for $\rho \rightarrow 0$.

Therefore, the field has the longest range ($\ell\sim 1/m_\varphi$)
in regions with lowest matter densities and shortest range in
local concentrations of matter (inside a galaxy or within the
solar system).

\subsection{Alternative potentials}

In what follows, we replace the symmetron potential (\ref{symmpotential}) with four popular scalar field theory potentials, extensively considered in the literature. These are the sine-Gordon (SG), double sine-Gordon (DSG), $\phi^{4}$ and $\phi^{6}$ systems.
The corresponding potentials of these systems are given by \cite{Pe,righ,Pey,gui,hosein}:
\begin{eqnarray}
V_{\rm SG}(\varphi)&=&\frac{a}{b}\left(1-\cos(b\varphi)\right),
\label{9a}
\end{eqnarray}
\begin{eqnarray}
V_{\rm DSG}(\varphi)&=&\frac{a}{b}\left(1+\varepsilon-\cos(b\varphi)-\varepsilon\cos(2b\varphi)\right),
\label{9b}
\end{eqnarray}
\begin{eqnarray}
V_{\varphi^{4}}(\varphi)&=&\frac{\beta^{2}}{2\alpha^{2}}(\varphi^{2}-\alpha^{2})^{2},  \label{9c}
\end{eqnarray}
\begin{eqnarray}
V_{\varphi^{6}}(\varphi)&=&\frac{\beta^{2}}{4\alpha^{2}}\varphi^{2}(\varphi^{2}-\alpha^{2})^{2},
\label{9d}
\end{eqnarray}
respectively.

\begin{figure*}
\epsfxsize=13.00cm\centerline{\epsfbox{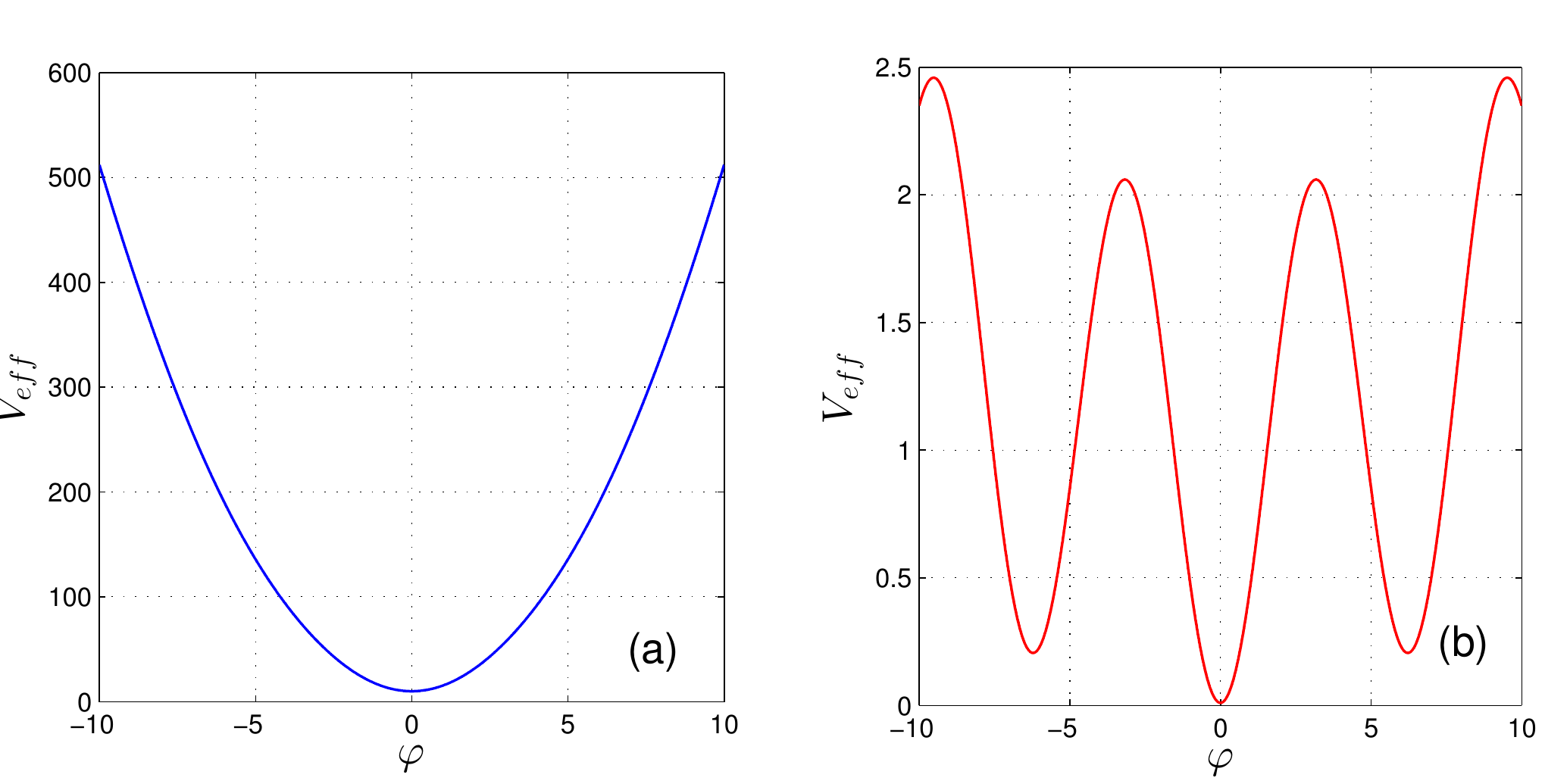}} \caption{The
effective potential of the SG system for $a=b=1$; (a) $\rho=10$ and (b) $\rho=0.01$. See the text for more details.}\label{SGeff}
\end{figure*}
\begin{figure*}
\epsfxsize=13.00cm\centerline{\epsfbox{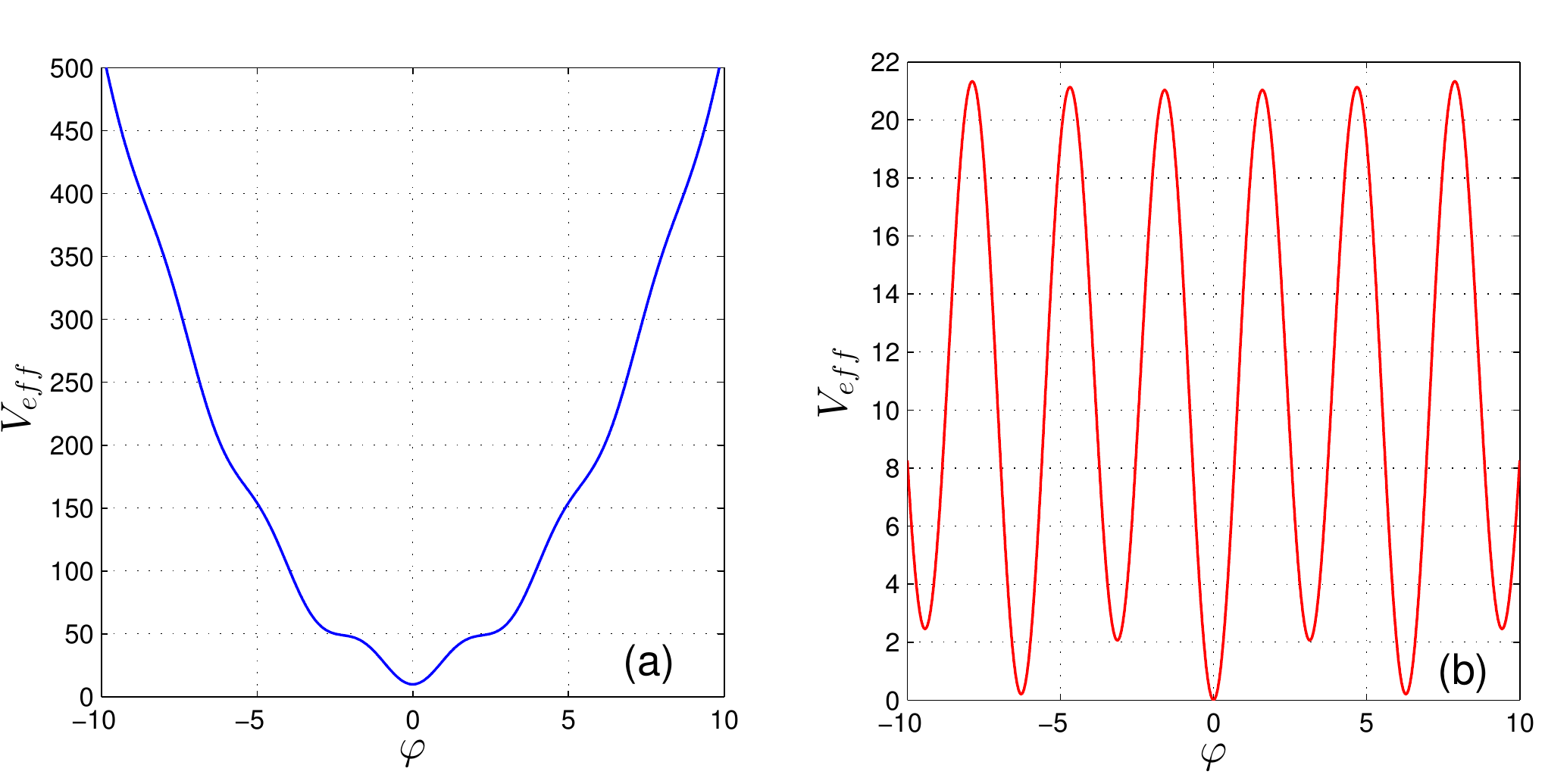}} \caption{The
effective potential of the DSG system for $a=b=1$ and $\varepsilon=10$; (a) $\rho=10$ and (b) $\rho=0.01$. See the text for more details.} \label{DSGeff}
\end{figure*}
\begin{figure*}
\epsfxsize=13.00cm\centerline{\epsfbox{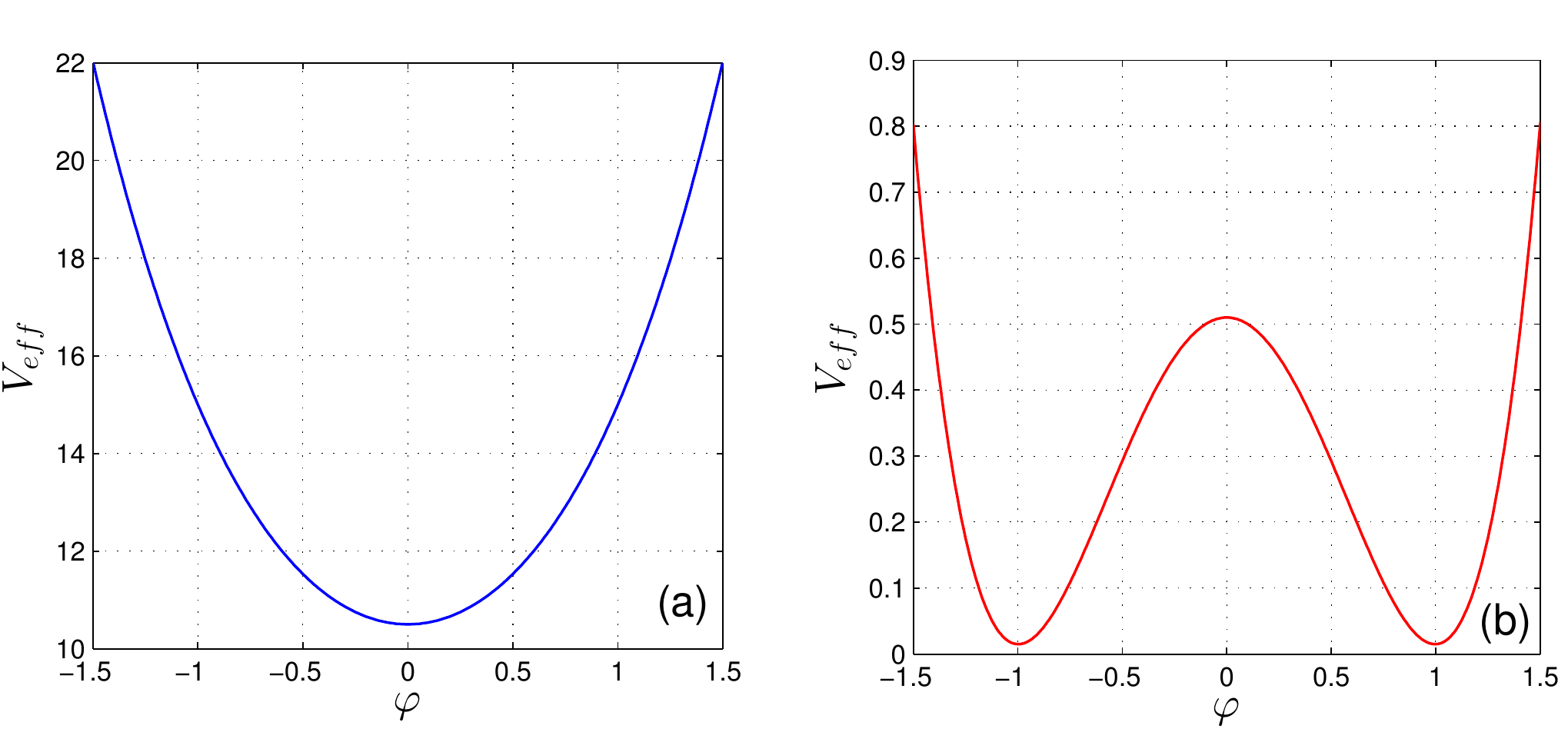}} \caption{The
effective potential of the $\varphi^{4}$ system for $\alpha=\beta=1$; (a) $\rho=10$ and (b) $\rho=0.01$. See the text for more details.} \label{phi4eff}
\end{figure*}
\begin{figure*}
\epsfxsize=13.00cm\centerline{\epsfbox{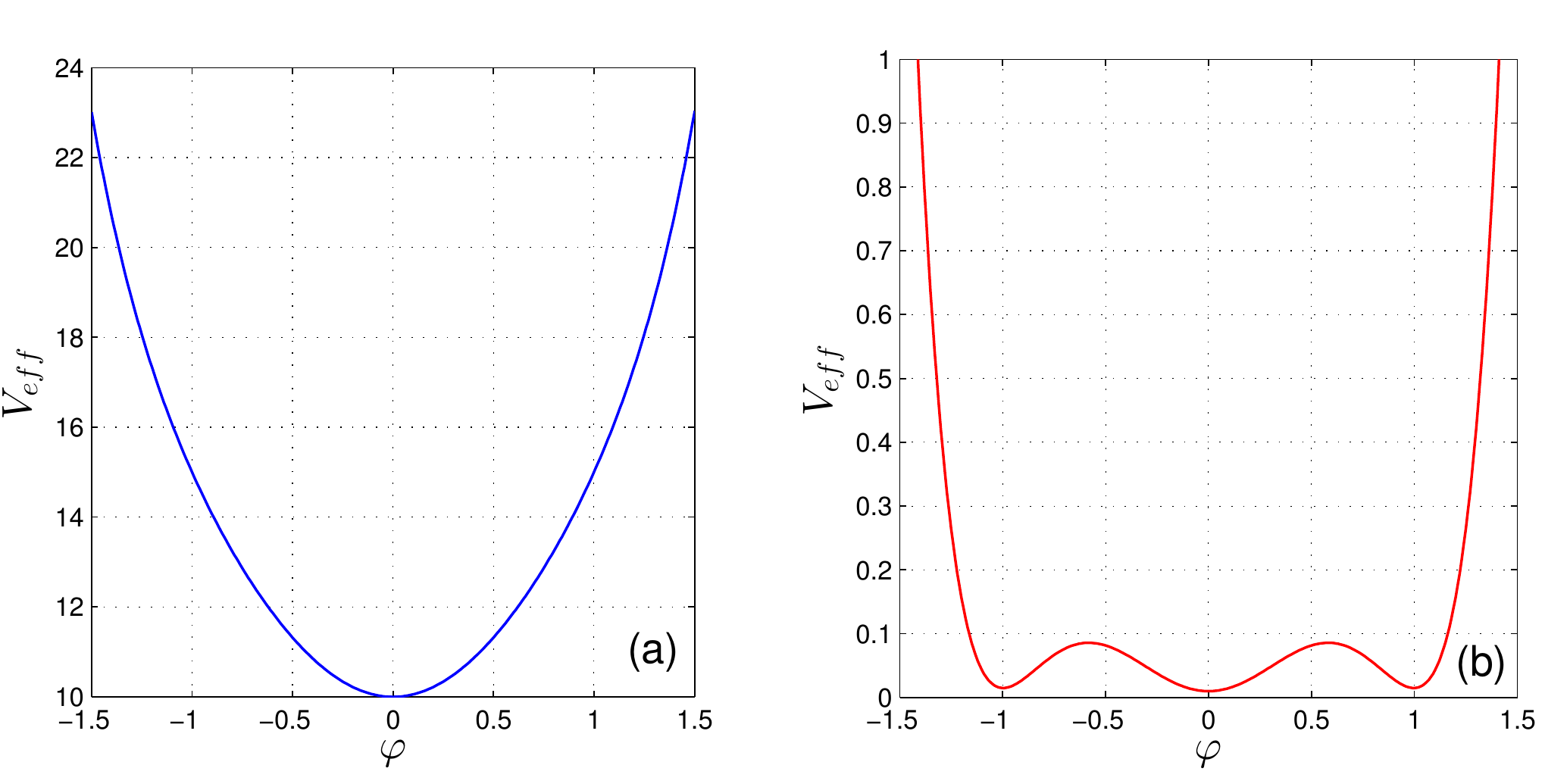}} \caption{The
effective potential of the $\varphi^{6}$ system for $\alpha=1$ and $\beta=\sqrt{2}$; (a) $\rho=10$ and (b) $\rho=0.01$. See the text for more details.} \label{phi6eff}
\end{figure*}

According to Eqs. (\ref{5}) and (\ref{6}) the effective potential for the above-mentioned systems are plotted in Figs. \ref{SGeff}-\ref{phi6eff}. The left plots (a), of these figures depict the effective potential for large $\rho$ ($\rho=10$) in which the symmetry is restored. The right plots (b) show the effective potential for small values of $\rho$ ($\rho=0.01$) with symmetry breaking. It should be emphasized that the minimum of these potentials can be degenerate or non-degenerate, depending on the value of $\rho$  in the second case. For instance, in the SG model with $\rho=0.01$, in Fig. \ref{SGeff}, the effective potential is similar to the DSG potential with non-degenerate vacua (Fig. \ref{SGeff}). However, as $\rho\rightarrow0$ these vacua become degenerate.

In what follows, we see that, by expanding each of these potentials around the location of the domain wall up to $\mathcal{O}(\varphi^{4})$, one may obtain a correspondence with the potential of the symmetron model, i.e.,
\begin{widetext}
\begin{eqnarray}
V_{\rm SG}(\varphi)&=&\frac{2a}{b}-\frac{1}{2}ab\left(\varphi-\frac{\pi}{b}\right)^{2}+\frac{1}{24}ab^{3}\left(\varphi-\frac{\pi}{b}\right)^{4}+\mathcal{O}(\varphi^{6}),
  \label{10a} \\
V_{\rm DSG}(\varphi)&=&\frac{1}{8}\frac{a\left(8\varepsilon b^{2}+16\varepsilon^{2}b^{2}+2b-1\right)}{\varepsilon b^{3}}+\frac{1}{4}\frac{a\sqrt{\frac{16\varepsilon^{2}b^{2}-1}{\varepsilon^{2}b^{2}}}\left(b-1\right)}{b}
\left[\varphi-\frac{\pi-\arccos\left(\frac{1}{4\varepsilon b}\right)}{b}\right]
\nonumber\\
&&
-\frac{1}{8}\frac{a\left(-2+16\varepsilon^{2}b^{2}+b\right)}{\varepsilon b}\left[\varphi-\frac{\pi-\arccos\left(\frac{1}{4\varepsilon b}\right)}{b}\right]^{2}
 -\frac{1}{24}a\sqrt{\frac{16\varepsilon^{2}b^{2}-1}{\varepsilon^{2}b^{2}}}{b}\left(-4+b\right)\left[\varphi-\frac{\pi-\arccos\left(\frac{1}{4\varepsilon b}\right)}{b}\right]^{3}
 \nonumber\\
&&+\frac{1}{96}\frac{ab\left(b-8+64\varepsilon^{2} b^{2}\right)}{\varepsilon}\left[\varphi-\frac{\pi-\arccos\left(\frac{1}{4\varepsilon b}\right)}{b}\right]^{4}+\mathcal{O}(\varphi^{5}), \label{10b}
     \\
V_{\varphi^{4}}(\varphi)&=&\frac{1}{2}\beta^{2}\alpha^{2}-\beta^{2}\varphi^{2}+\frac{1}{2}\frac{\beta^{2}}{\alpha^{2}}\varphi^{4}, \label{10c}
     \\
V_{\varphi^{6}}(\varphi)&=&\frac{1}{27}\beta^{2}\alpha^{4}-\frac{1}{3}\beta^{2}\alpha^{2}\left(\varphi-\frac{\sqrt{3}}{3}\alpha\right)^{2}-\frac{\sqrt{3}}{9}\beta^{2}\alpha\left(\varphi-\frac{\sqrt{3}}{3}\alpha\right)^{3}+\frac{3}{4}\beta^{2}\left(\varphi-\frac{\sqrt{3}}{3}\alpha\right)^4+\mathcal{O}(\varphi^{5}),\label{10d}
\end{eqnarray}
respectively.
\end{widetext}

Here, we note that the appearance of odd terms in $\varphi$ indicate that these models break the $Z_{2}$ symmetry around the domain wall.
Note that the sign of the second-order term is very important, since it
ensures symmetry breaking and the occurrence of degenerate vacua
necessary for the formation of domain walls \cite{KJAA}. As one can see, the negative sign appears in the second term of all of these models, except in the DSG model. In particular, for this special case, the sign of the second-order term depends on $\varepsilon$. Moreover, the type and position of the minima in the potential vary according to the value of $\varepsilon$ and as a result, various domain walls with different values of vacuum energy densities will appear. For instance, for $\varepsilon > 0.25$, there are two kinds of vacua [local minima at $\varphi=(2n+1)\pi$ and global minima at $\varphi=2n\pi$] which result in the appearance of kink domain walls with two subkinks \cite{Pe,Pey}. On the other hand, if $-0.25<\varepsilon<0.25$, false vacua of the potential disappear and the system tends to the SG system with true vacuum at zero \cite{Pe,Pey}. However, the most important case occurs for $\varepsilon<-0.25$. One can see that for each $\varepsilon$, the potential contains two kinds of maxima [local at $\varphi=2n\pi$ and global at $\varphi=(2n+1)\pi$], while the minima are all degenerate. Remarkably, the structure of the potential in this case, provides two different pathways to connect absolute degenerate minima, which means that we encounter two types of domain walls. The surface energy density of these two types of domain walls are not the same \cite{Pe}.

Furthermore, based on Eqs. (\ref{10a})--(\ref{10d}), while the SG potential can satisfy the symmetron model conditions for positive free parameters ($a,b>0$), the $\varphi^{6}$ potential fulfils this model for both positive and negative parameters ($\alpha$ and $\beta$). Besides, in the $\varphi^{6}$ system, domain walls are not located at $\varphi=0$ and as a result its potential expansions involve odd terms of $\varphi$ as well as even terms. It means that for this potential, the symmetric phase corresponds to complex alpha [i.e. $\varphi^{2}(\varphi^{2}+|\alpha|^{2})^{2}$].
Symmetry breaking occurs when we have real alpha [i.e., $\varphi^{2}(\varphi^{2}-|\alpha|^{2})^{2}$].
Since it seems plausible to apply these models to large scale structures present in the late time universe, we interpret $V_{0}$ in Eq. (\ref{symmpotential}) as a positive cosmological constant $\Lambda$ which may be responsible for the accelerated expansion of the universe \cite{ann}.

\section{Free spherical domain walls}\label{wom}

In this section, we start with investigating the evolution of spherical domain walls of the systems (\ref{9a})--(\ref{9d}) in the absence of
matter density. In Section \ref{wm}, we will study the evolution of these topological defects in the presence of an inner matter aggregation. On the other hand, for simplicity, in the SG, DSG, $\varphi^{4}$ and $\varphi^{6}$ models, we choose $a=b=1$, $\alpha=\beta=1$ and $\alpha=1,\beta=\sqrt{2}$, respectively.

\subsection{Free collapse of spherical domain walls}

In the absence of matter, by omitting $\rho$ in Eq. (\ref{5}), and assuming spherical symmetry, the general form of the equations in natural units, for the SG, DSG, $\varphi^{4}$ and $\varphi^{6}$ models, reduce to the following simple forms:
\begin{equation}
\begin{split}
\frac{\partial^{2}\varphi}{\partial
t^{2}}+\sin(\varphi)=\frac{\partial^{2}\varphi}{\partial
r^{2}}+\frac{2}{r}\frac{\partial \varphi}{\partial r},\\
\frac{\partial^{2}\varphi}{\partial
t^{2}}+\sin(\varphi)+2\varepsilon\sin(2\varphi)=\frac{\partial^{2}\varphi}{\partial
r^{2}}+\frac{2}{r}\frac{\partial \varphi}{\partial r},\\
\frac{\partial^{2}\varphi}{\partial
t^{2}}+2\varphi^{3}-2\varphi=\frac{\partial^{2}\varphi}{\partial
r^{2}}+\frac{2}{r}\frac{\partial \varphi}{\partial r},\\
\frac{\partial^{2}\varphi}{\partial
t^{2}}+3\varphi^{5}-4\varphi^{3}+\varphi =\frac{\partial^{2}\varphi}{\partial
r^{2}}+\frac{2}{r}\frac{\partial \varphi}{\partial r},
\end{split}
\end{equation}
respectively. The planar approximation static solutions (large spherical walls) of these equations, as outlined in the literature \cite{Pe,righ,Pey,gui,hosein}, for the SG, DSG, $\varphi^{4}$ and $\varphi^{6}$ models, are given by:
\begin{eqnarray}
\begin{split}
&\varphi(r)=4\arctan[\exp(r-R_{0})],\\
&\varphi(r)=2\arccos\left[\pm
\frac{\sinh\sqrt{4\varepsilon+1}(r-R_{0})}{\sqrt{4\varepsilon+\cosh^{2}\sqrt{4\varepsilon+1}(r-R_{0})}}\right],\\
& \varphi(r)=\tanh(r-R_{0}),\\
& \varphi(r)=\left\{1+\exp[-2(r-R_{0})]\right\}^{-\frac{1}{2}},
\end{split}
\label{16}
\end{eqnarray}
respectively, in which $R_{0}$ is the location of the domain wall. Note that this approximation is valid as long as the radius of the spherical domain wall is much larger than the thickness of the wall.
The thickness of these domain walls \cite{man,Va,Vi,RLP} is given by $\delta_{SG}\sim 1/(2\sqrt{ab})=1/2$ (with $a=b=1$), $\delta_{DSG}\sim 1$, $\delta_{\varphi^{4}}\sim 1/\beta=1$ (for $\alpha=\beta=1$), and $\delta_{\varphi^{6}}\sim 1/(\sqrt{2}\alpha\beta)=1/2$ (for $\alpha=1,\beta=\sqrt{2}$), respectively.

The thin wall approximation breaks down when \cite{Va}
\begin{equation}\label{18}
\frac{R}{R_{0}}\sim\left(\frac{\delta}{R_{0}}\right)^{1/3},
\end{equation}
where $R_0$ is the initial radius of the bubble (spherical wall) and $\delta$ is the wall thickness. Here $R_{0}$ is chosen to be 25, for simplicity. Thus, the quantity $\delta/R_{0}$ is equal to $0.02, 0.04, 0.04, 0.02$ for the SG, DSG, $\varphi^{4}$ and $\varphi^{6}$ systems, respectively. We emphasize that when this condition is not valid the domain wall will be thick and in this case one can consider $R$ as an average radius, as outlined in \cite{thick}. We will use static solutions of Eq. (\ref{16}) as the initial conditions for the numerical investigation of the spherical wall collapse.

Note that in $1 + 1$ dimensions, the kink behaves and moves like a massive
particle. The action of a kink, by ignoring its internal structure, can be written as $S_{1+1}=-M\int d\tau$,
where $M$ and $d\tau$ are the mass of the kink and the invariant line element,
respectively \cite{Va}. The latter $d\tau$ may also be written as
$d\tau=dt \left(g_{\mu\nu}\frac{dx^{\mu}}{dt}\frac{dx^{\nu}}{dt}\right)^{1/2}$,
where $g_{\mu\nu}$ is the metric of the spacetime background and
$x^{\mu}(t)$ is the location of the kink at time $t$.
By extending the $1+1$ kink in two more spacelike dimensions, one can
construct domain walls. Indeed, the dynamics in $3 + 1$ dimensions is
considerably richer, where for instance, a domain wall can bend, oscillate and move in more complicated ways.

\begin{figure*}
\epsfxsize=17cm
\centerline{\epsfbox{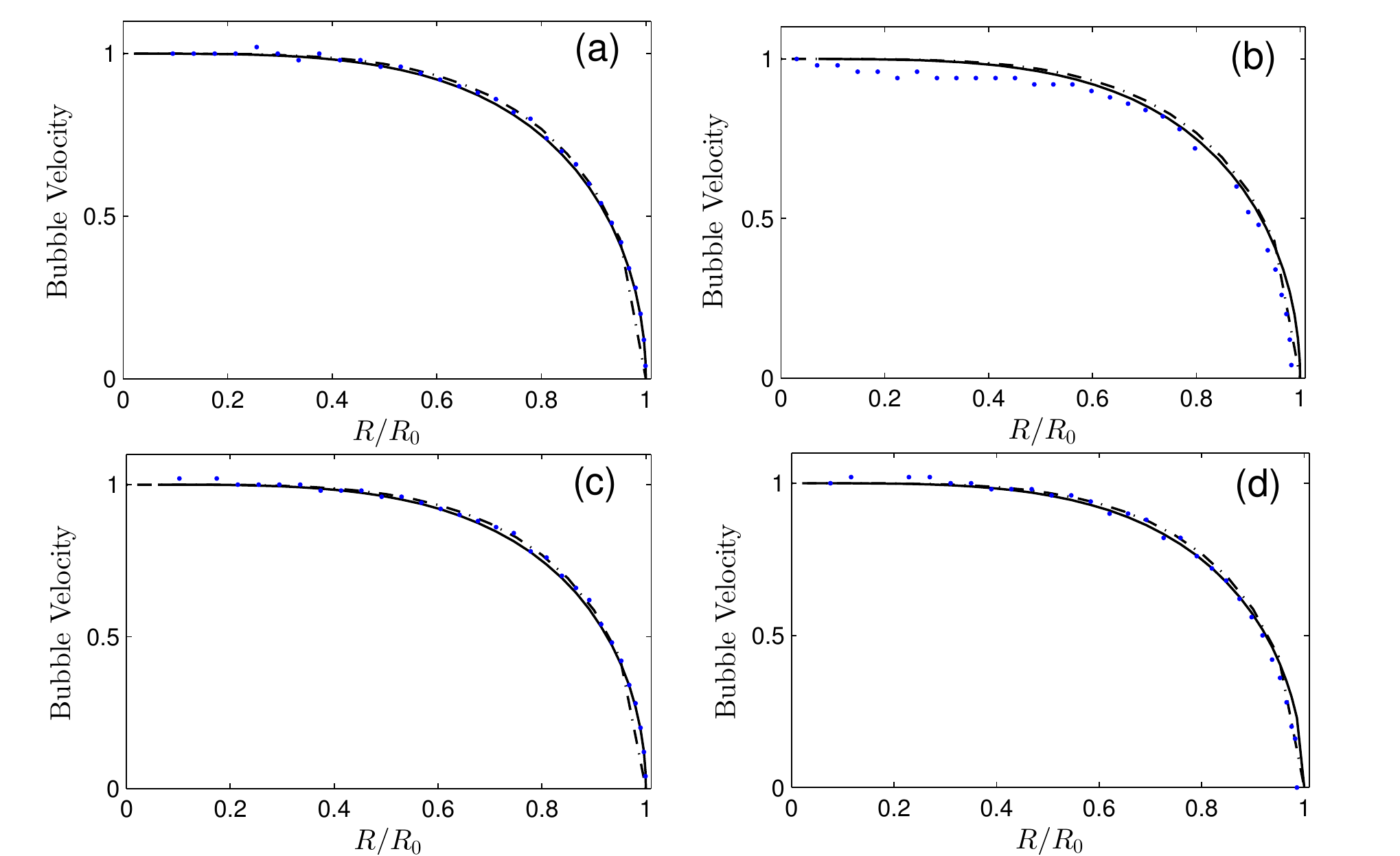}}
\caption{The numerical results are shown by the dotted curve depicting the bubble collapse velocity curve in terms of the radius of the bubble for the (a) SG ($a=b=1$),
(b) DSG ($a=b=1$ and $\varepsilon=10$), (c) $\varphi^{4}$ ($\alpha=\beta=1$) and (d) $\varphi^{6}$ ($\alpha=1$ and $\beta=\sqrt{2}$) systems. The analytical results of Eq. (\ref{vbk}) and Eq. (\ref{vbr}) for $n=0.3$ are shown in the dash-dotted and solid curves, respectively.}\label{vr}
\end{figure*}

For instance, consider non-relativistic domain wall solutions with a spherical symmetry ansatz: $X^{\mu}(t,\theta,\varphi) = [\tau, R(\tau)\mathbf{\hat{r}}]$,
where $\tau = t$, $\mathbf{\hat{r}} = (\sin\theta \cos\varphi, \sin\theta
\sin\varphi, \cos\theta)$
and $\theta$, $\varphi$ are the standard spherical angular
coordinates \cite{Va}. The spacetime metric is $\eta_{\mu\nu} = {\rm diag}(1, -1,
-1, -1)$. From the Nambu-Goto action, one can derive the equation of motion for domain walls, i.e., $S_{0}=-\sigma\int d\Sigma \sqrt{|h|}$,
where the integral is over the wall world volume $\Sigma$, $\sigma$ is the tension of the domain wall (the energy per unit area) and $h={\rm det}(h_{ab})$\footnote{This quantity is positive for the kink in $1+1$ dimensions and for domain wall in $3+1$ dimensions, as well \cite{Va}.} \cite{Va}. Note that the induced metric on the wall is given by $h_{ab}={\rm diag} (1-\dot{R}^{2},-R^{2},-R ^{2}\sin^{2}\theta )$,
where $\dot{R}=v=dR/d\tau$. The approximate equation of motion turns out to be \cite{Va,thick}:
\begin{equation}
\ddot{R} =-\frac{2}{R} (1 - \dot{R}^{2}).
\end{equation}
By integrating this equation one arrives at the following relation for the velocity of the bubble:
\begin{equation}\label{vbk}
v=\sqrt{1-\left(\frac{R}{R_{0}}\right)^{4}}.
\end{equation}
This is depicted in Fig. \ref{vr} (see details below).

In the next section, we will present a specific model for the speed of the collapsing bubble.
Note that the behaviour of the collision of kinks and antikinks in 1+1 dimensions differs in integrable and non-integrable systems. For the integrable SG system, the kink and antikink keep their form after their collision and continue to move with the same velocity, although a phase shift results \cite{righ}. For non-integrable
systems such as the DSG, $\phi^4$ and $\phi^6$ systems, certain
scattering windows appear between which the pair annihilate each
other \cite{hosein}. This situation holds almost (but not exactly)
the same for a collapsing spherical domain wall \cite{Va}. Numerical
simulations show that a collapsing spherical SG domain
wall does not radiate scalar waves as long as the radius is larger
than the wall thickness [see Eq. (\ref{18})]. In the final stages of the collapse,
however, it emits strongly and oscillates for a while \cite{Va,thick}.
In the next section, we will present a more accurate, relativistic model for the spherical wall collapse.

\subsection{A simple analytical model of spherical domain wall collapse}

The flat version of the domain wall has an energy per unit surface
$\sigma_{0}\equiv\int_{-\infty}^{+\infty}T_{0}^{0}dx$, where $x$ is
a coordinate normal to the wall. For the SG system with the potential $V = 1
-\cos(\varphi)$ (recall $a=b=1$), we obtain $\sigma_{0} = 8$.
This quantity is found to be about $50$ (for $\varepsilon=10$), $4/3$ and $1/4$ for the DSG, $\varphi^{4}$ and $\varphi^{6}$ systems, respectively. Here, we consider a specific model for the energy per unit surface area of a spherical domain wall according to the energy per unit surface, given by the following equation\footnote{Note that this surface energy density is a function of time implicitly through $R$.}
\begin{equation}
\sigma(R)=\sigma_{0}\left[1+\left(\frac{R_0}{R}\right)^{n} \right],
\end{equation}
where the second term is related to the curvature effect and $n$ is to be
determined by comparison with numerical calculations. As before, $R_{0}$ is the initial radius of the bubble and $R$ is the (time-dependent) radius at any arbitrary time before full collapse.

Using the conservation of the total energy, we have
\begin{equation}
\gamma 4\pi R^{2}\sigma(R)=\gamma M(R) c^{2}=E_{0}= {\rm const},
\end{equation}
where $\gamma = (1- \dot{R}^{2} /c^{2})^{ -1/2}$ and $M$ is the
total rest mass of the domain wall. Solving for $\gamma$, we
obtain
\begin{equation}
\gamma=\frac{E_{0}}{4\pi R^{2}\sigma(R)}.
\end{equation}
For $R \gg R_{0}$, so that $\sigma\simeq\sigma_{0}$, we have the planar
approximation
\begin{equation}
\gamma\simeq \left(\frac{E_{0}}{4\pi\sigma_{0}}\right)\frac{1}{R^{2}}.
\end{equation}
For the case $R \ll R_{0}$, we have
\begin{equation}\label{gam}
\gamma\simeq\frac{E_{0}}{4\pi\sigma_{0}R_{0}^{n}R^{2-n}}\propto
R^{n-2}.
\end{equation}
Our numerical calculations show that the bubble starts collapsing initially and therefore $R$ is
always less than $R_{0}$. We therefore expect the bubble velocity to approach a constant
value if $n = 2$. For $n < 2$, $\gamma$ tends to infinity, meaning
that in the relativistic regime the bubble approaches the speed
of light if enough time is available before full collapse. For $n > 2$, we expect that the contracting bubble stops at some stage and begins expanding.

Moreover, Eq. (\ref{gam}) can be solved for the collapse speed of the bubble, which is given by
\begin{equation}\label{vbr}
v=\sqrt{1-\frac{1}{4}\left(\frac{R}{R_{0}}\right)^{4}-\frac{1}{4}\left(\frac{R}{R_{0}}\right)^{4-2n}-\frac{2}{4}\left(\frac{R}{R_{0}}\right)^{4-n}},
\end{equation}
where the resulting dynamics is shown in Fig. \ref{vr}, in the solid curve.
Thus, in order to investigate the evolution of the collapsing bubble, we plot the analytical results of Eqs. (\ref{vbk}) and (\ref{vbr}), along with the results of the numerical calculations for the SG, DSG, $\varphi^{4}$ and $\varphi^{6}$ systems, respectively, in Fig. \ref{vr}. The numerical results are also plotted in order to compare  with the simple analytical model. As is transparent from the figure, in all cases the bubbles start to collapse slowly. The velocity of the bubble surface increases steadily (until full collapse) as it shrinks to the center. The analytical and numerical results match well for the SG, $\varphi^{4}$ and $\varphi^{6}$ models. Note that for the DSG model, however, there is a slight mismatch, which can be attributed to the existence of sub-kinks in this system.

\subsection{Numerical investigation and comparison with the analytical models}

\begin{figure*}
\begin{center}
{\label{fig:scalarSGw}
\includegraphics[width=0.48\textwidth,height=0.3\textheight]{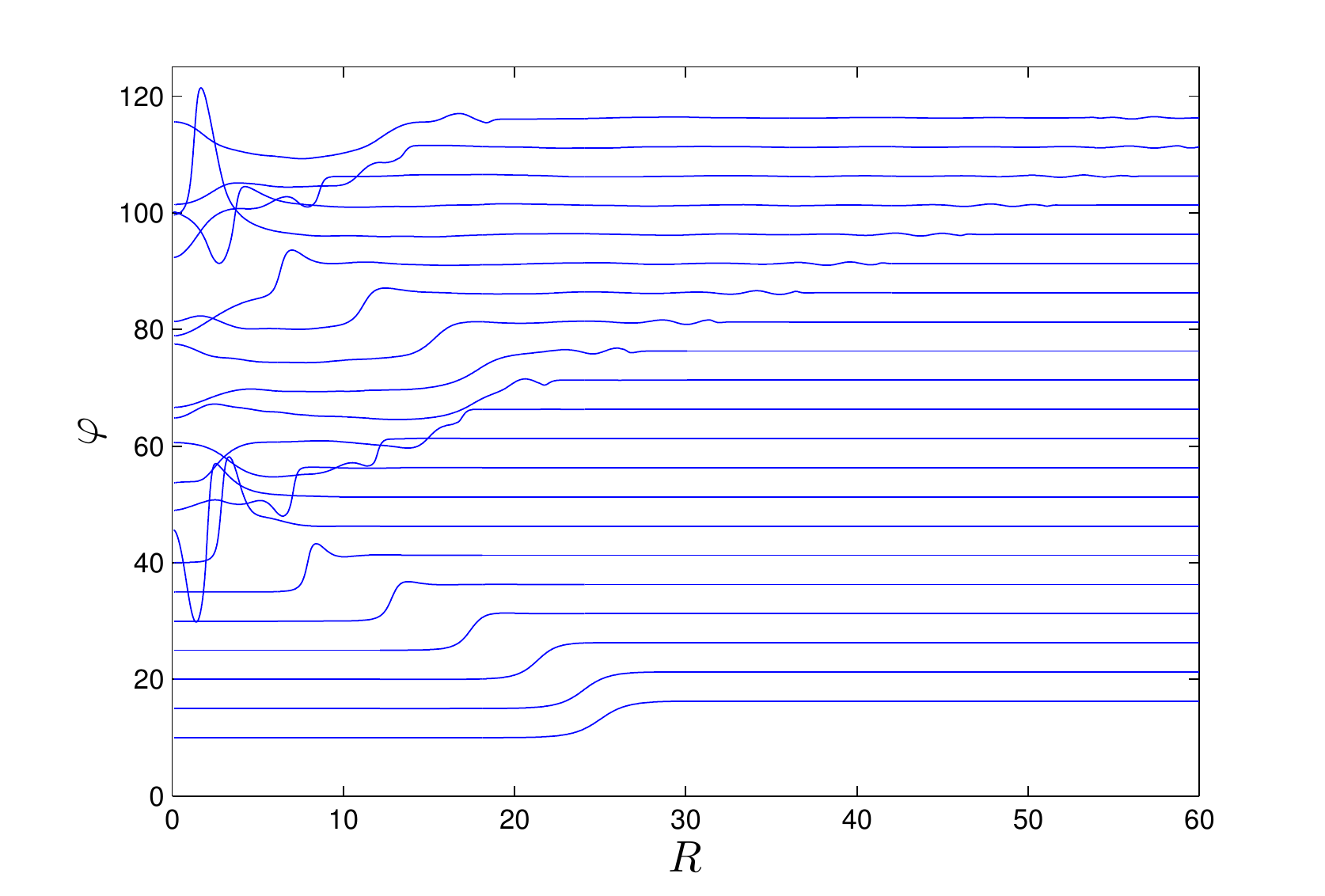}} \text{%
\hspace{0cm}}
{\label{fig:energySGw}
\includegraphics[width=0.48\textwidth,height=0.3\textheight]{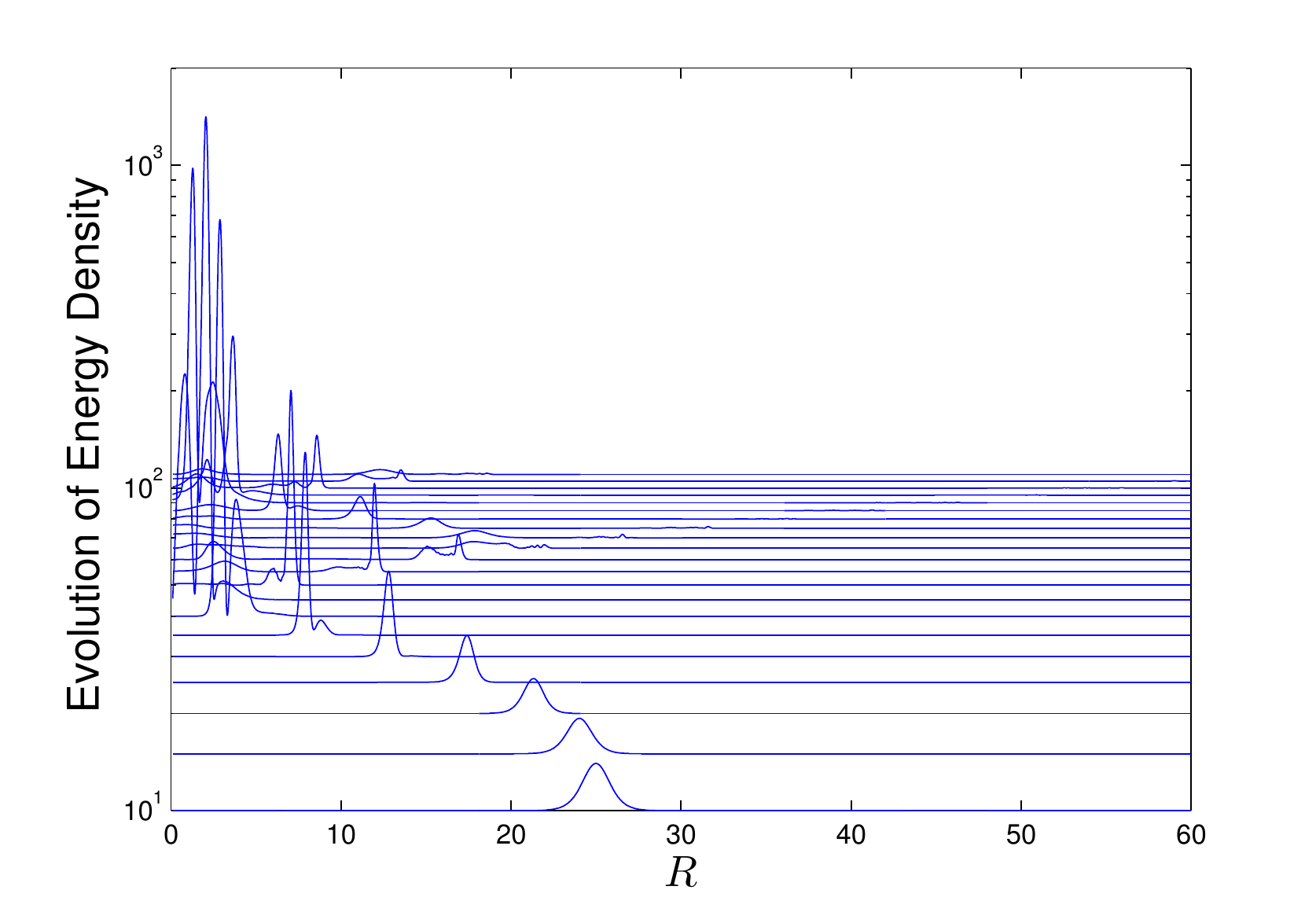}}
\end{center}
\caption{Evolution of the scalar field (left plot) and the energy density (right plot) of the SG domain wall, where we have considered the parameter values $a=b=1$. See the text for more details.}
\label{fig:SGw}
\end{figure*}
%
\begin{figure*}
\subfigure[]{\includegraphics[trim = 0.5in 0.6in 0.6in 0.6in,
clip,width=0.65\columnwidth]{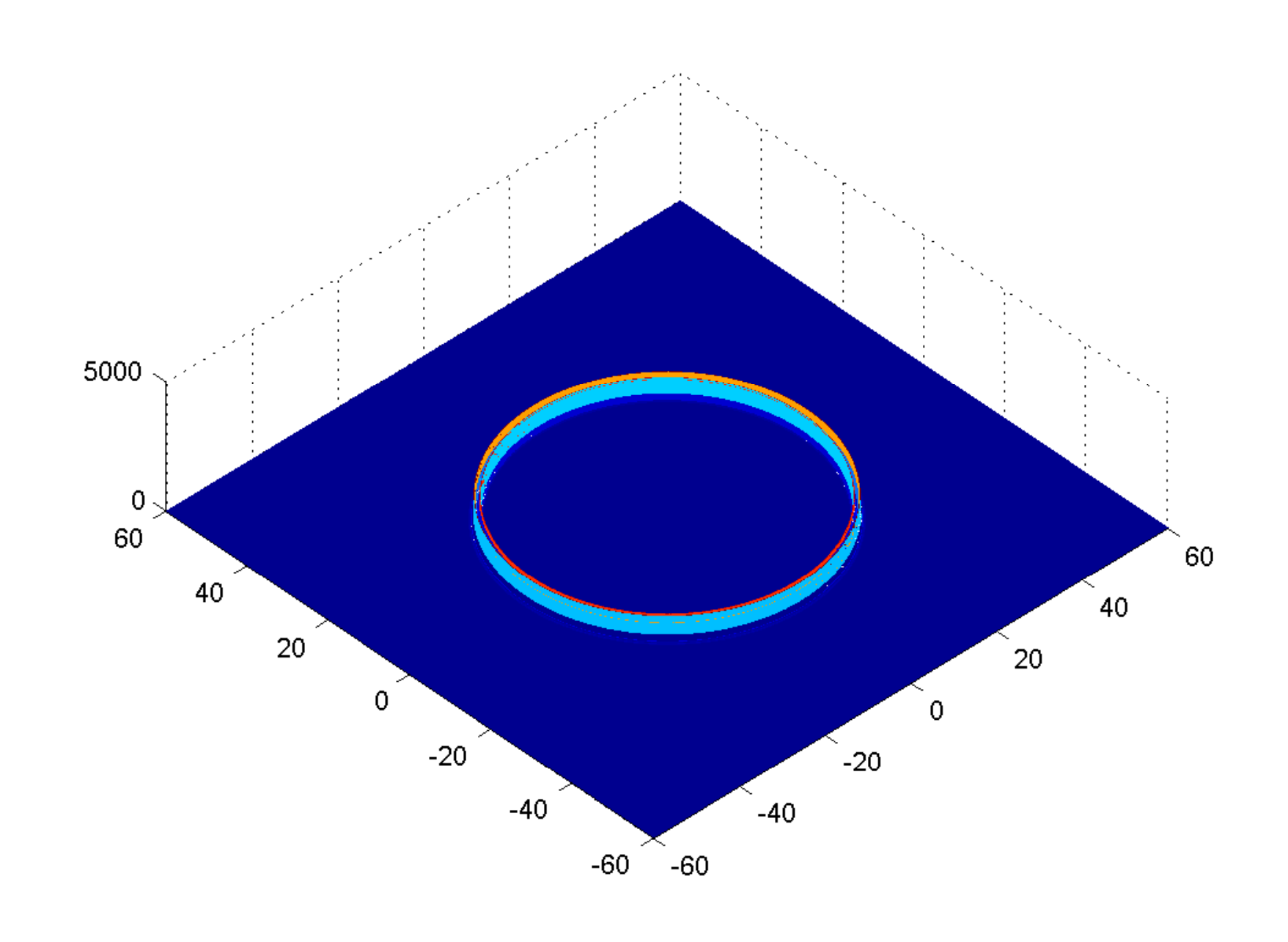}\label{fig:r25s02e10t1060}}
\subfigure[]{\includegraphics[trim = 0.5in 0.6in 0.6in 0.6in,
clip,width=0.65\columnwidth]{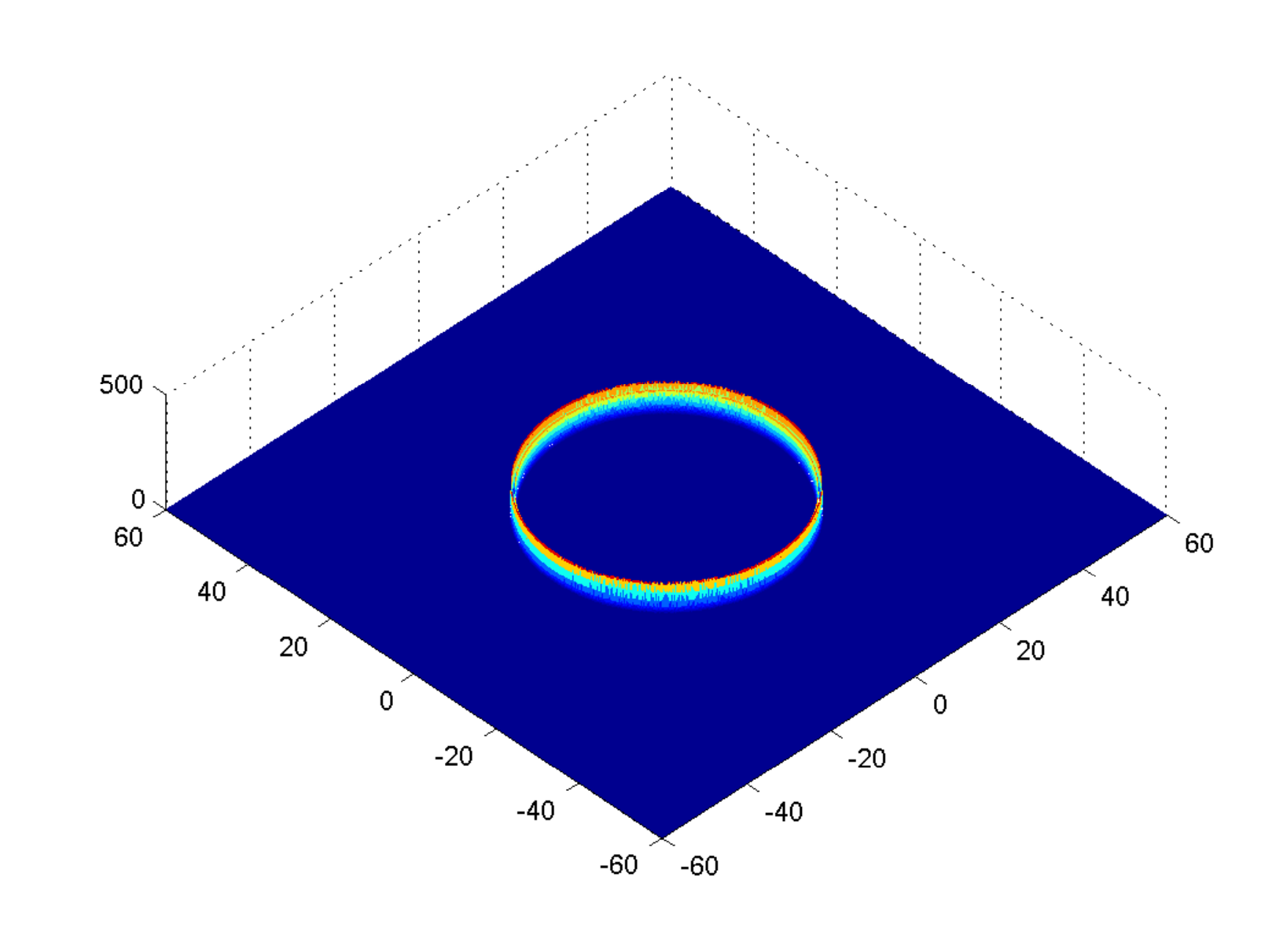}\label{fig:r25s02e10t1560}}
\subfigure[]{\includegraphics[trim = 0.5in 0.6in 0.6in 0.6in,
clip,width=0.65\columnwidth]{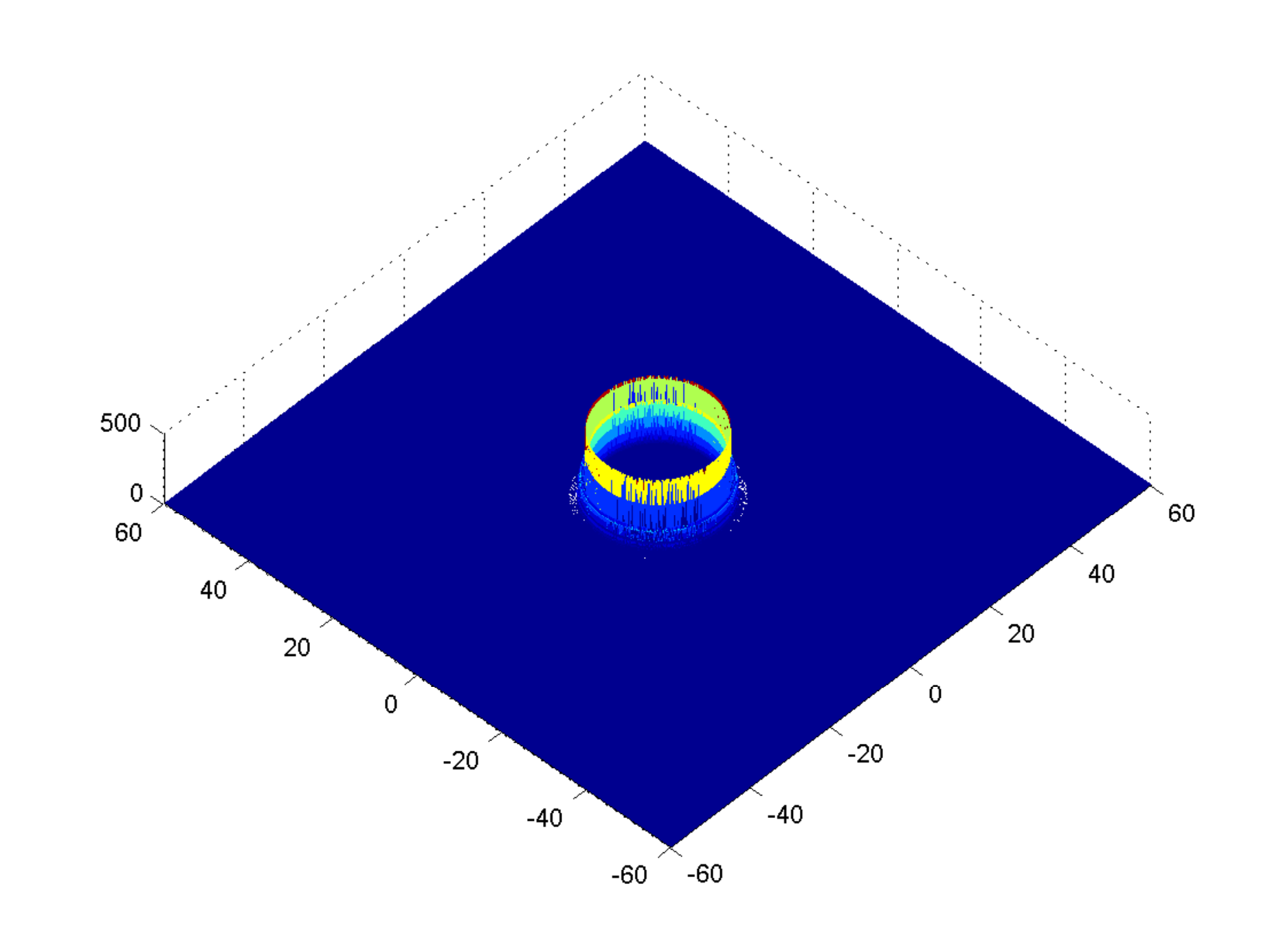}\label{fig:r25s02e10t2560}}
\subfigure[]{\includegraphics[trim = 0.5in 0.6in 0.6in 0.6in,
clip,width=0.65\columnwidth]{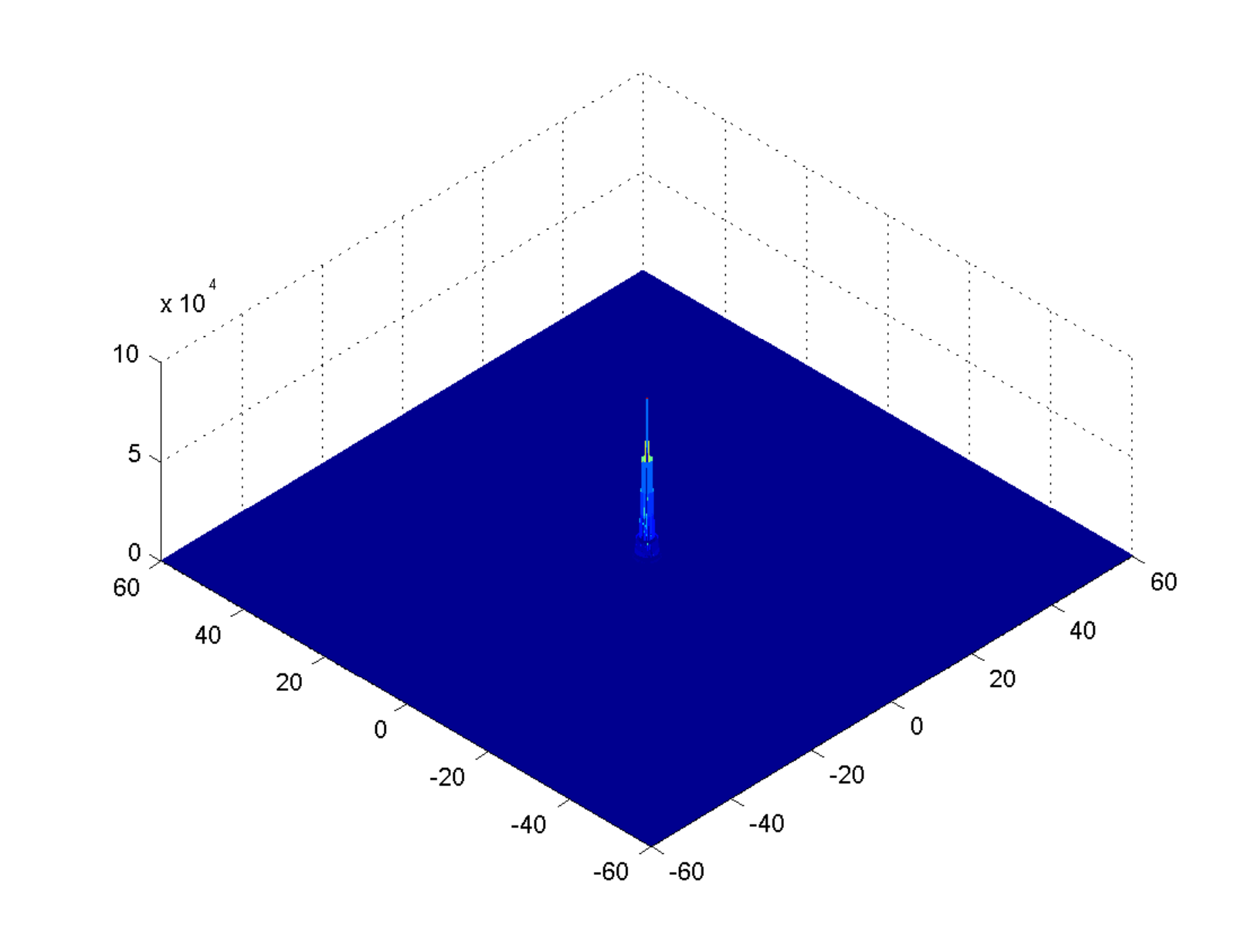}\label{fig:r25s02e10t3560}}
\subfigure[]{\includegraphics[trim = 0.5in 0.6in 0.6in 0.6in,
clip,width=0.65\columnwidth]{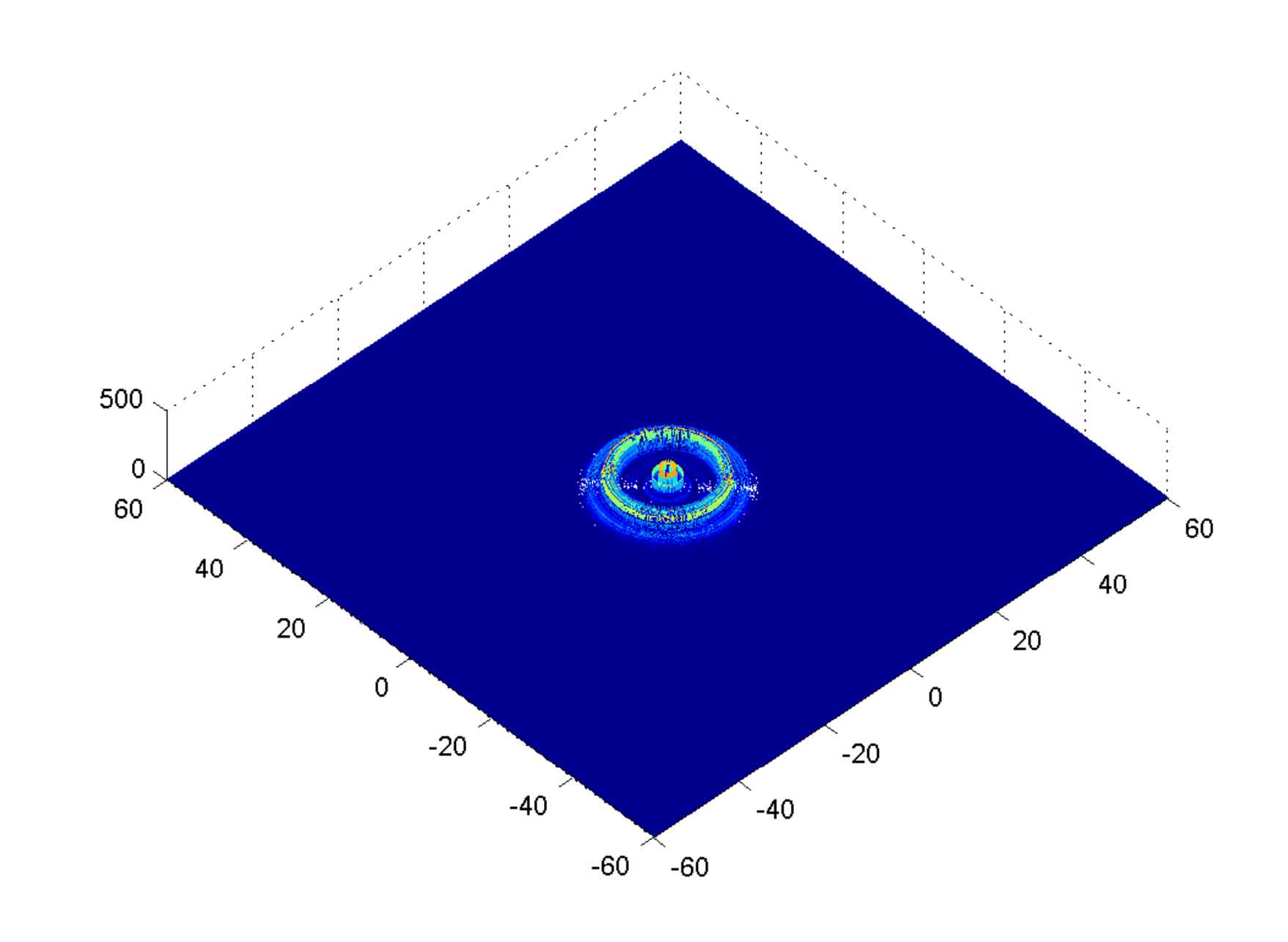}\label{fig:r25s02e10t4560}}
\subfigure[]{\includegraphics[trim = 0.5in 0.6in 0.6in 0.6in,
clip,width=0.65\columnwidth]{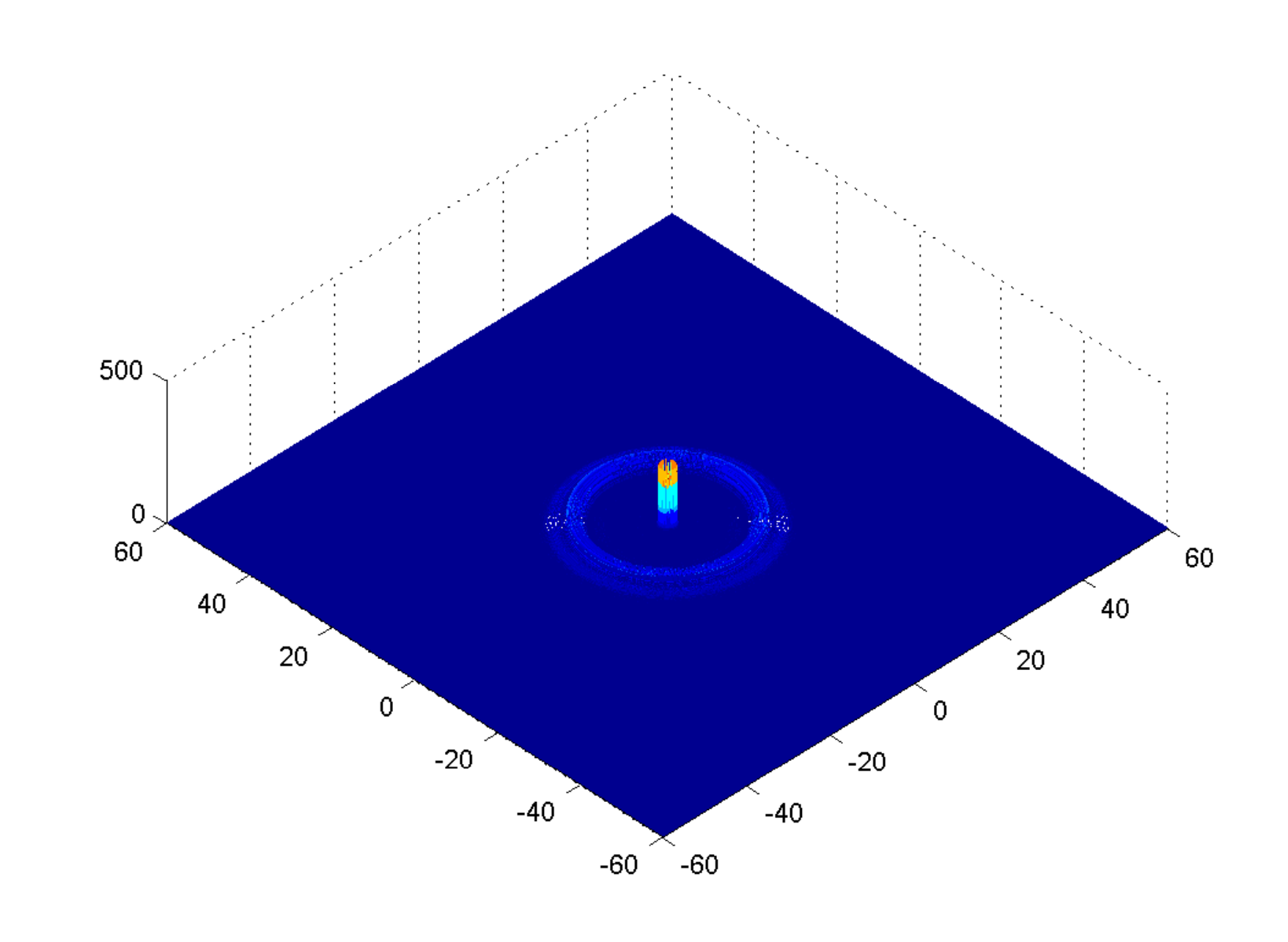}\label{fig:r25s02e10t5060}}
\caption{Evolution of the energy density of the DSG domain wall ($a=b=1$ and $\varepsilon=10$) from $t=10$
to $t=50$.}
\end{figure*}
%
%
\begin{figure*}
\subfigure[]{\includegraphics[trim = 0.6in 0.6in 0.6in 0.6in,
clip,width=0.65\columnwidth]{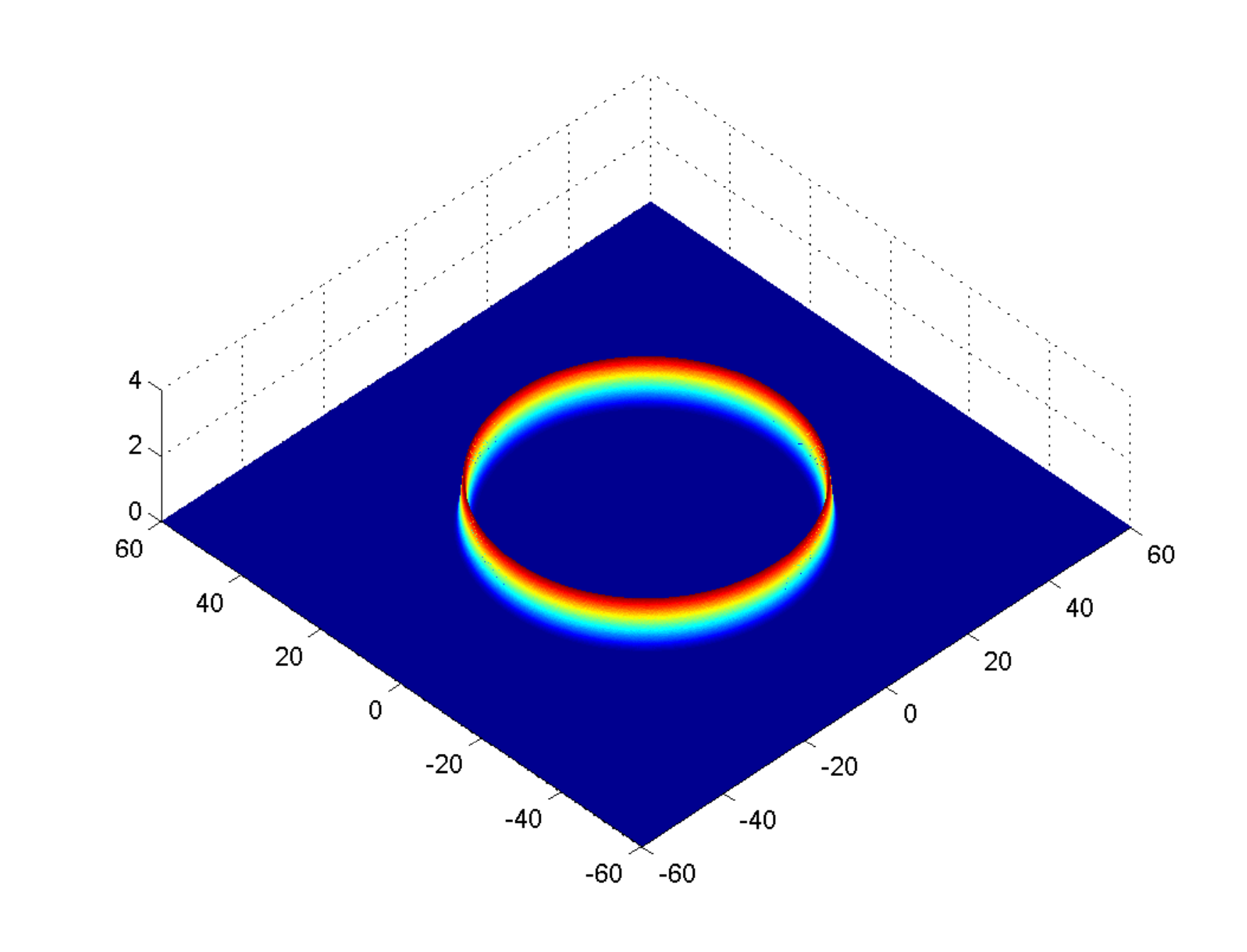}\label{fig:pr25s02t1060}}
\subfigure[]{\includegraphics[trim = 0.6in 0.6in 0.6in 0.6in,
clip,width=0.65\columnwidth]{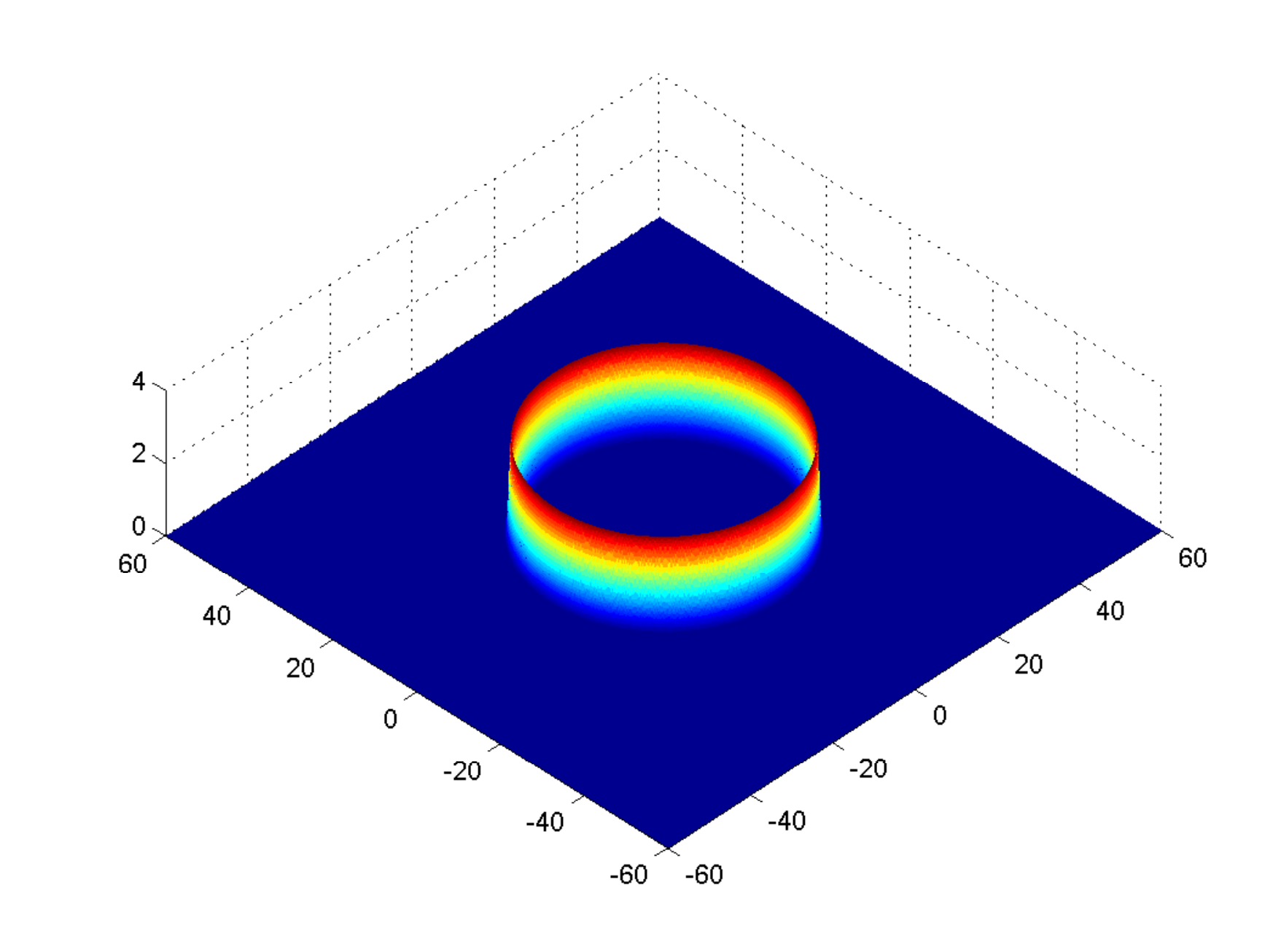}\label{fig:pr25s02t1560}}
\subfigure[]{\includegraphics[trim = 0.6in 0.6in 0.6in 0.6in,
clip,width=0.65\columnwidth]{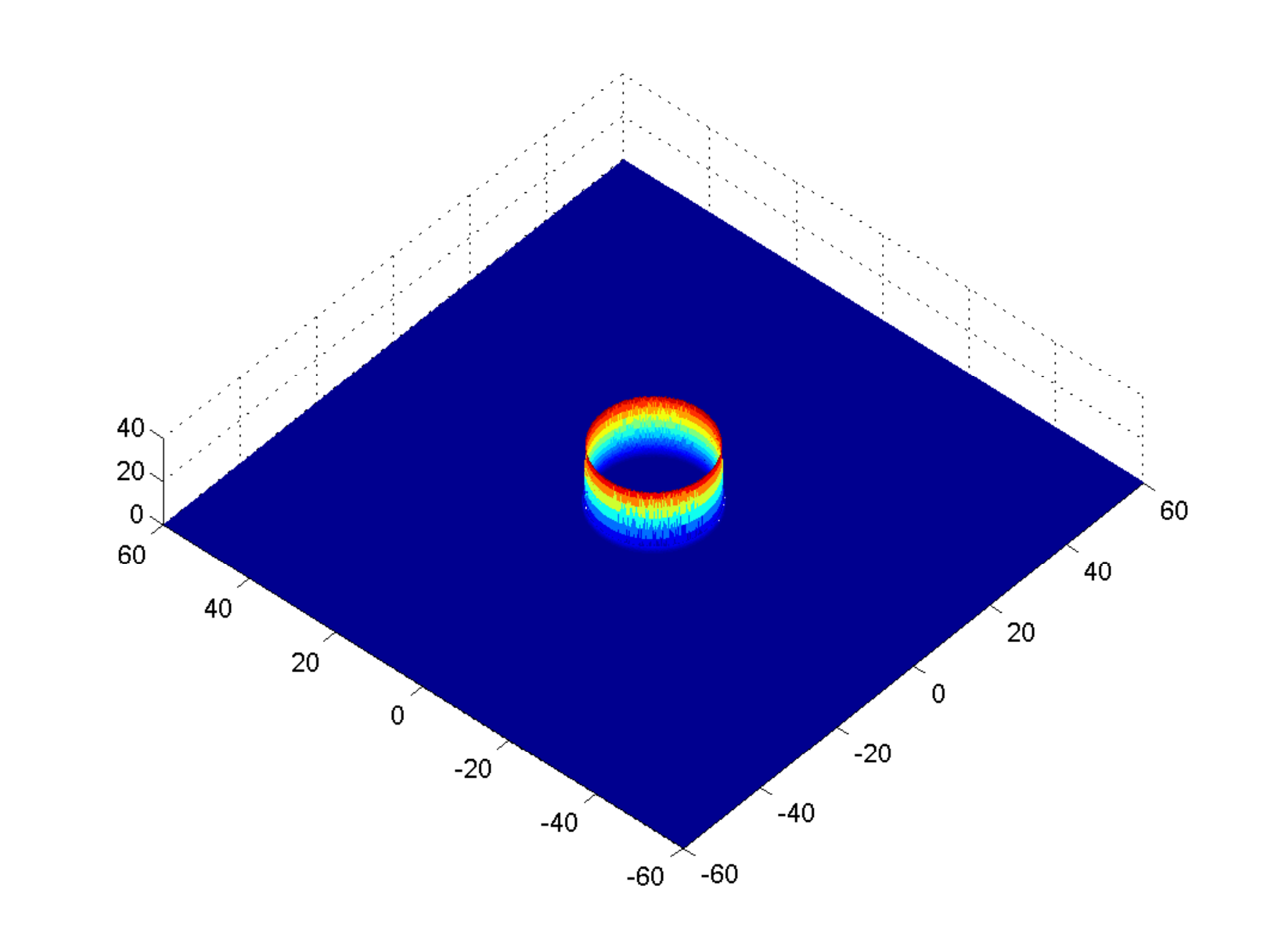}\label{fig:pr25s02t2560}}
\subfigure[]{\includegraphics[trim = 0.5in 0.6in 0.6in 0.6in,
clip,width=0.65\columnwidth]{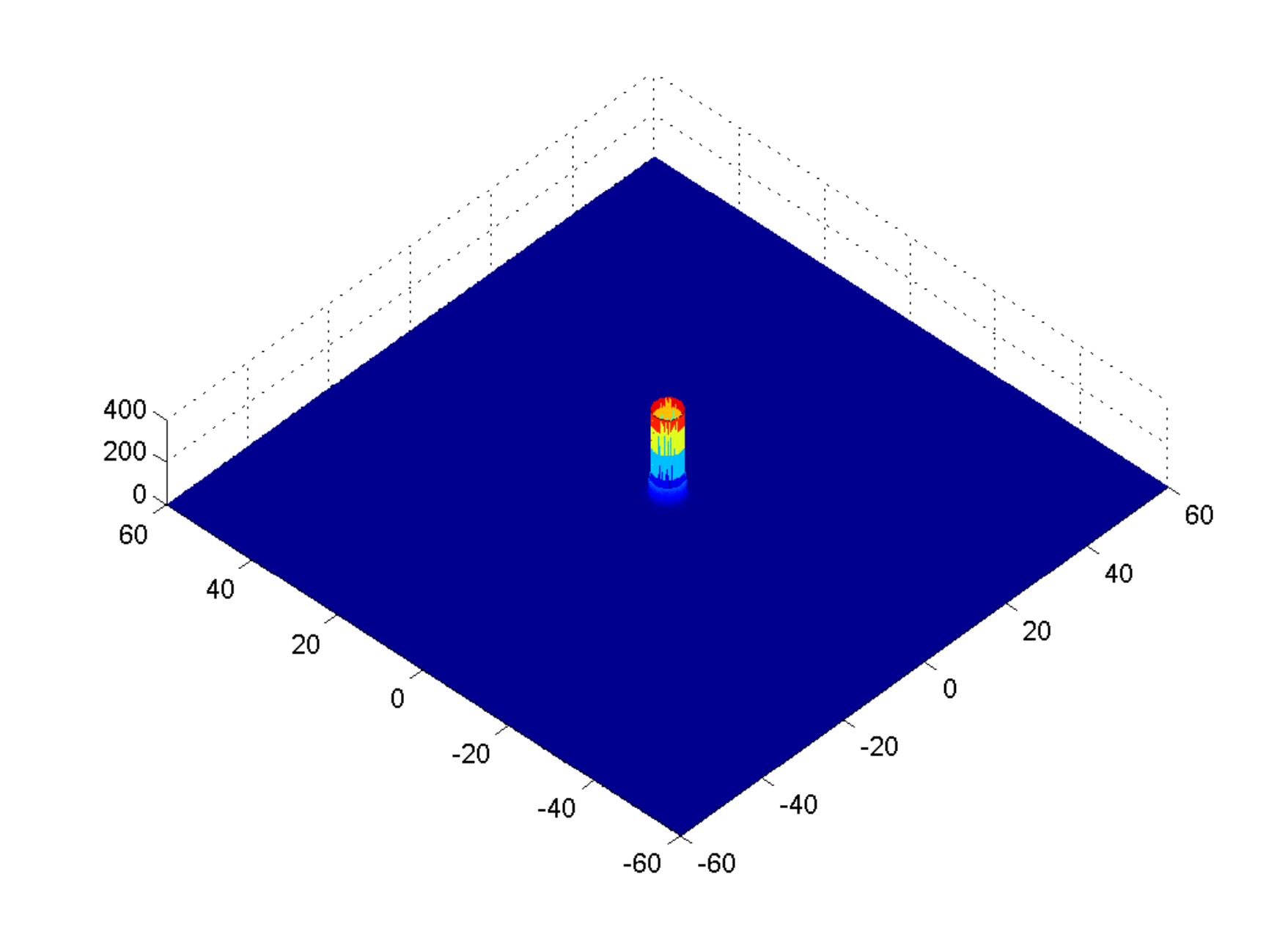}\label{fig:pr25s02t3560}}
\subfigure[]{\includegraphics[trim = 0.5in 0.6in 0.6in 0.6in,
clip,width=0.65\columnwidth]{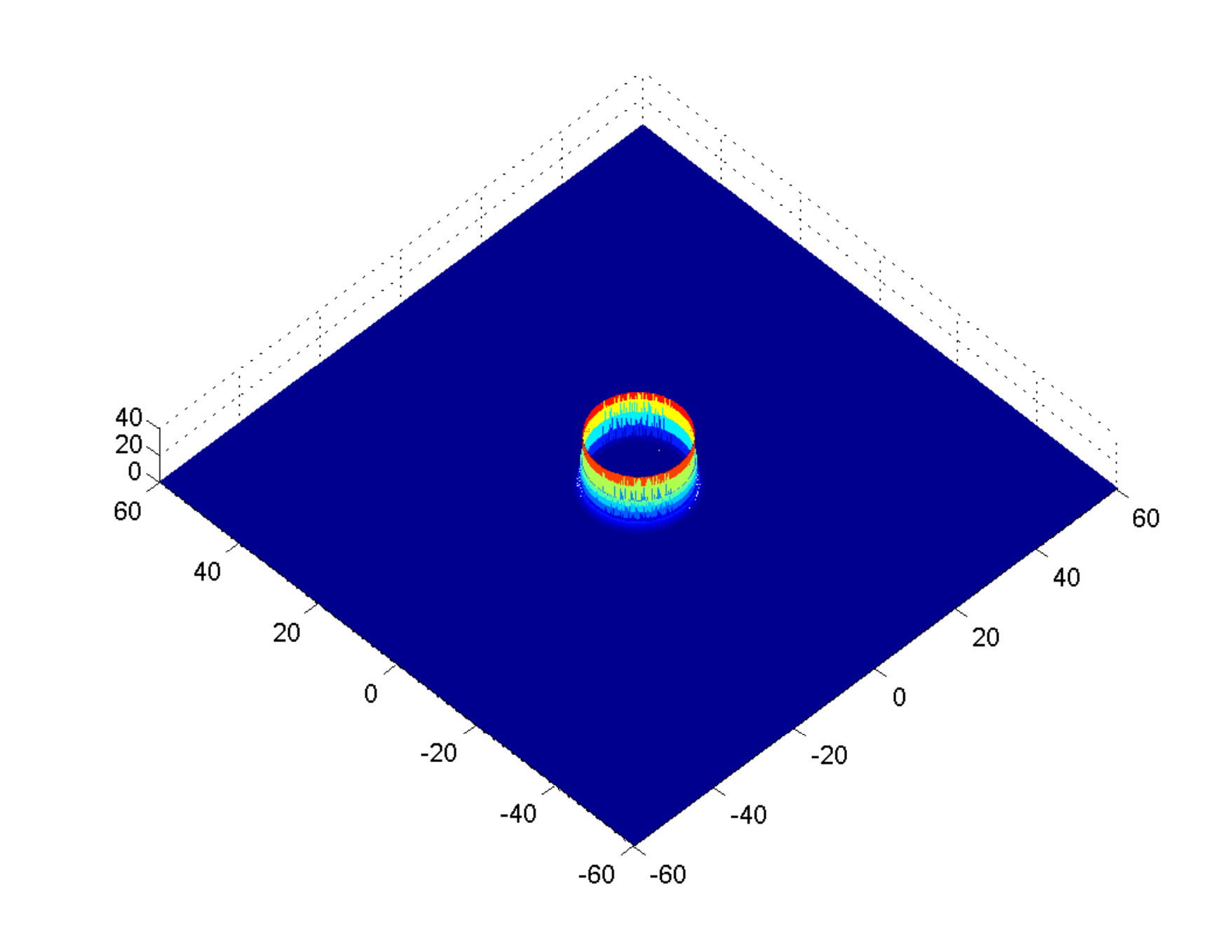}\label{fig:pr25s02t4060}}
\subfigure[]{\includegraphics[trim = 0.6in 0.6in 0.6in 0.6in,
clip,width=0.65\columnwidth]{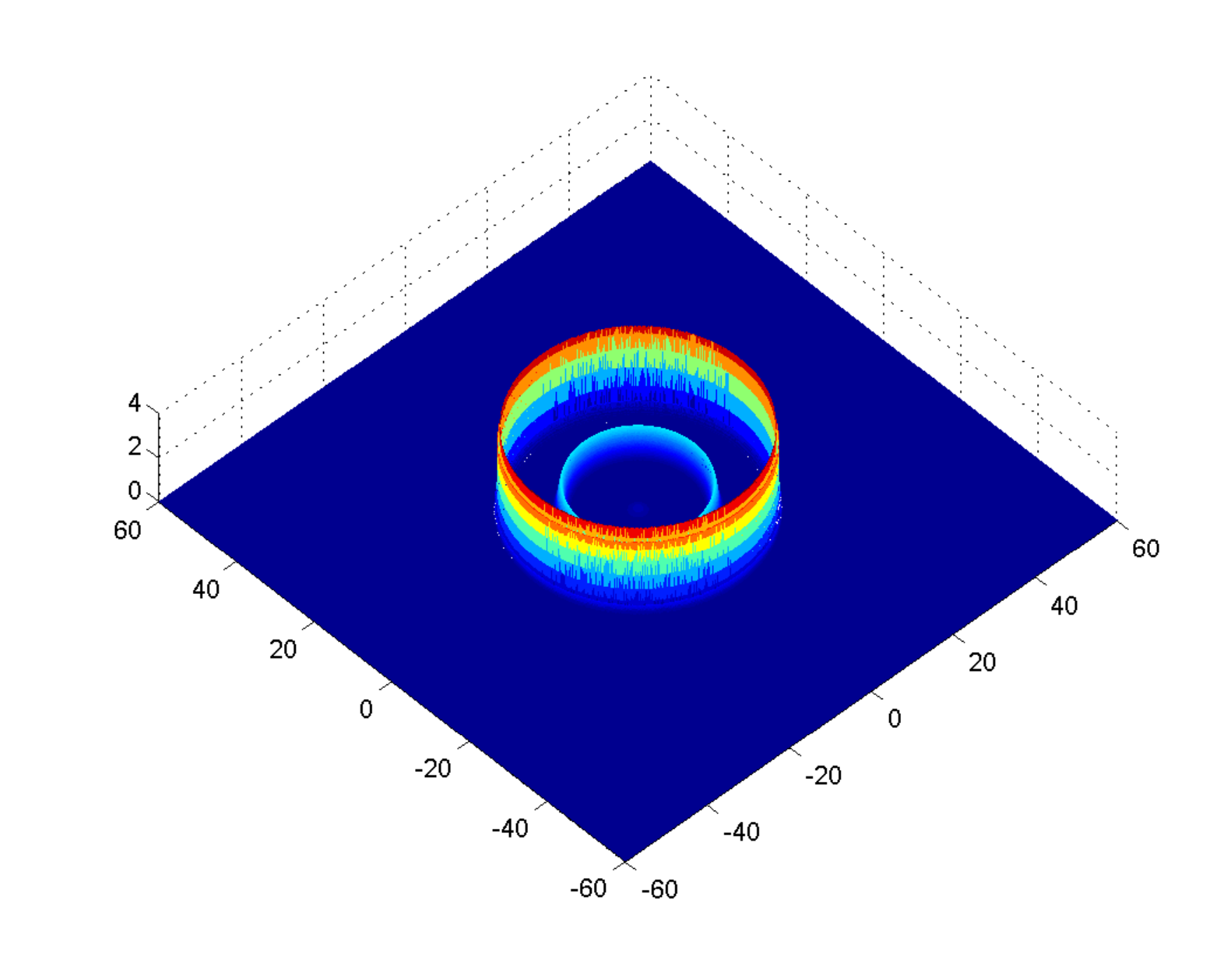}\label{fig:pr25s02t5060}}
\caption{Evolution of the energy density of the $\varphi^{4}$ domain wall ($\alpha=\beta=1$) from $t=10$ to $t=50$.}
\label{fig:phi4w}
 \end{figure*}
%
\begin{figure*}
\begin{center}
{\label{fig:fieldphi6w}
\includegraphics[width=0.48\textwidth,height=0.3\textheight]{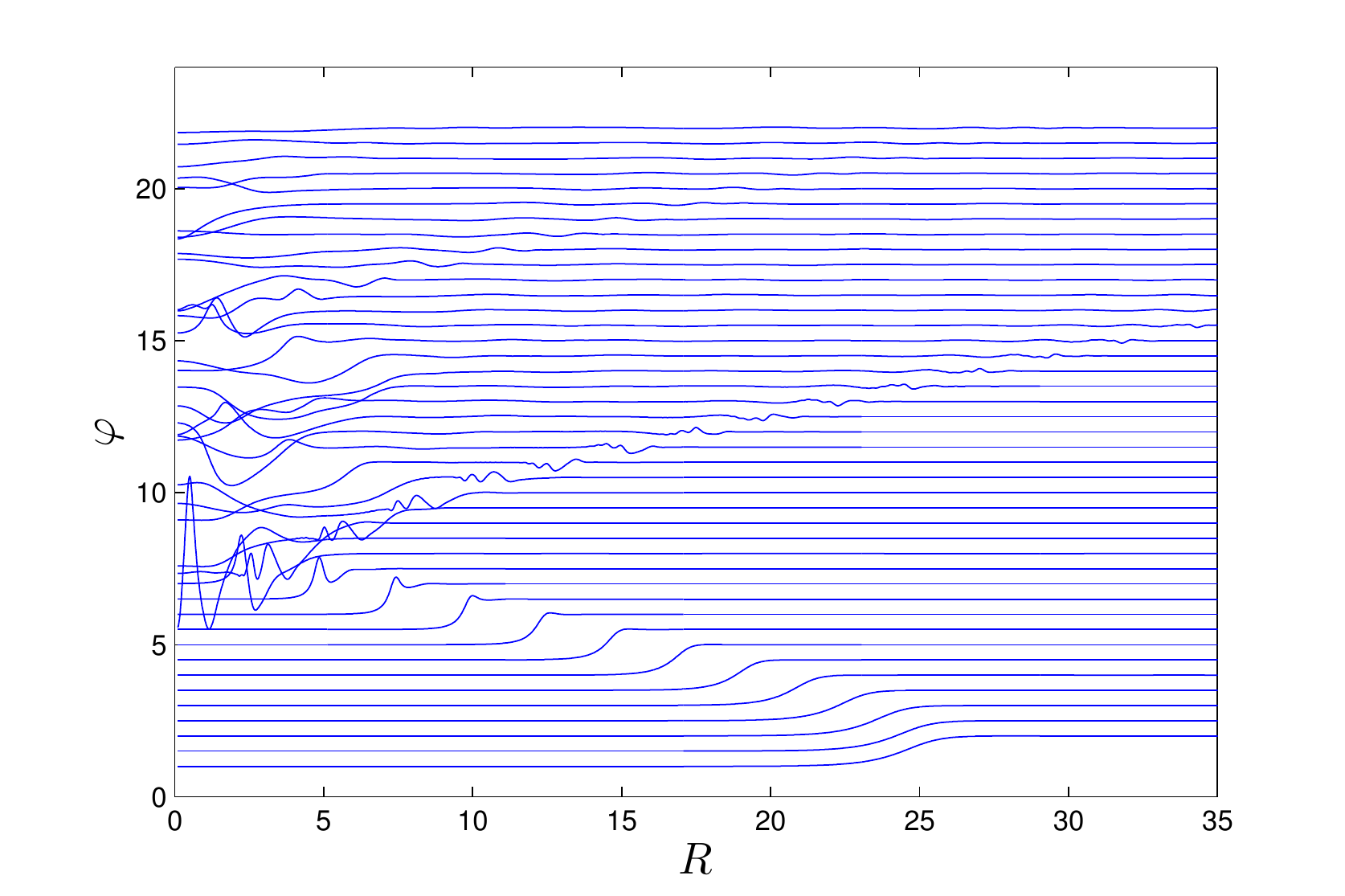}} \text{%
\hspace{0cm}}
{\label{fig:energydensityphi6w}
\includegraphics[width=0.48\textwidth,height=0.3\textheight]{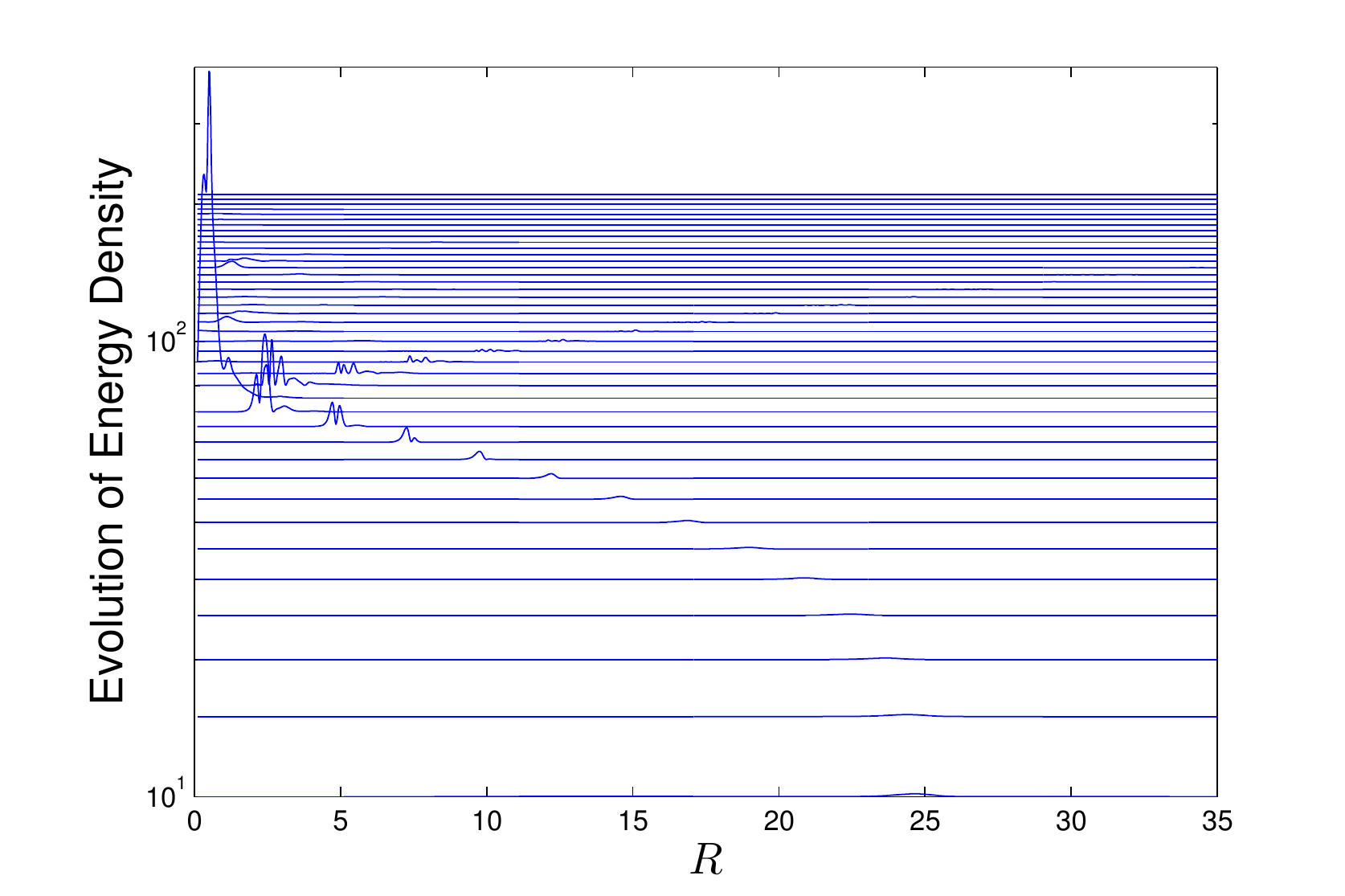}}
\end{center}
\caption{Evolution of the scalar field (left plot) and the
energy density (right plot) of the $\varphi^{6}$ domain wall, for the parameter values $\alpha=1$ and $\beta=\sqrt{2}$. See the text for more details.}
\label{fig:phi6w}
\end{figure*}

Spherical domain walls and bubbles may have been formed during inflation \cite{vink}, and in this case, they will have dynamical effects both on themselves and on the matter distribution in the universe. In some cases, the expansion and the collapse of the spherical domain walls is due to the difference in energy density in the interior and exterior of the bubble \cite{vink}. The collapse of thin spherical domain walls in an expanding background is explored in \cite{Va}. However, in this work, we consider domain wall creation in the symmetron inspired model based on the above-mentioned potentials in a flat spacetime background, and where the force which leads to the collapse of the domain wall is solely due to the self-interaction of the bubble. As shown in the previous section, the velocity of the collapsing wall approaches the speed of light as the bubble approaches the center. Likewise, the collision between the kink and antikink pair near the center occurs at velocities of approximately the speed of light. Moreover, the  kink and antikink scatter each other leading to an expanding bubble and the emission of scalar waves.

Consider, for instance, the $\varphi^{4}$ system, where there are several approaches to calculate the interaction energy between a kink-antikink pair. One of them consists in calculating the energy of the static solution based on the Dirac delta function \cite{man}. Moreover, it is possible to calculate this energy through the interaction force between solitons on the semi infinite interval with the rate of change of the momentum given by \cite{man,Va}
\begin{equation}
P=-\int_{-\infty}^{b}\dot{\varphi}\varphi^{\prime} dx,
\end{equation}
where $b$ is the end point of the interval, and lies between the kink and antikink.
Accordingly, the force is given by
\begin{eqnarray}
F=\dot{P}
&=&\left[-\frac{1}{2}\left(\dot{\varphi}^{2}+\varphi^
{\prime2}\right)+V(\varphi)\right]^{b}_{-\infty}.
   \label{fo}
\end{eqnarray}
This equation shows that it is possible to calculate the force based on the difference between the pressure at the end points
\cite{man,Va}.

In addition to this, one can calculate the interaction force between the kink and antikink, by considering the asymptotic form of the solitonic solutions in Eq. (\ref{fo}) \cite{man}:
\begin{equation}
F=32e^{-2R}=\frac{dE_{int}}{dR},
\end{equation}
where $R=2a$ and $a$ ($-a$) represents the position of the kink (antikink) and $-a\ll b\ll a$ \cite{man}. Then the interaction energy is giving by \cite{man}
\begin{equation}
E_{int}=-16e^{-2R}.
\end{equation}
This equation is compatible with the numerical simulations and it means that when separated, the static kink and antikink pair, which are located near each other, start to move toward each other and annihilate into radiation \cite{man}.

Here, we present the numerical results for the spherical
collapse in the SG, DSG (for $\varepsilon=10$), $\varphi^{4}$ and $\varphi^{6}$ systems.
These evolving solutions are obtained by numerically solving the dynamical field
equations in spherical coordinates. To this end, for each model we use the static approximation, namely Eqs. (\ref{16}), and the corresponding boundary conditions as initial conditions.
Figures \ref{fig:SGw}-\ref{fig:phi6w} show the evolution of the energy
density  for these systems. The energy density is given by the $00$ component of the energy-momentum tensor in spherical coordinates\footnote{Note that this energy density is calculated in a coordinate system which is co-moving with
the domain wall surface.} \cite{Mukh}:
\begin{eqnarray}
T^{0}{}_{0}=\frac{1}{2}\left({\varphi^{\prime}}\right)^{2}+V(\varphi)
=\frac{1}{2}\left(\frac{\partial\varphi}{\partial r}\right)^{2}+V(\varphi).
\end{eqnarray}
The total energy is given by $E=\int T^{0}{}_{0} \, d\bar{V}$,
where $d\bar{V}$ is the volume element of the spherical domain wall.
One can interpret $E$ as the rest mass of a domain wall \cite{man}.

Our results show that solutions of the four systems have a similar general behavior, but differ in some specific details. For instance, as depicted in Fig. \ref{fig:SGw}, the SG bubble starts to collapse from $t=0$ and $R=25$ to $R\simeq 0$ at $t=35$ (dimensionless units). During this process, the energy density of the domain wall increases steadily, reaching a factor of 30 at $t=35$. The fact that the bubble starts to collapse (instead of expanding) can be understood in terms of the tension of the bubble surface. Of course, when the radius of the bubble becomes comparable to the thickness of the wall, the kink-antikink interaction becomes important (note that the two facing sides of the bubble behave like a kink-antikink pair). As we indicated in the previous section, the final collapse near the center of the bubble will occur at very high velocities, i.e., near to the speed of light. After full collapse, the bubble starts to oscillate, while radiating scalar waves. As can be seen from the figure, after full expansion, the bubble radius will never reach the same initial bubble radius again. These results are in agreement and consistent with the analysis carried out in \cite{Va,thick}.

It can be seen from the plots in Figs. \ref{fig:r25s02e10t1060}-\ref{fig:r25s02e10t5060} for the DSG system that the bubble is double-layered due to the presence of sub-kinks. After reaching the center, the bubble of the DSG system develops a sharp peak of energy density which is relatively long lived. This peak results from the strong interaction between the two sub-kinks (the two layers of the bubble). For this case, it can be shown that the rate of scalar radiation is stronger than that of the SG system Figs. \ref{fig:r25s02e10t1060}-\ref{fig:r25s02e10t5060}. Similar results for the  $\varphi^{4}$ and $\varphi^{6}$ systems are shown in Figures \ref{fig:phi4w} and \ref{fig:phi6w}. However, in these cases, more energy is radiated away in the form of spherical waves, in agreement with the results in the literature. Numerical results are compared with the simple analytical models of the previous section in Figure \ref{vr}.

It is interesting to note that the SG equation is the best sample of the completely integrable system in 1+1 dimensions \cite{Va,1304}. In other words, when two solitons of this system collide with each other, they completely pass or scatter from each other without any dissipation in energy. However, based on the interaction force between them, this process will happen with a time delay \cite{man,Va}, which can be interpreted as a phase shift \cite{Va}. On the other hand, there is another option for a kink and antikink pair, which is a {\it breather}\footnote{For kink bearing nonlinear systems like the SG system, the breather is a solution which can be visualized as a kink and antikink bound together and oscillating about their common center of mass \cite{Va}.} \cite{Va}. But, it should be emphasized that the occurrence of the above-mentioned cases critically depend on the collision velocity \cite{man,Va}. For non-integrable systems like the $\phi^4$, the formation of a breather is predictable at low collision velocities and the kink-antikink scattering will occur at high velocities, although the exact behavior is quite complicated \cite{man,Va}. Note that in more than $1+1$ dimensions, the SG system is not integrable and the dissipation of energy is expected. These kinds of dissipation are interpreted as propagating excitations of small amplitude and they will appear as radiation in the system \cite{Va}.

In general, static domain walls and moving domain walls at constant speed do not emit any radiation, while deformed and accelerated domain walls can emit radiation. It is interesting to note that domain walls can be accelerated based on their own tension or due to some external force. For curved domain walls the radiation has been calculated numerically \cite{Va}.

The collision of solitons for the $\varphi^{4}$ system is more complicated in comparison with the previous cases. Since this system is not integrable, there is no possibility for two kinks to approach and as a result they interact strongly with each other \cite{Va}, which implies that this model only contains the scattering and annihilation of kinks-antikinks. Moreover, the kink-antikink collision is completely chaotic, in other words, even at high collision velocities, one can expect breather formations \cite{man,Va}. See the center of the bubbles in  Figs. \ref{fig:pr25s02t5060} and \ref{fig:phi6w} (and Fig. \ref{fig:al1r010pr25s02t1060} and \ref{fig:energydensityphi6}, given below). The numerical results obtained in this work are in agreement with these theoretical considerations.

\section{Spherical domain walls around matter overdensities}\label{wm}

\begin{figure*}
\begin{center}
{\label{fig:scalarSGmatter}
\includegraphics[width=0.45\textwidth,height=0.25\textheight]{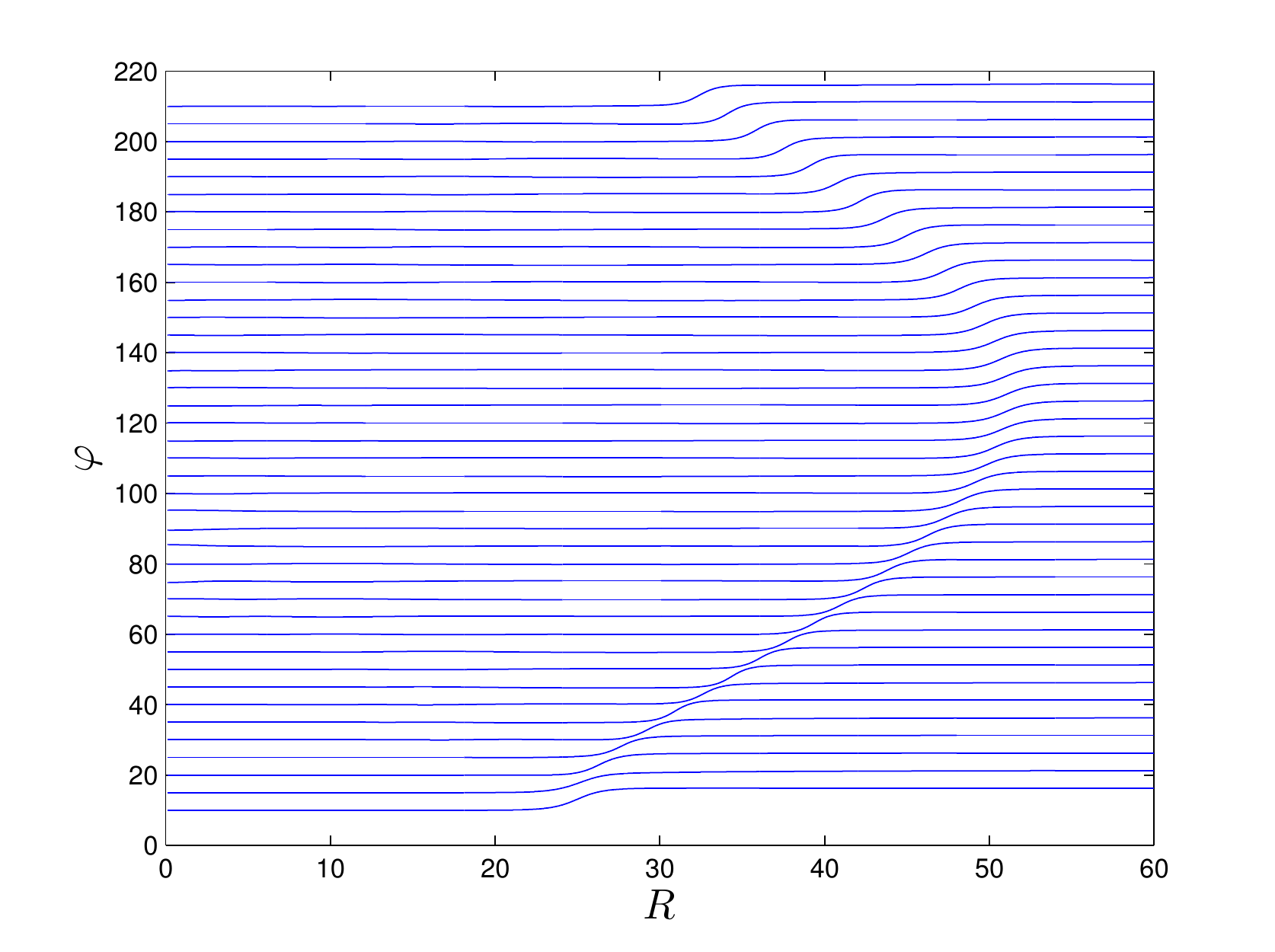}} \text{%
\hspace{0cm}}
{\label{fig:energySGmatter}
\includegraphics[width=0.45\textwidth,height=0.25\textheight]{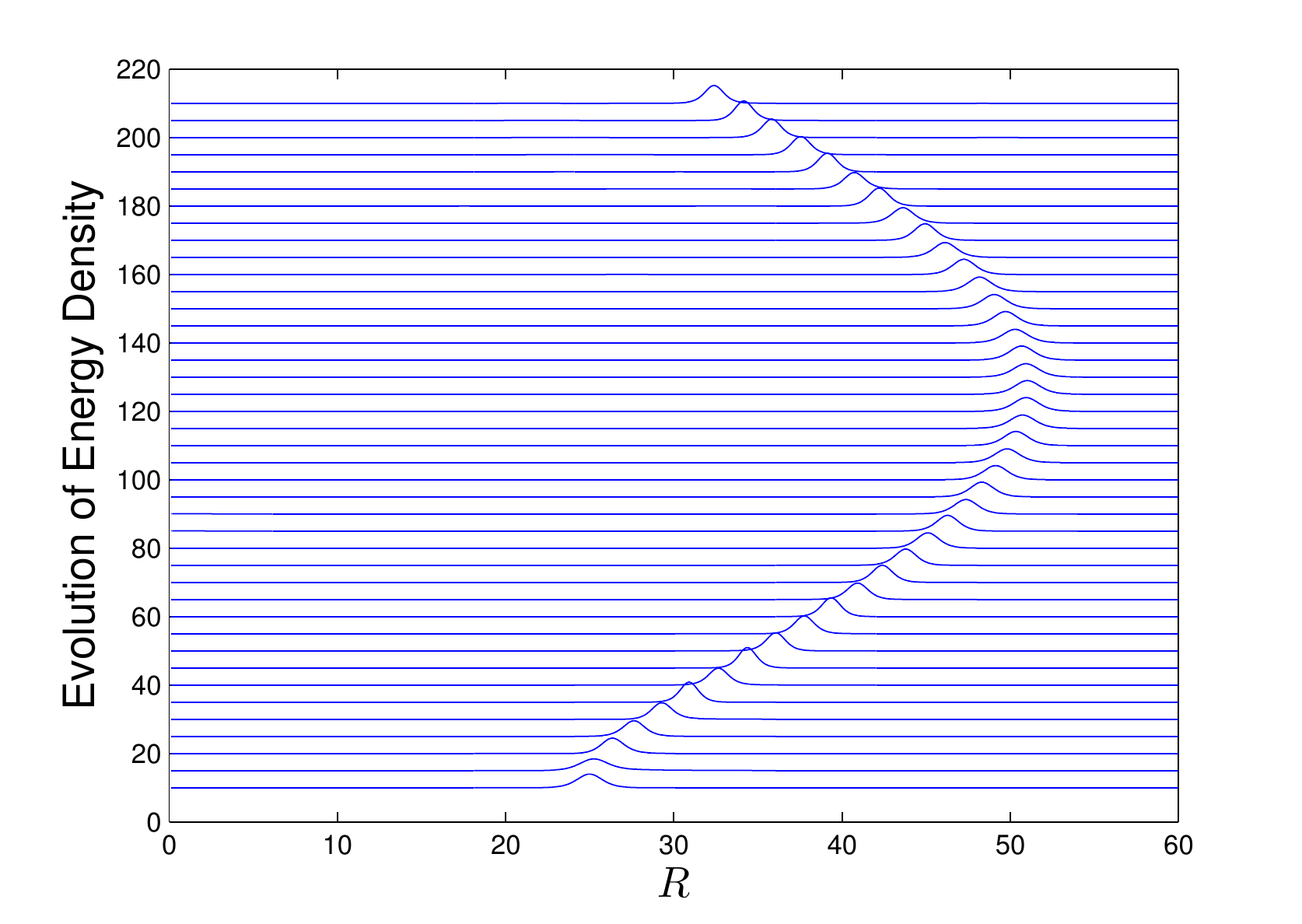}}
\end{center}
\caption{Evolution of the scalar field (left plot) and the energy density (right plot) of the SG domain wall with an internal matter density ($\rho=\rho_{0}e^{-r/r_{0}}$), and for the parameter values $a=b=1$. See the text for more details.}
\label{fig:SGmatter}
\end{figure*}
\begin{figure*}
\begin{center}
{\label{fig:scalarDSGn}
\includegraphics[width=0.48\textwidth,height=0.25\textheight]{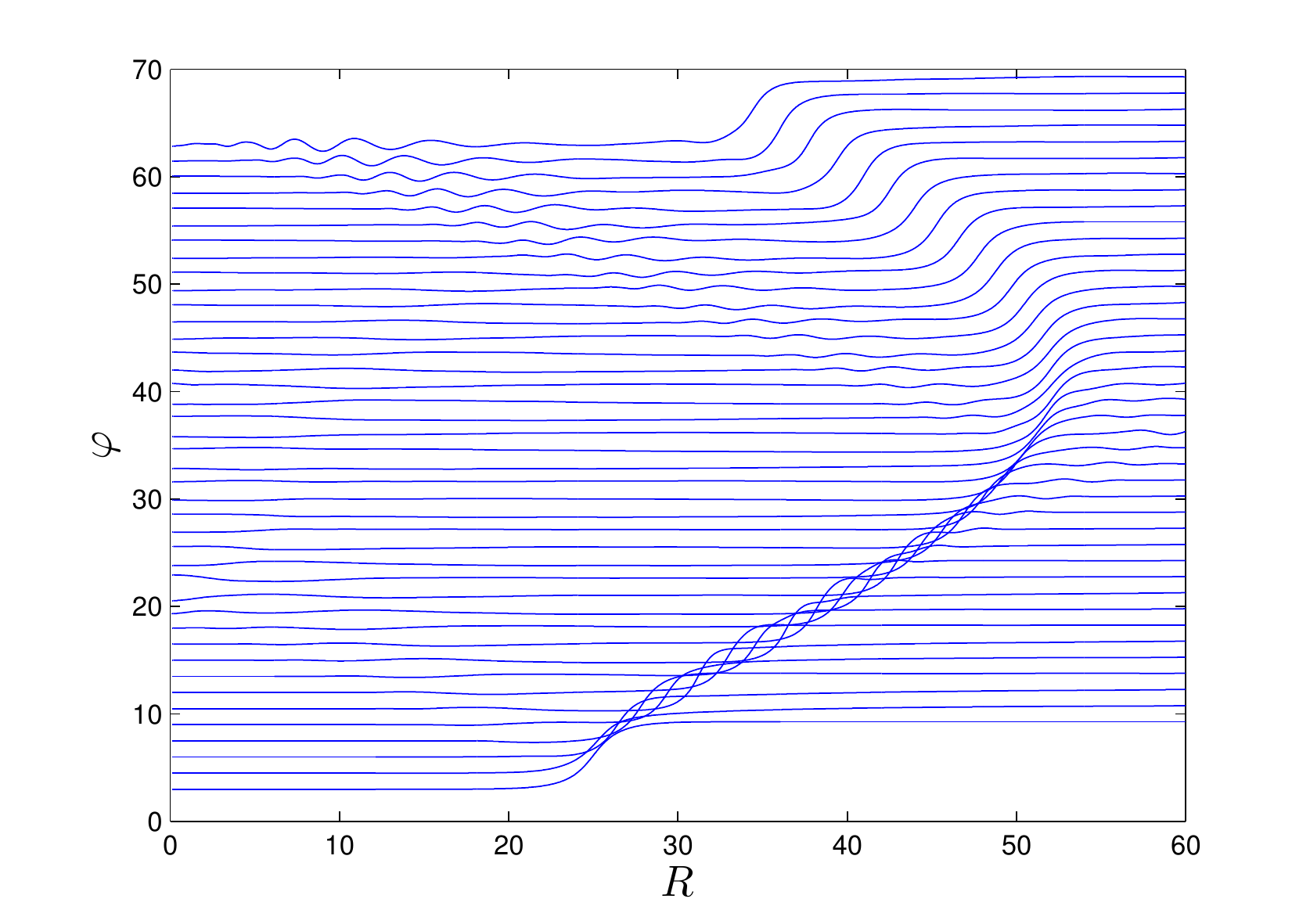}} \text{%
\hspace{0cm}}
{\label{fig:energyDSGn}
\includegraphics[width=0.48\textwidth,height=0.25\textheight]{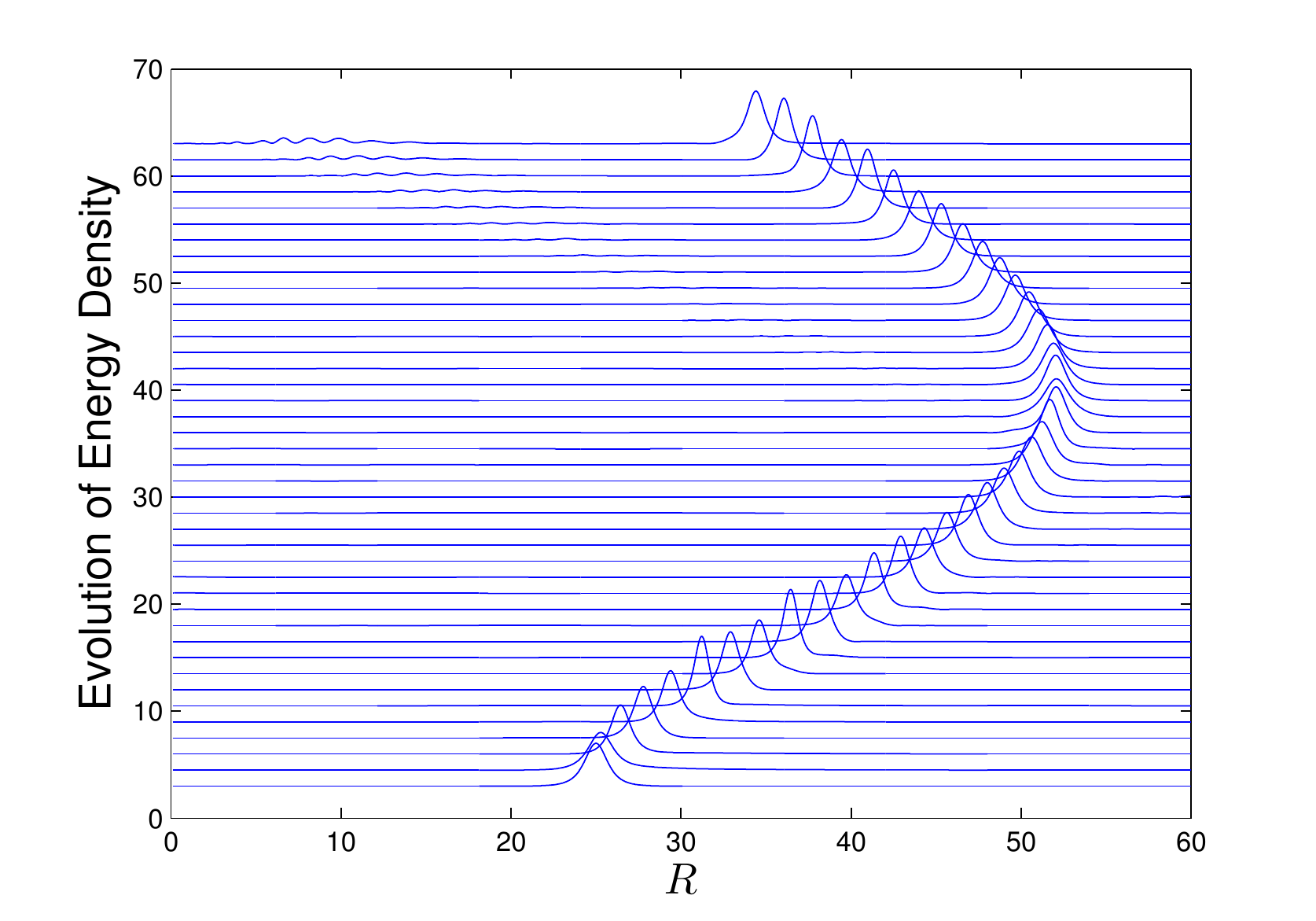}}
\end{center}
\caption{Evolution of the scalar field and the energy density of DSG domain wall with an internal matter density ($\rho=\rho_{0}e^{-r/r_{0}}$), with $\varepsilon=-0.1$  and for the parameter values $a=b=1$. See the text for more details.}
\label{fig:DSGn}
\end{figure*}
\begin{figure*}[tbp]
\begin{center}
{\label{fig:al1r010sr25s02t1060}
\includegraphics[width=0.48\textwidth,height=0.25\textheight]{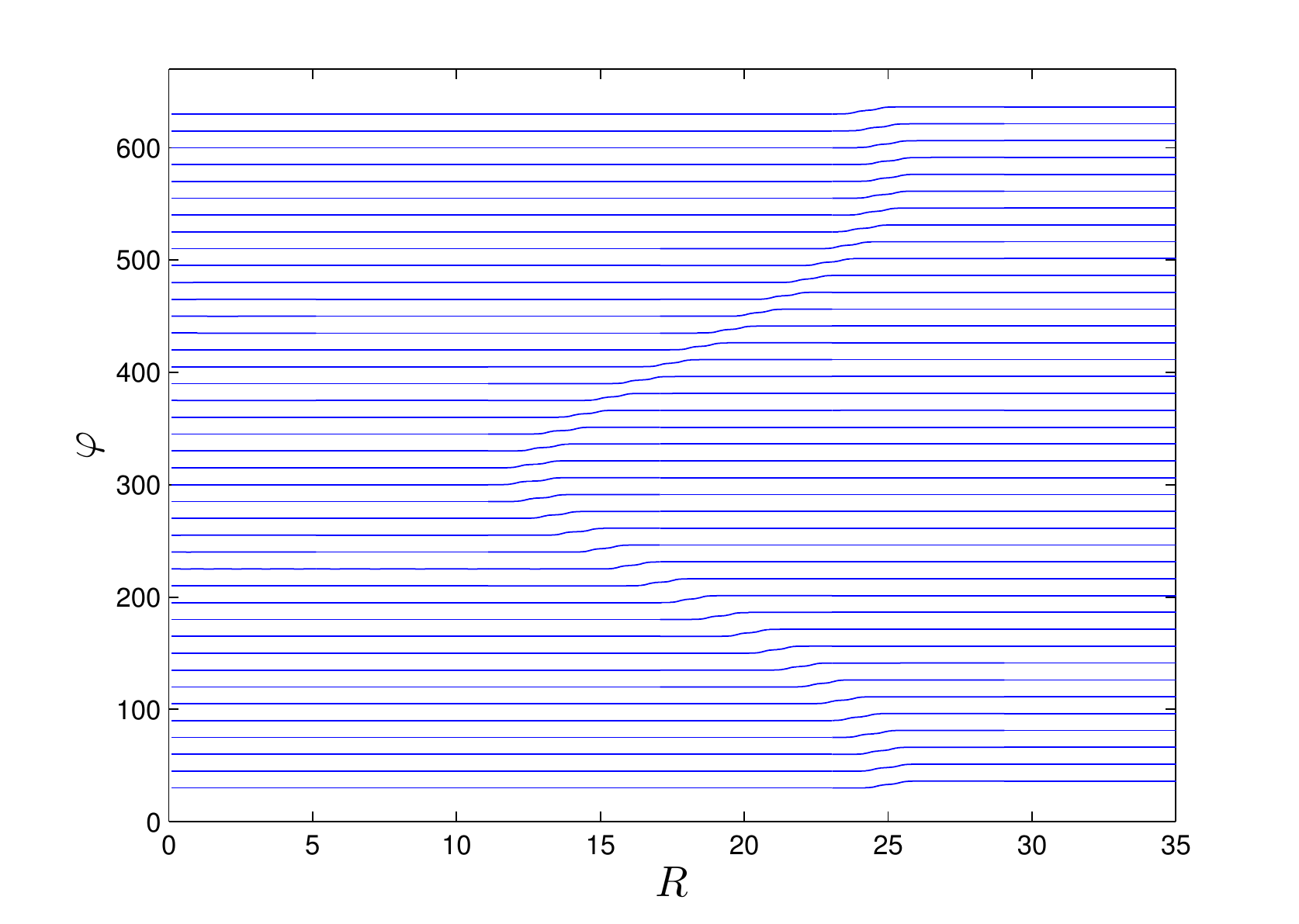}} \text{%
\hspace{0cm}}
{\label{fig:al1r010sr25s02t9060}
\includegraphics[width=0.48\textwidth,height=0.25\textheight]{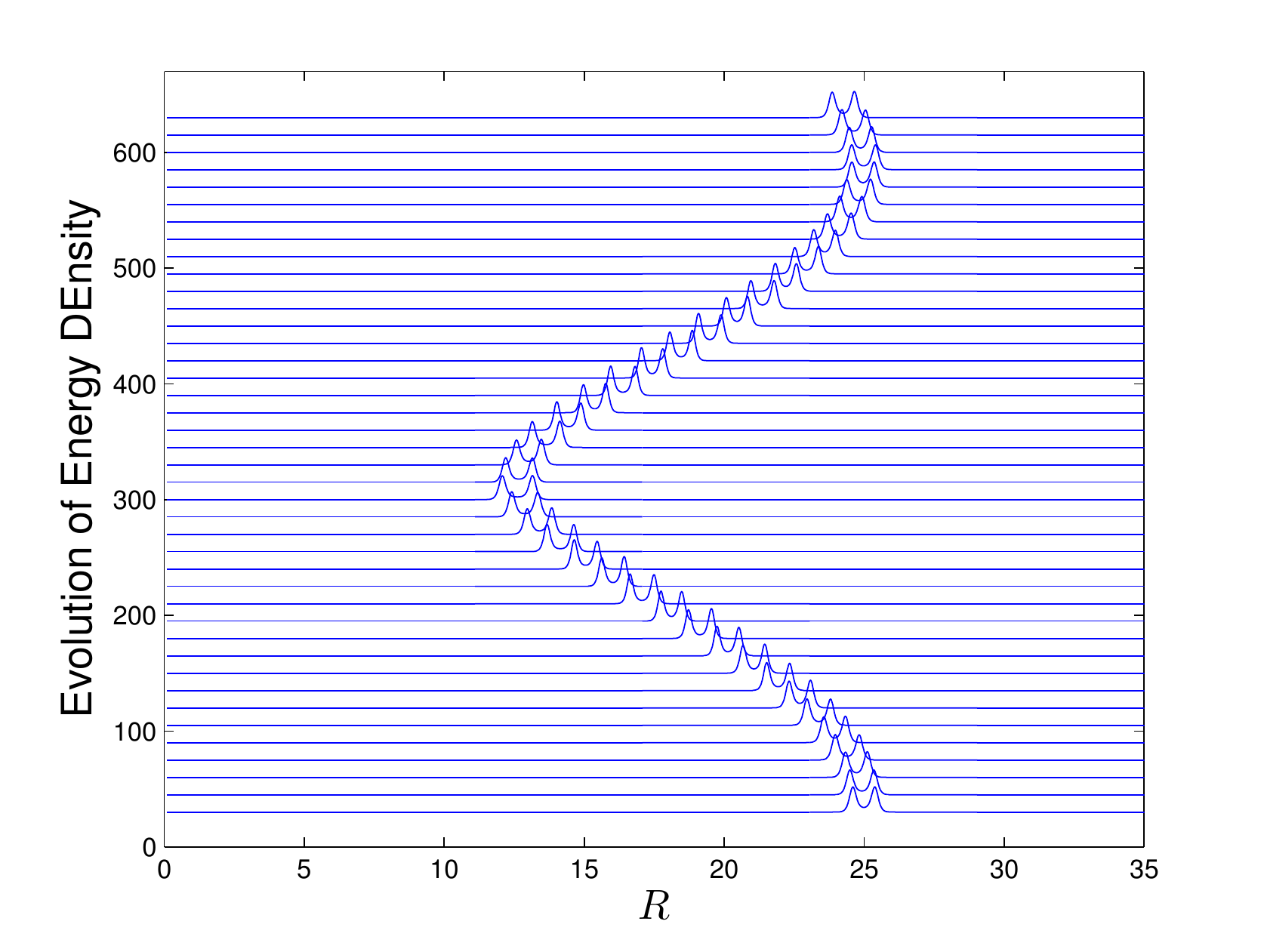}}
\end{center}
\caption{Evolution of the scalar field and the energy density of DSG  domain wall with an internal matter density ($\rho=\rho_{0}e^{-r/r_{0}}$) for $\varepsilon=10$ and for the parameter values $a=b=1$. See the text for more details.}
\label{fig:DSGp}
\end{figure*}

In this section, we consider the symmetron field in the presence of a central static matter lump, in order to investigate the effect of matter on the dynamics of the domain wall. In this case, due to the coupling with matter, the energy trapped in a bubble not only depends on the configuration of the symmetron field but also on the matter density, $\rho$ \cite{CL} . Moreover, for simplicity the boundary condition is taken as $\varphi(r\rightarrow\infty)=\varphi_{0}$ \cite{KJAA,Kh}. To model this situation, we choose the following matter density
\begin{equation}
\rho_{m}=\rho_{0}e^{-\frac{r}{r_{0}}},
\end{equation}
where $\rho_{0}$ is the central density and $r_{0}$ is a scale radius. Note that the matter density decreases by increasing the radius. In other words, the coupling between matter and the symmetron field is weak around the center, however, it can be perceptible around the surface \cite{KJAA, Kh}.

The system is now governed by the evolution equation:
\begin{equation}
\Box\varphi-V_{,\varphi}-\frac{\varphi}{M^{2}}\rho_{0}e^{-\frac{r}{r_{0}}}=0.
\end{equation}
In order to solve this equation numerically, we have set the strength of the fifth force in the symmetron model such that $\rho_{0}/M^{2} \approx \mathcal{O}(1)$.
We investigate the bubbles of the four above-mentioned systems around the central matter density. As seen in Fig. \ref{fig:SGmatter}, unlike the previous situation, the SG bubble starts expanding until it reaches a maximum radius and then re-contracts. It should be noted that in this process (expansion and contraction of the bubble), the peak energy density of the bubble remains almost constant and there is  little energy loss due to radiation. Unlike the results of the previous section, here we verify that in the presence of matter the symmetron domain wall is stable. This result is consistent with \cite{CL}.

In general, it is interesting to note that the gravitational field of the domain wall is completely different from a huge heavy plate.  In this regard, by writing the energy momentum tensor of the domain wall and considering the Newtonian limit of Einstein's equation for a static mass distribution, one can show that the gravitational effect of the domain wall is negative \cite{Vi,vink}. In the simulations performed here, the gravitational effects of the wall on itself and on the matter distribution have not been taken into account. In other words, we have not used the Einstein equations nor the Poisson equation in calculating the dynamics. So, the force between the matter and the wall is due to the non-minimal coupling between them, while the force which leads to the collapse of the domain wall is solely due to the self-interaction of the bubble as in the previous section.

In order to study DSG bubbles in the presence of the central matter density, we consider two different kinds of bubbles with $\varepsilon=10$ and $\varepsilon=-0.1$. Based on our earlier discussion in Section \ref{Sym}, it is clear that these bubbles should have different properties. While the bubble which is formed in the first case ($\varepsilon=10$) is double layered, the bubble of the second case ($\varepsilon=-0.1$) does not have any sub-layers. So, one can expect different behaviours for these bubbles due to the interaction between the sub-layers (or subkinks). Figures \ref{fig:DSGn} and \ref{fig:DSGp} show the evolution of the DSG bubble with two subkinks. Figure \ref{fig:DSGn} displays the evolution of the DSG bubble for the case $\varepsilon=-0.1$. In this situation, the domain wall behaves like the SG system with a central matter lump. However, in Fig. \ref{fig:DSGp}, for $\varepsilon=10$, the bubble starts to contract, reaches a minimum radius, and then expands.

%
\begin{figure*}
\subfigure[]{\includegraphics[trim = 0.6in 0.6in 0.6in 0.6in,
clip,width=0.65\columnwidth]{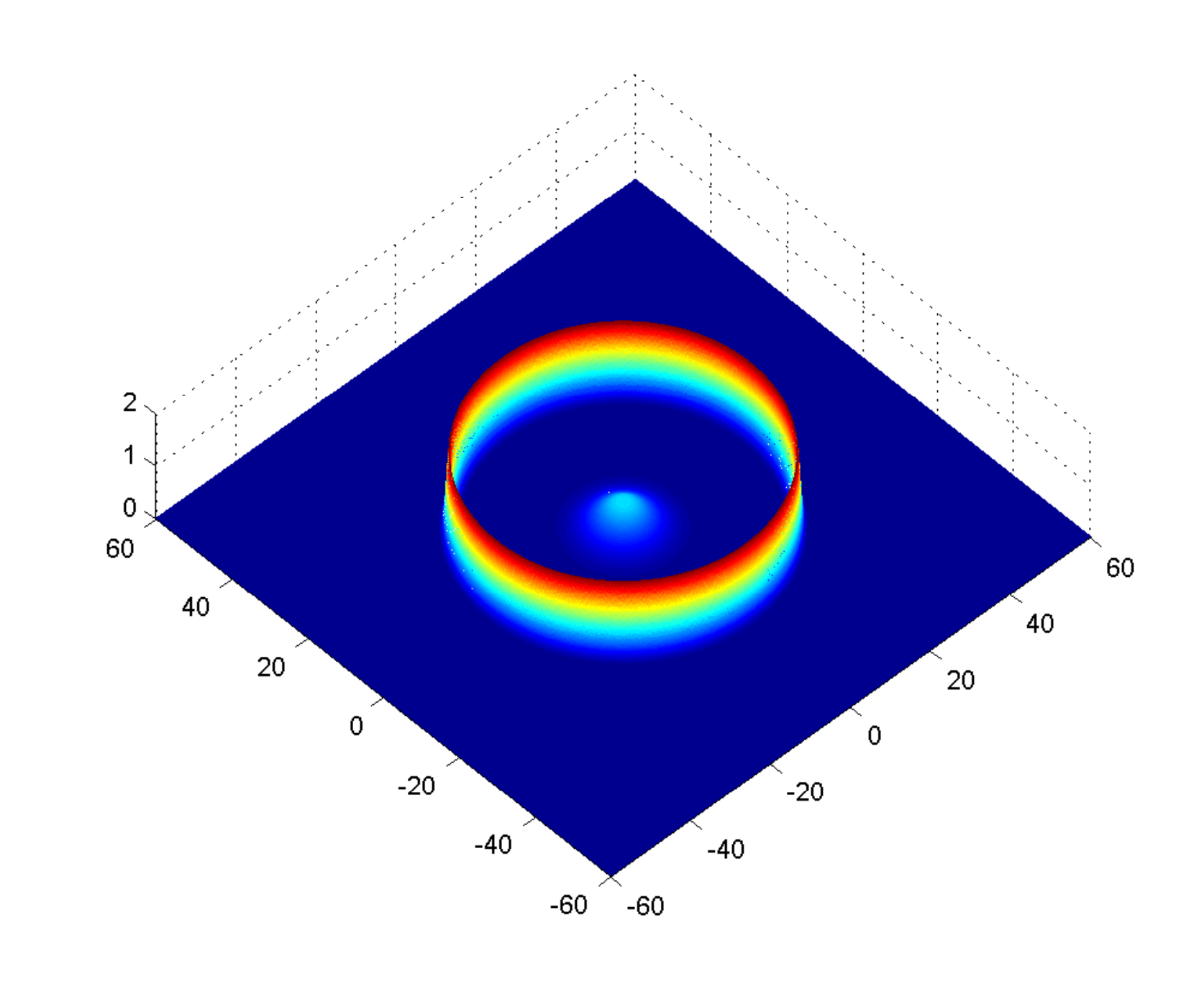}\label{fig:al1r010pr25s02t1060}}
\subfigure[]{\includegraphics[trim = 0.6in 0.6in 0.6in 0.6in,
clip,width=0.65\columnwidth]{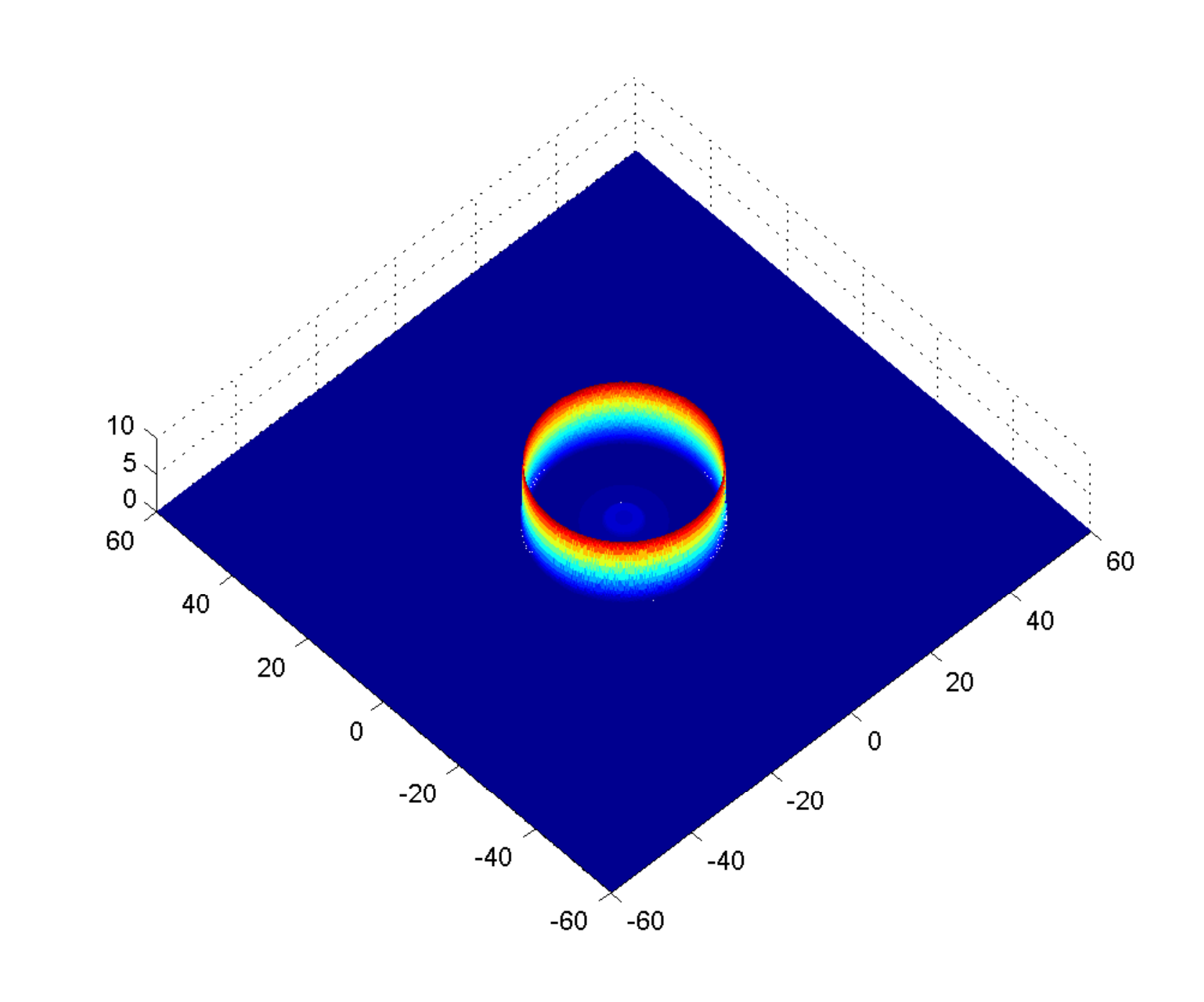}\label{fig:al1r010pr25s02t2060}}
\subfigure[]{\includegraphics[trim = 0.5in 0.6in 0.6in 0.6in,
clip,width=0.65\columnwidth]{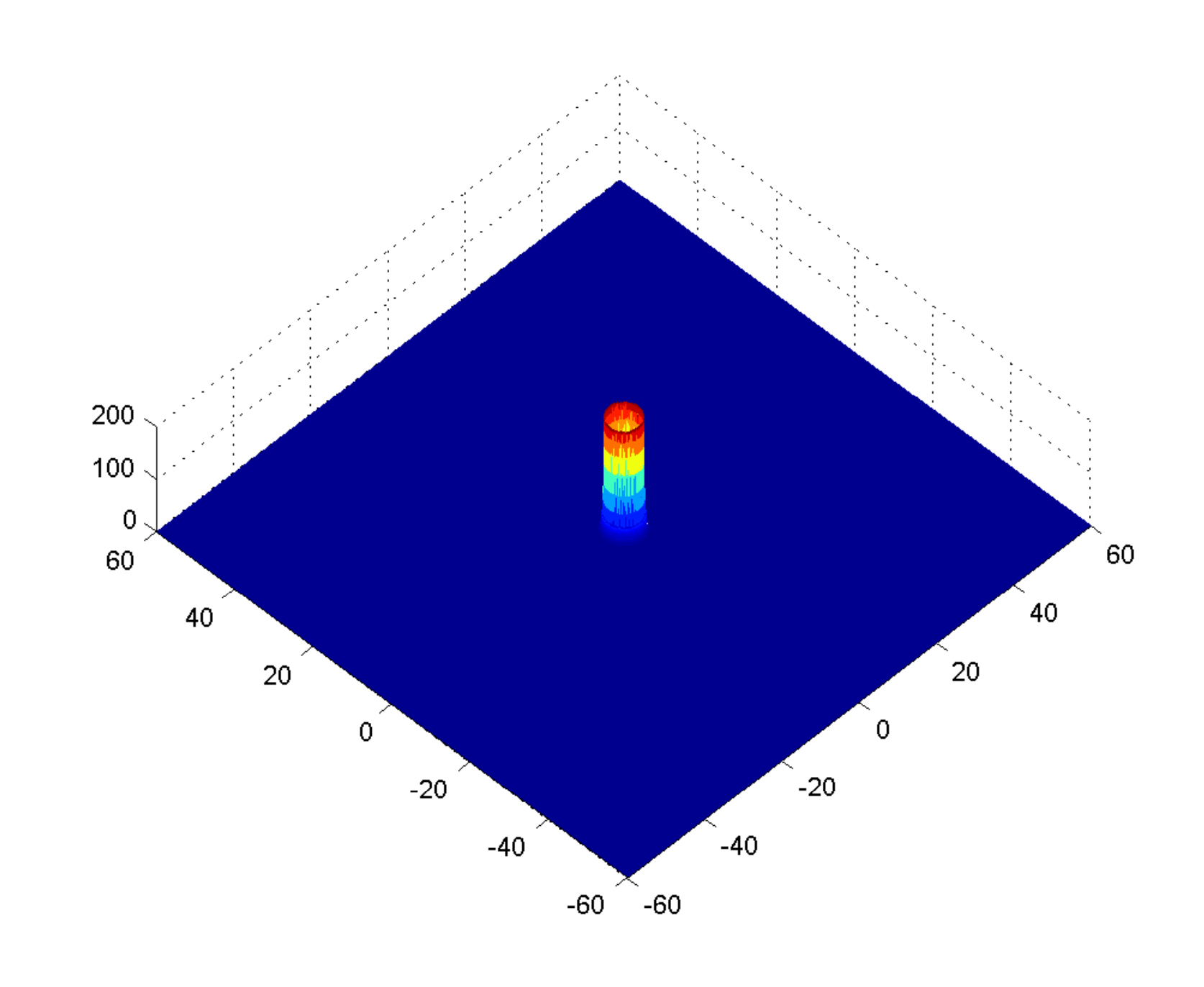}\label{fig:al1r010pr25s02t3060}}
\subfigure[]{\includegraphics[trim = 0.5in 0.6in 0.6in 0.6in,
clip,width=0.65\columnwidth]{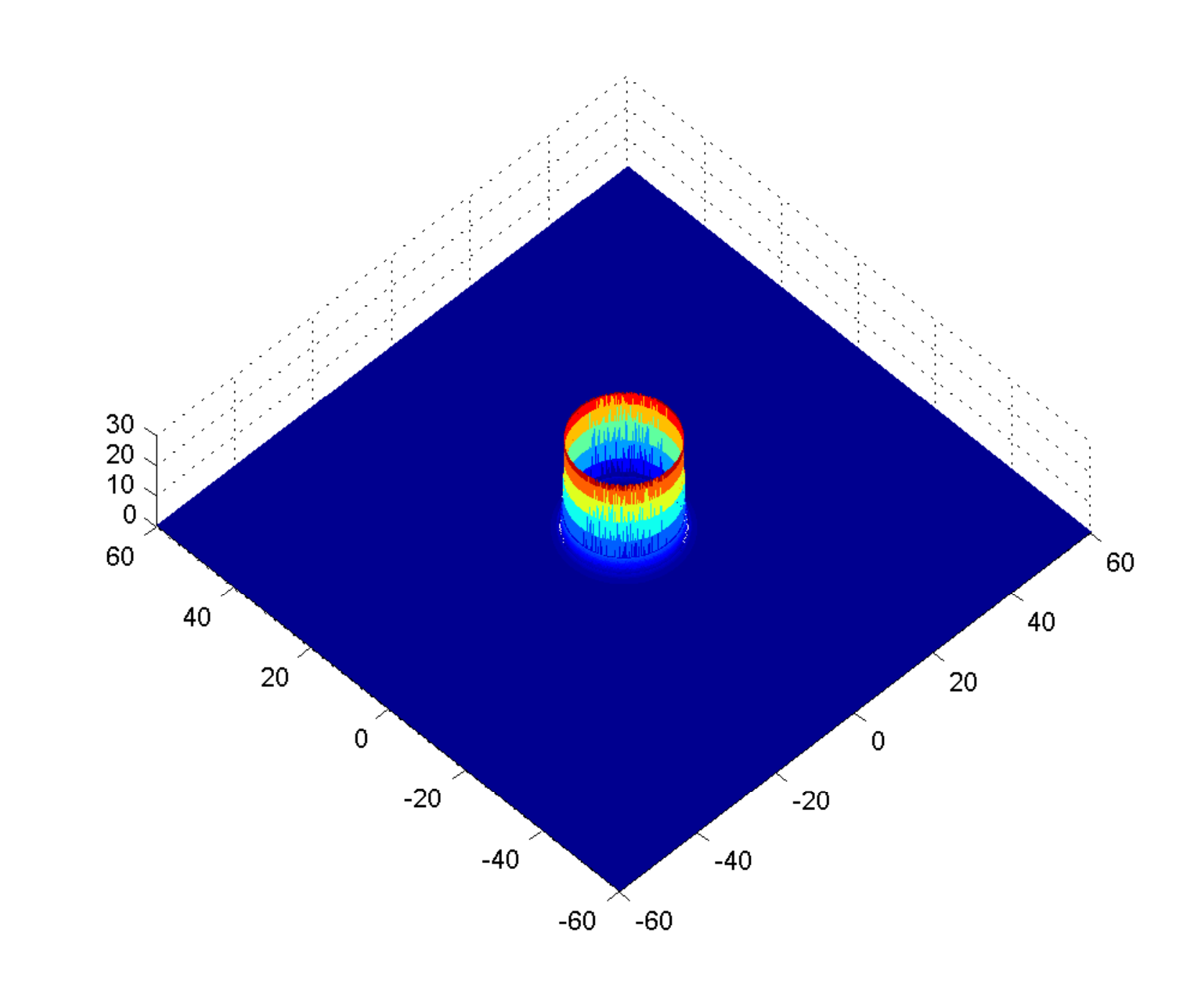}\label{fig:al1r010pr25s02t4060}}
\subfigure[]{\includegraphics[trim = 0.6in 0.6in 0.6in 0.6in,
clip,width=0.65\columnwidth]{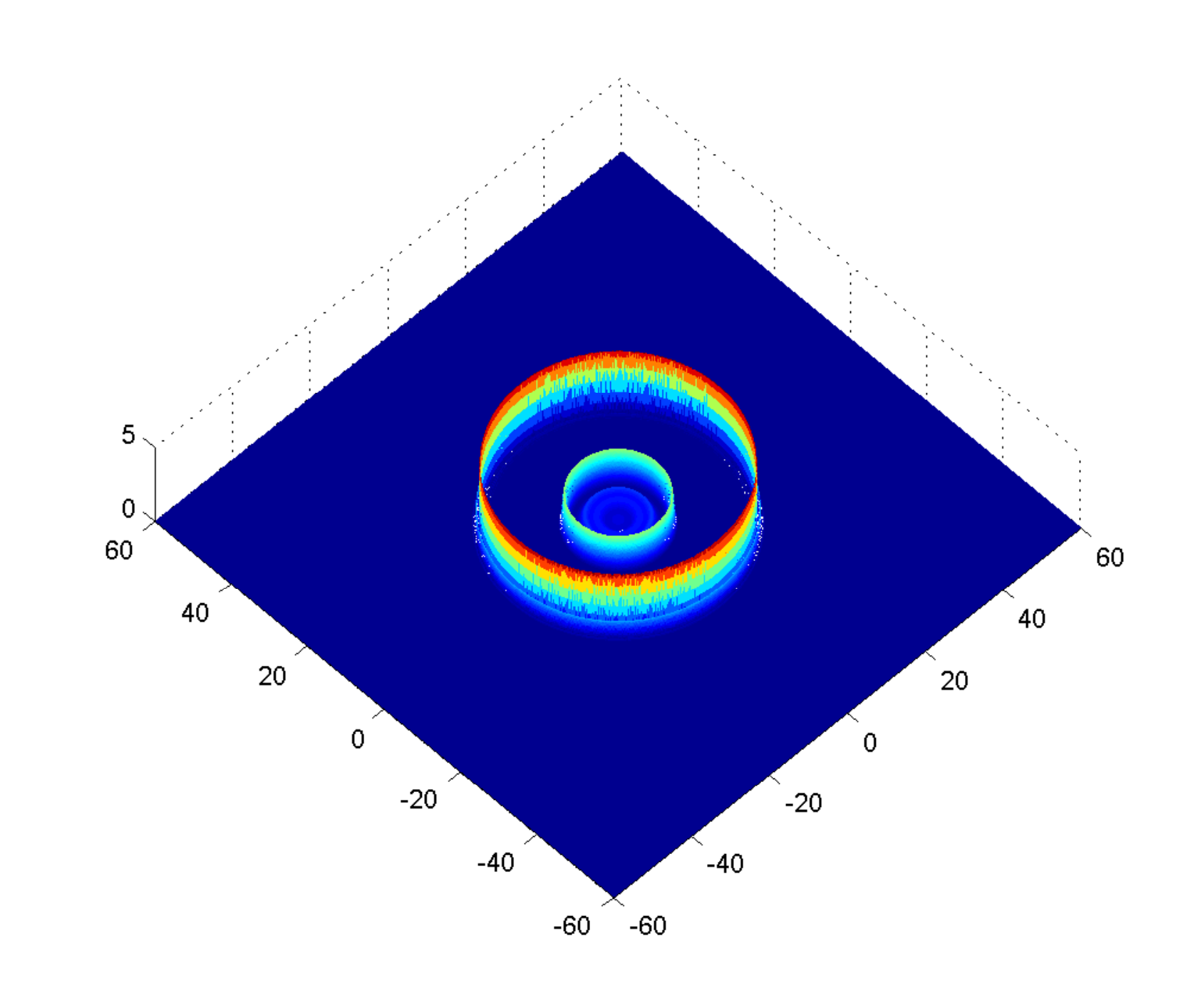}\label{fig:al1r010pr25s02t5060}}
\subfigure[]{\includegraphics[trim = 0.6in 0.6in 0.6in 0.6in,
clip,width=0.65\columnwidth]{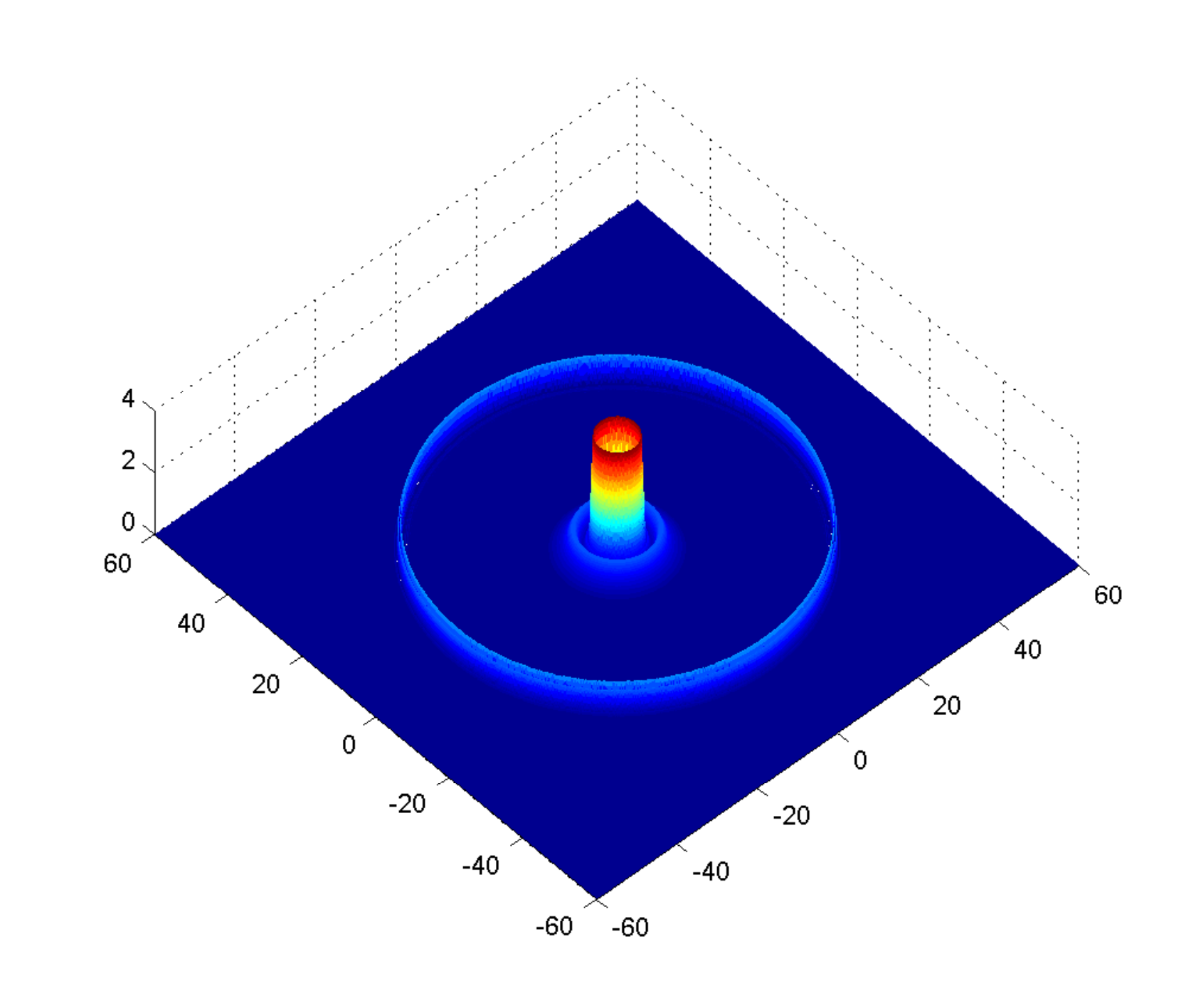}\label{fig:al1r010pr25s02t6060}}
\subfigure[]{\includegraphics[trim = 0.6in 0.6in 0.6in 0.6in,
clip,width=0.65\columnwidth]{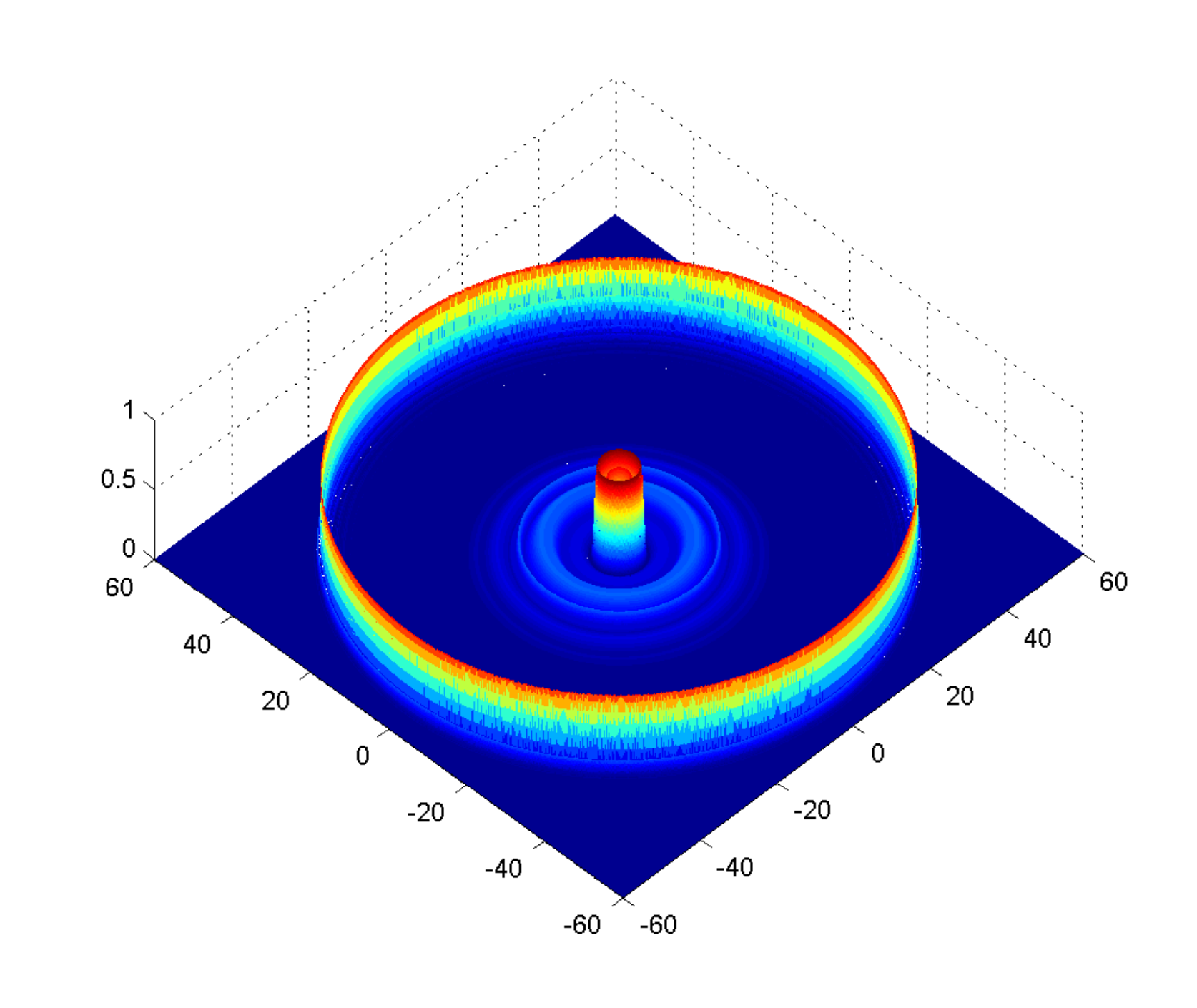}\label{fig:al1r010pr25s02t7060}}
\subfigure[]{\includegraphics[trim = 0.6in 0.6in 0.6in 0.6in,
clip,width=0.65\columnwidth]{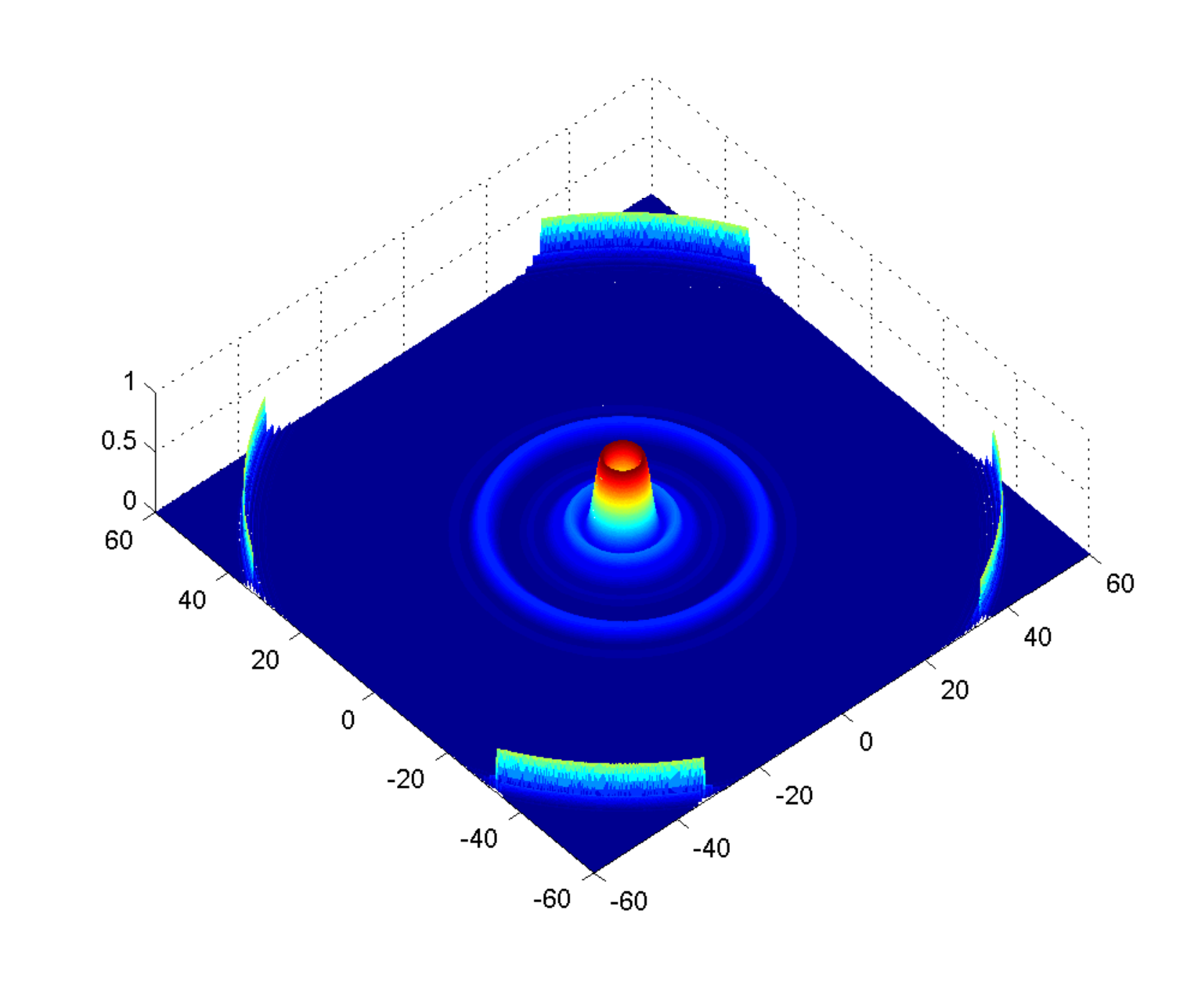}\label{fig:al1r010pr25s02t8060}}
\subfigure[]{\includegraphics[trim = 0.6in 0.6in 0.6in 0.6in,
clip,width=0.65\columnwidth]{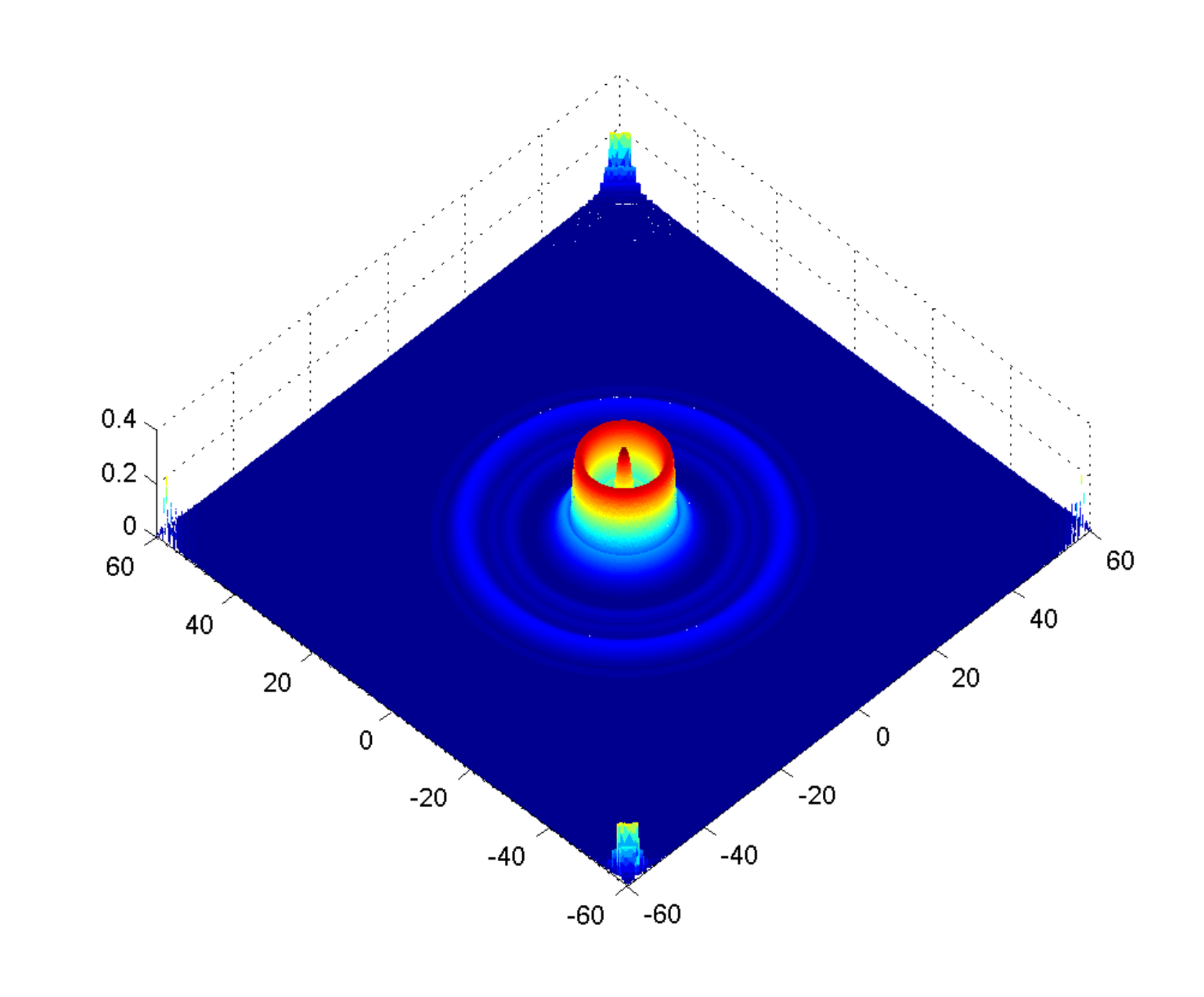}\label{fig:al1r010pr25s02t9060}}
\caption{Evolution of the energy density of the $\varphi^{4}$ domain wall ($\alpha=\beta=1$) with an internal matter density ($\rho=\rho_{0}e^{-r/r_{0}}$) from $t=10$ to
$t=90$.}
\end{figure*}
\begin{figure*}
\begin{center}
{\label{fig:fieldphi6}
\includegraphics[width=0.48\textwidth,height=0.3\textheight]{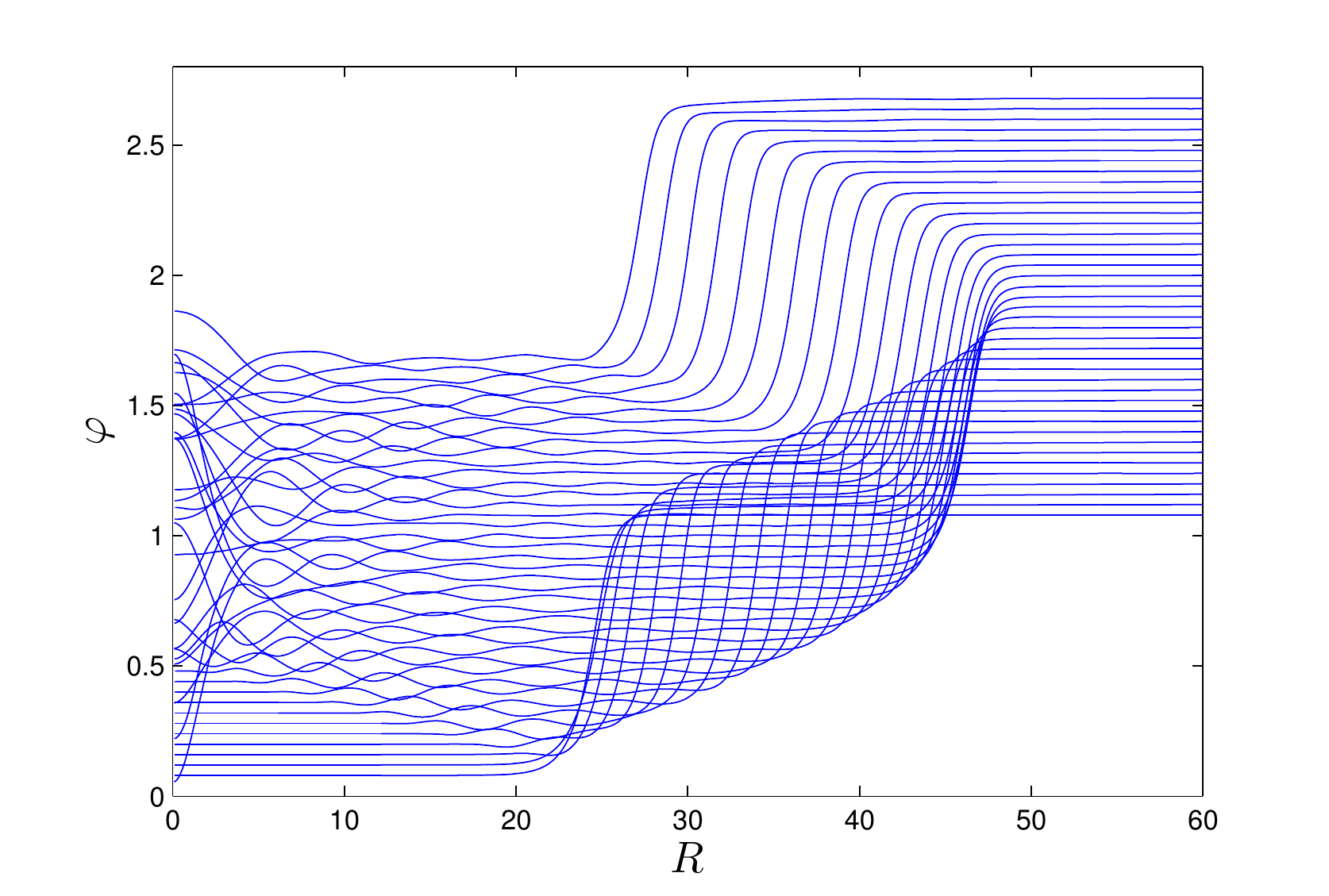}} \text{%
\hspace{0cm}}
{\label{fig:energyphi6matter}
\includegraphics[width=0.48\textwidth,height=0.3\textheight]{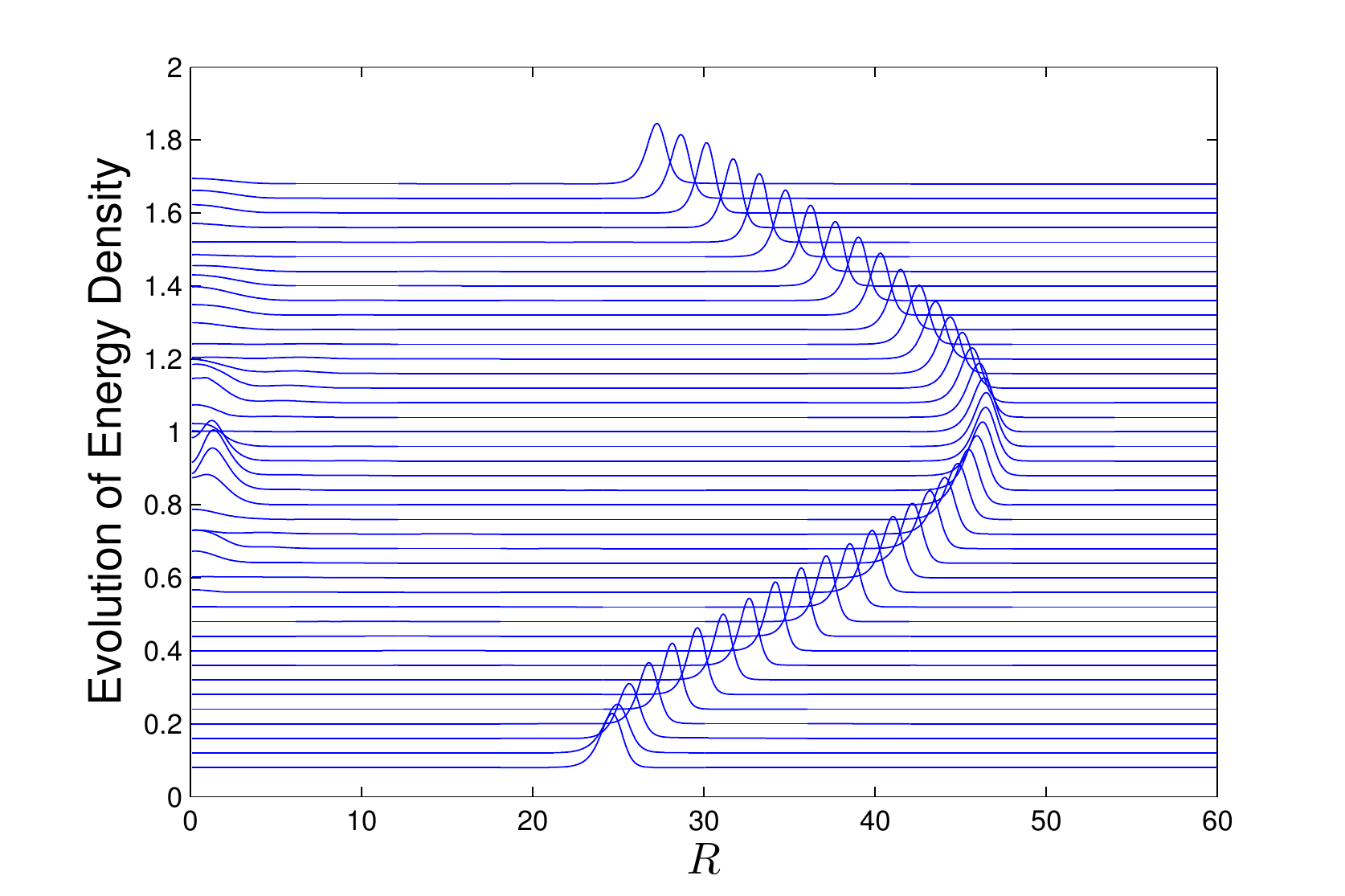}}
\end{center}
\caption{Evolution of the scalar field (left plot) and the energy density (right plot) of the $\varphi^{6}$ domain wall with an internal matter density, given by $\rho=\rho_{0}e^{-r/r_{0}}$, for the parameter values $\alpha=1$ and $\beta=\sqrt{2}$. See the text for more details.}
\label{fig:energydensityphi6}
\end{figure*}

The evolution of the bubble energy density for $\varphi^{4}$ is presented in Fig. \ref{fig:al1r010pr25s02t1060}--\ref{fig:al1r010pr25s02t9060}. The bubbles contract toward the center with increasing energy density like the DSG bubble with $\varepsilon=10$.  However, in this case, a breather appears in the center of the bubble where matter resides and it remains there for a long time. By comparing the evolution of the $\varphi^{4}$ bubble in this case with the previous one, one can recognize that despite collapsing, the change in energy density is less and the radiation via scalar field emission is higher. Moreover, it seems that there is an interesting similarity between the evolution of the $\varphi^{6}$ bubble without considering matter density and evolution of $\varphi^{4}$ spherical domain wall with a matter core.

Bubbles of the $\varphi^{6}$ system, like those of SG and DSG with  $\varepsilon=-0.1$, expand to a maximum radius and then start to contract, as depicted in Fig. \ref{fig:energydensityphi6}. It is worth noting that a breather appears in the center of the bubble for a short period of time.
Our results are consistent with those of \cite{CL}, regarding domain wall stability in the present of matter.

\section{Collective coordinate approach and the effect of Gravity}\label{gravity}

\subsection{Newtonian approach}

In the previous sections, we neglected the gravitational effects of the spherical domain wall and the central matter.
Here, we describe a collective coordinate approach which enables us to include the gravitational effects in the approximation that the wall
thickness is much smaller than its radius and the central mass is spherically symmetric. In order to investigate the collapse of the bubble under the combined influence of bubble tension, the bubble self-gravity, and the gravitational field of the central matter, we begin with a simplified Newtonian calculation, followed by a thin shell calculation in the framework of GR.

It is easy to show that for the initially stationary bubble at  $r=R_{0}$  with mass $M_{b}$ and surface tension $\sigma$ which surrounds central mass $M_{i}$,  the total energy will be:
\begin{equation}\label{1}
E=4\pi\sigma R_{0}^{2}-\frac{GM_{b}M_{i}}{R_{0}}-\frac{1}{2}\frac{GM_{b}^{2}}{R_{0}}.
\end{equation}
This Newtonian calculation is rather simple and we present it as a first estimate of the collapse including the effect of gravity, provided that the gravitational field is weak and the wall velocity is much smaller than the velocity of light. The first term in the RHS of Eq. (\ref{1}) comes from the tension of the bubble, while the remaining terms are caused by the gravitational field. The gravitational field  thus affects (accelerates) the collapse in a Newtonian description. 

At any later time, we have
\begin{equation}\label{2}
E=4\pi\sigma R^{2}-\frac{GM_{b}M_{i}}{R}-\frac{1}{2}\frac{GM_{b}^{2}}{R}+\frac{1}{2}M_{b}\dot{R}^2.
\end{equation}
Equating (\ref{1}) and (\ref{2}), leads to the following ODE for $R(t)$:
\begin{eqnarray}
\dot{R}^{2}=\frac{8\pi\sigma}{M_{b}}\left(R_{0}^{2}-R^{2}\right)+\frac{2GM_{i}}{RR_{0}}\left(R_{0}-R\right)
   \nonumber  \\
+\frac{GM_{b}}{RR_{0}}\left(R_{0}-R\right).
\end{eqnarray}
Since $M_{b}=4\pi R_{0}^{2}\sigma/c^{2}$, we get
\begin{equation}
\begin{split}
\dot{R}^{2}=\frac{2c^{2}}{R_{0}^{2}}\left(R_{0}^{2}-R^{2}\right)+\frac{2GM_{i}}{RR_{0}}\left(R_{0}-R\right)\left(1+\frac{M_{b}}{2M_{i}}\right),
\end{split}
\end{equation}
by introducing $a(t)\equiv R/R_{0}$,
\begin{equation}
\begin{split}
\dot{a}^{2}=\frac{2c^{2}}{R_{0}^{2}}\left(1-a^{2}\right)+\frac{2GM_{i}}{R_{0}^{3}}\left(\frac{1}{a}-1\right)\left(1+\frac{M_{b}}{2M_{i}}\right),
   \label{Neweq39}
\end{split}
\end{equation}
which leads to
\begin{equation}
\frac{da}{dt}=-\frac{c}{R_{0}}\sqrt{2\left(1-a^{2}\right)+\frac{2GM_{i}}{R_{0}c^{2}}\left(\frac{1}{a}-1\right)\left(1+\frac{M_{b}}{2M_{i}}\right)}.
\end{equation}

\begin{figure}
\epsfxsize=7.50cm
\centerline{\epsfbox{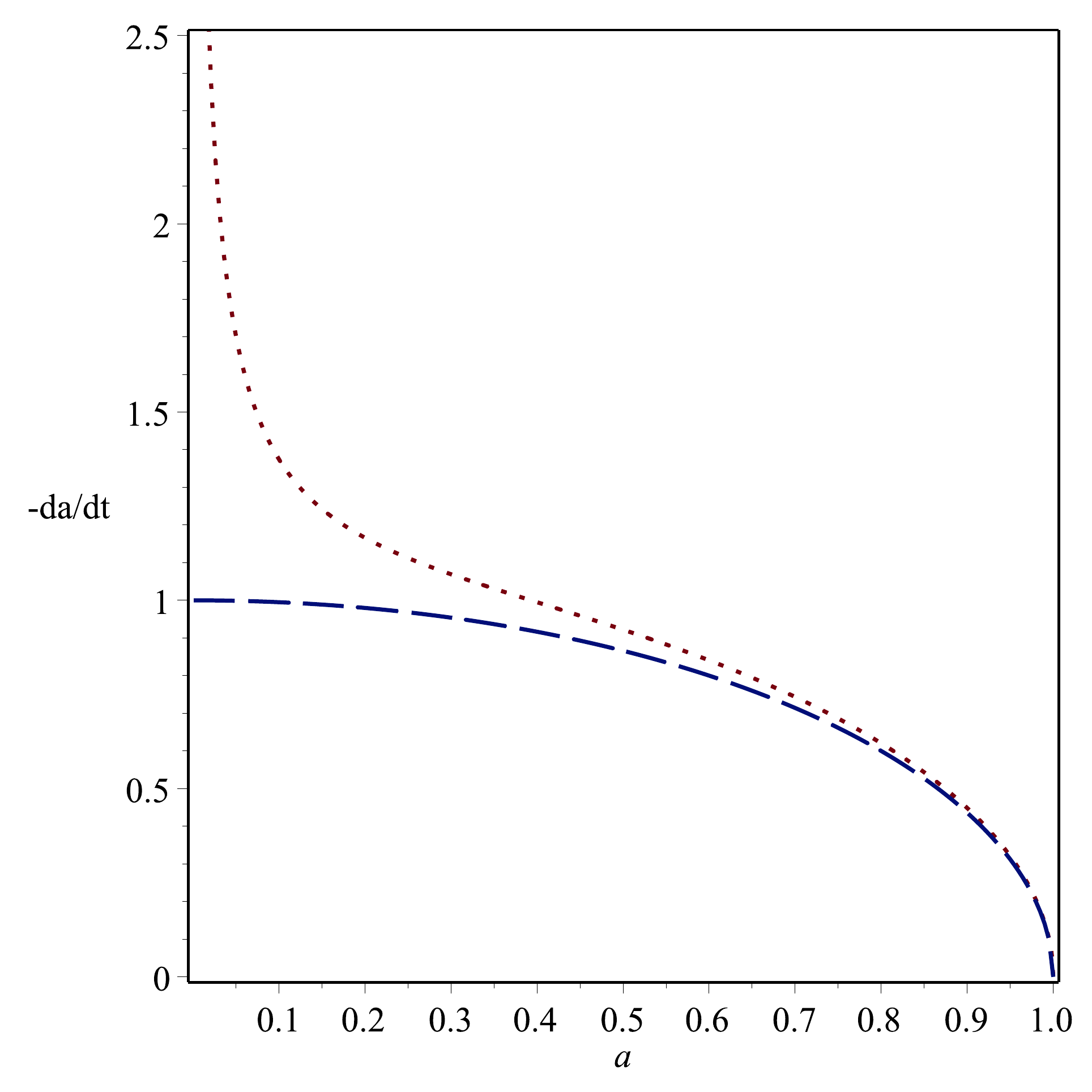}} \caption{The plot depicts the result when the gravitational field is switched off (dashed curve) and switched on (dotted curve), using the Newtonian result (\ref{Neweq39}). Note that the Newtonian gravitational force causes the collapse velocity of the bubble to diverge as $R\rightarrow 0$. In the absence of the gravitational force, the collapse becomes similar to the analysis depicted in Fig. \ref{vr}.}\label{fig:Newton}
\end{figure}

Note that the Newtonian gravitational force causes the collapse velocity of the bubble to diverge as $R\rightarrow 0$. If we switch off the gravitational force, the collapse becomes similar to what we had in Fig. \ref{vr}. Figure \ref{fig:Newton} shows the result for gravitational field switched off (dashed curve) and switched on (dotted curve), using the Newtonian result (\ref{Neweq39}).

The approximate time for collapsing will be:
\begin{equation}
t=\frac{R_{0}}{c}\int_{0}^{1}\frac{da}{\sqrt{2\left(1-a^{2}\right)+\epsilon\left(\frac{1}{a}-1\right)\left(1+\frac{M_{b}}{2M_{i}}\right)}} \,,
\end{equation}
where $\epsilon=r_{\rm Sch}(i)/R_{0}\ll1$.
This result shows that the bubble collapses to zero radius in a finite time, with bubble velocity diverging at zero radius (Fig. \ref{fig:Newton}).

\subsection{General relativistic approach}

For the GR calculation, we follow the thin shell formalism described in \cite{JDD}. We can also follow starting from the metric \cite{JDD}:
\begin{eqnarray}
\begin{split}
ds_{i}^{2}&=&-f_{i}(r)dt_{i}^{2}+\frac{1}{f_{i}(r)}dr^{2}+r^{2}d\Omega^{2},\\
ds_{o}^{2}&=&-f_{o}(r)dt_{o}^{2}+\frac{1}{f_{o}(r)}dr^{2}+r^{2}d\Omega^{2},
\end{split}
\end{eqnarray}
for the inside and the outside of the bubble; where $f_{i}(R)$ and $f_{o}(R)$ are the metric functions for the inner and outer regions of the bubble respectively. 
The inner and outer metrics are static, spherically symmetric vacuum solutions of the Einstein equations and are therefore forced to be Schwarzschild $f_{i,o}=1-2M_{i,t}/r$ due to the Birkhoff theorem. Only the mass parameters differ: $M_{i}$ for the inner region and $M_{t}$ for the outer region.
Besides, the metric for the transition region leads \cite{JDD}:
\begin{equation}
ds^{2}_{\rm wall}=-d\tau^{2}+R(\tau)^{2}d\Omega^{2}.
\end{equation}
where $r=R(\tau)$. Then the equation of motion of the bubble can be written as\footnote{Here $c=G=1$ and $[\sigma]=1/[kg]$.}:
\begin{equation}
\sqrt{\dot{R}^{2}+f_{i}(R)}-\sqrt{\dot{R}^{2}+f_{o}(R)}=4\pi\sigma R
\end{equation}
where $\sigma$ is the surface tension of the bubble. Solving for $\dot{R}$, we obtain
\begin{equation}
\frac{dR}{dt}=\sqrt{-V_{eff}},
\end{equation}
where the effective potential is defined as
\begin{equation}
\begin{split}
V_{eff}(R)=f_{0}(R)-\frac{\left(f_{i}(R)-f_{0}(R)-16\pi^{2}\sigma^{2}R^{2}\right)^{2}}{64\pi^{2}\sigma^{2}R^{2}}
\end{split}
\end{equation}
or
\begin{equation}
\begin{split}
\frac{da}{dt}=-\frac{c}{R_{0}}\sqrt{4\pi^{2}\sigma^{2}-1+\frac{1}{2}\left(1+\varsigma\right)\frac{\epsilon}{a}+\frac{1}{64}\left(\varsigma-1\right)^{2}\frac{\epsilon^{2}}{a^{2}}}
\end{split}
\end{equation}
with $\varsigma\equiv M_{t}/M_{i}$.
Note that this equation is consistent with the total mass of the domain wall bubble, which is initially at rest, $M_{b}=M_{t}-M_{i}=4\pi\sigma R_{0}^{2}$, as long as $M_{i},M_{t}\ll R_{0}$.

\begin{figure}
\epsfxsize=9.90cm
\centerline{\epsfbox{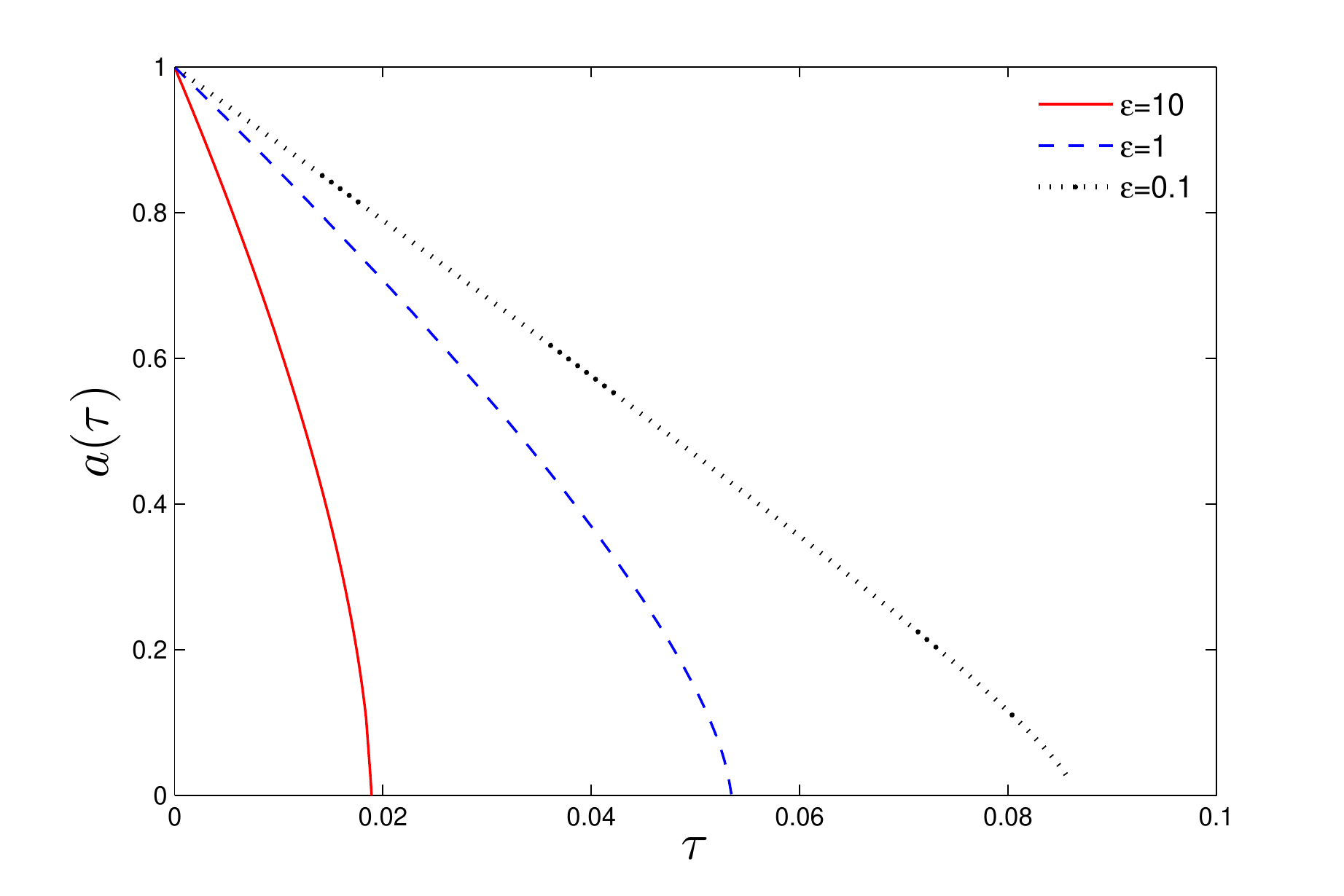}} \caption{ Collapse of the domain wall, tacking into account the gravitational effects (see Eq. (\ref{Tau})).}\label{tau}
\end{figure}

This equation can be cast into the following dimensionless form:
\begin{equation}\label{Tau}
\begin{split}
\frac{da}{d\tau}=-\sqrt{\Sigma+32\left(1+\varsigma\right)\frac{\epsilon}{a}+\left(\varsigma-1\right)^{2}\frac{\epsilon^{2}}{a^{2}}}\,,
\end{split}
\end{equation}
where $\Sigma\equiv256\pi^{2}\sigma^{2}-64$ and $\tau\equiv ct/(8R_{0})$. Note that one expects $\epsilon\ll1$ (i.e., the Schwarzschild radius of the inner mass much less than the initial bubble radius) and $\varsigma=o(1)$. We have plotted $a(\tau)$ for a few typical values of $\Sigma=100$ and $\varsigma=2$, in Figure.\ref{tau}.
For more realistic values of $\Sigma$ and $\varsigma$ and $\epsilon$ inspired by the symmetron model the $\Sigma$ term in (\ref{Tau}) is dominant and $a(t)$
is linear with negative slope $-\sqrt{\Sigma}$. This result is entirely consistent with our Lagrangian approach to be described later in this section.

We now follow a collective coordinate approach, by taking care of gravitational effects. To this end, we start with the action
\begin{equation}\label{act}
S=\int d^{4}x \sqrt{-g} \left(\frac{{\cal R}}{16\pi G}-\frac{1}{2}\partial^{\mu}\varphi\partial_{\mu}\varphi-V(\varphi)\right),
\end{equation}
where ${\cal R}$ is the Ricci scalar.
For the metric, we use
\begin{equation}\label{metr}
ds^{2}=-c^{2}f(r,t)dt^{2}+\frac{1}{f(r,t)}dr^{2}+r^{2}d\Omega^{2}
\end{equation}
with $f(r,t)=1-2GM(r,t)/c^{2}r$.

\begin{figure}
\epsfxsize=9.90cm
\centerline{\epsfbox{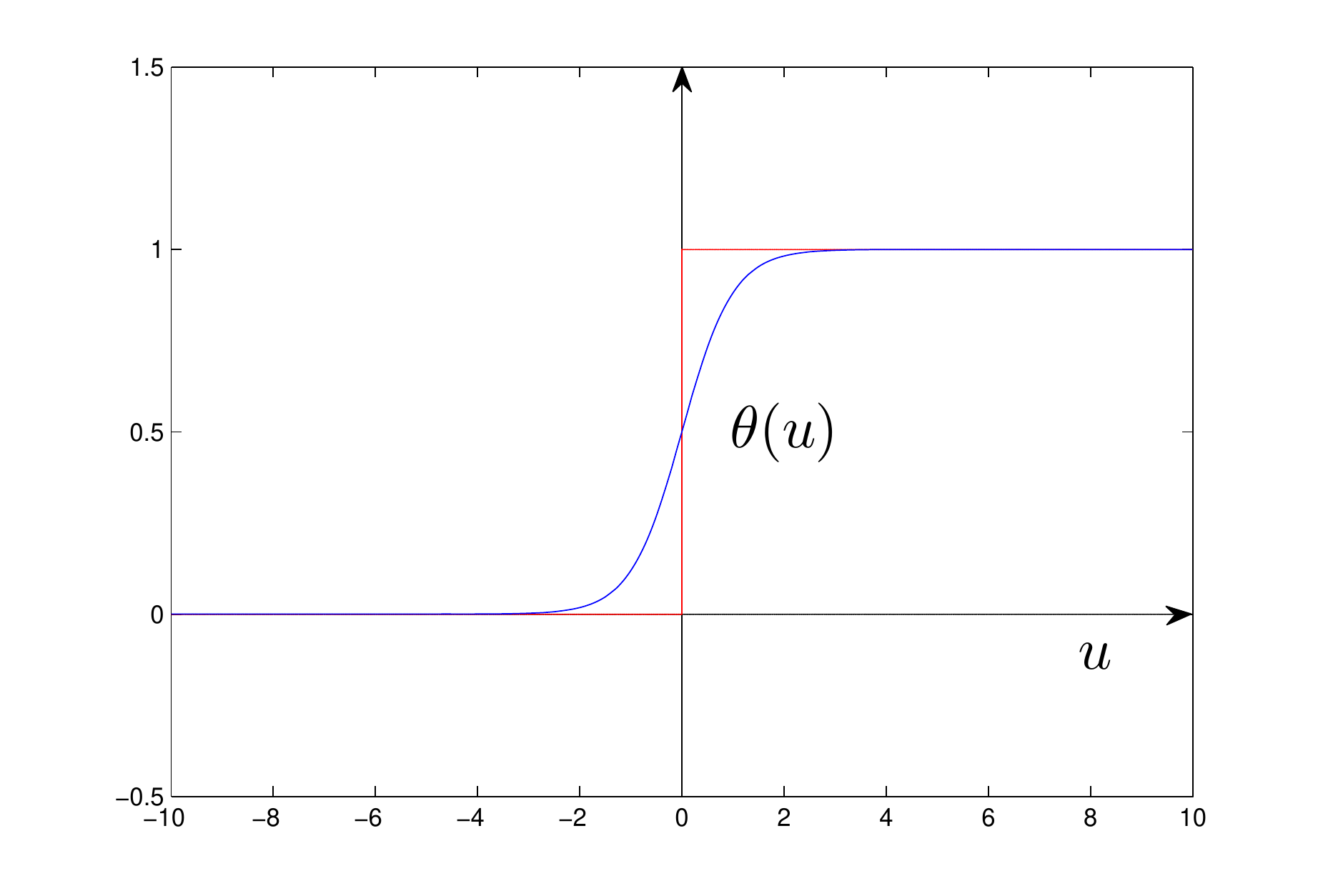}} \caption{Smoothing out the step function (red lines) using $\tanh$ function (\ref{47})}.\label{ta}
\end{figure}

In order to determine $M(r,t)$, we note that inside the spherical wall, we only have the gravitational field of the central mass $M_{i}$, which according to Birkhoff's theorem gives:
\begin{equation}\label{mi}
M(r,t)=M_{i}  \qquad {\rm for} \qquad r<R(t)\,.
\end{equation}
Outside the spherical bubble we have
\begin{equation}\label{mt}
M(r,t)=M_{t}  \qquad {\rm for}  \qquad R(t)>r \,,
\end{equation}
where $M_{t}$ is the total mass of the central object and the collapsing shell.
The relations (\ref{mi}) and (\ref{mt}) can be unified using the step function $\theta\left(r-R(t)\right)$:
\begin{equation}\label{m}
M(r,t)=M_{i}\theta\left(R(t)-r\right)+M_{t}\theta\left(r-R(t)\right),
\end{equation}
In order to avoid a discontinuity and make the metric jump smooth for the computation of the Ricci scalar and thus implementing
the collective coordinate approach, we replace Eq. (\ref{m}) with the smoothed version (see Fig.\ref{ta}):
\begin{equation}\label{47}
\theta(u)\rightarrow\tilde{\theta}(u)=\frac{1}{2}\left[1+\tanh(u)\right],
\end{equation}
which provides
\begin{equation}\label{55}
f(u)=1-\frac{G}{r}\left[M_{i}\left(1-\tanh(\beta u)\right)+M_{t}\left(1+\tanh(\beta u)\right)\right].
\end{equation}

Now, by starting with the Einstein-Hilbert action (\ref{act}), using the metric function ansatz (\ref{55}) and integrating over spatial coordinates for a bounded shell which includes the bubble, one arrives at a specific action involving the Lagrangian $L\left(R,\dot{R},\ddot{R}\right)$. More specifically, assume that the gravitational field does not have appreciable effects on the internal structure of the wall across it. The $t$ and $r$ dependence of the scalar field is accordingly assumed to be in the solitonic form given by Eq. (\ref{16}). For example, consider the $\varphi^{4}$ model (including dimensional parameters), i.e.,
\begin{equation}
\varphi=\alpha\tanh\big[\beta\left(r-R(t)\right)\big].
\end{equation}
Integration of the various terms over spatial coordinates in the action then proceeds easily. For instance, the potential term of the scalar field $V_{\varphi^{4}}(\varphi)=\frac{\beta^{2}}{2\alpha^{2}}(\varphi^{2}-\alpha^{2})^{2}$ yields:
\begin{eqnarray}
&&\int\sqrt{-g} d^{4}x V(\varphi)=
    \nonumber  \\
&=& \frac{\alpha^{2}\beta^{2}}{2}\int dt \; 4\pi \int dr r^{2}\big(\tanh^{2}\beta\left(r-R(t)\right)-1\big)^{2} 
	\nonumber\\
&=&2\pi\alpha^{2}\beta^{2}\int dt \int dr \; r^{2}{\rm sech}\,^{4}\big(\beta\left(r-R(t)\right)\big)
	\nonumber\\
&=&2\pi\alpha^{2}\beta\int dt \int \left(R+\frac{\xi}{\beta}\right)^{2}{\rm sech}\,^{4}\xi \; d\xi
	\nonumber\\
&=&2\pi\alpha^{2}\beta \int dt R(t)^{2}\int {\rm sech}\,^{4}\xi \; d\xi
	\nonumber\\
&&+\frac{2\pi\alpha^{2}}{\beta}\int dt \int \xi^{2} {\rm sech}\,^{4}\xi \; d\xi
	\nonumber\\
&&+4\pi\alpha^{2}\int dt  \; R(t) \int \xi \,{\rm sech}\,^{4}\xi \; d\xi.
\end{eqnarray}
in which $\xi\equiv \beta\left(r-R(t)\right)$.
The three integrals over $\xi$ are $\frac{4}{3}$, $-\frac{2}{3}+\frac{\pi^{2}}{9}$ and $0$, respectively. We therefore have
\begin{equation}
\int\sqrt{-g} d^{4}x V(\varphi)\approx \frac{8\pi\alpha^{2}\beta}{3}\int \left[R^{2}(t)+{\rm constant}\right]dt \,.
\end{equation}
The other terms in the action (\ref{act}) can be evaluated in an analogous manner.

Thus, after a tedious, but straightforward, calculation, the following action is obtained:
\begin{equation}
S=\int L\left(R,\dot{R},\ddot{R}\right) dt,
\end{equation}
where:
\begin{eqnarray}
&&L(R,\dot{R},\ddot{R})=\frac{\left(M_{t}-M_{i}\right)}{2c\left[\left(M_{i}+M_{t}\right)G-c^{2}R(t)\right]^{3}}\nonumber\\
&&\Big[-2\left(M_{t}-M_{i}\right)\dot{R}^{2}(t)R^{3}(t)c^{2}G\beta\nonumber\\
&&-\left(M_{t}+M_{i}\right)\ddot{R}(t)R^{3}(t)c^{2}G+\ddot{R}(t)R^{4}(t)c^{4}\nonumber\\
&&-\frac{4}{5}\left(-3M_{i}^{3}-M_{i}^{2}M_{t}+M_{i}M_{t}^{2}+3M_{t}^{3}\right)R(t)G^{3}\beta\nonumber\\
&&+4\left(M_{t}^{2}-M_{i}^{2}\right)R^{2}(t)c^{2}G^{2}\beta-2\left(M_{t}-M_{i}\right)R^{3}(t)c^{4}G\beta\nonumber\\
&&+4\left(M_{i}^{3}+M_{i}^{2}M_{t}+M_{i}M_{t}^{2}+M_{t}^{3}\right)G^{3}\nonumber\\
&&-8\left(M_{i}^{2}+M_{i}M_{t}+M_{t}^{2}\right)R(t)c^{2}G^{2}\nonumber\\
&&+6\left(M_{t}+M_{i}\right)R^{2}(t)c^{4}G-2R^{3}(t)c^{6}\Big]\nonumber\\
&&+\frac{8\pi}{3}cR^{2}(t)\alpha^{2}\beta \Bigg[-\frac{R(t)\dot{R}^{2}(t)}{\left(M_{i}+M_{t}\right)G-c^{2}R(t)}\nonumber\\
&&-\frac{\left(M_{i}+M_{t}\right)G-c^{2}R(t)}{c^{2}R(t)}-1\Bigg].
\end{eqnarray}
We already know that the symmetron field within the central mass is screened, tending to its $Z_{2}$ symmetric vacuum $\left<\varphi \right>=0$, except for the skin depth which is of the order of \cite{KJAA}
\begin{equation}
\Delta \mathfrak{a}\sim\frac{M_{i}^{2}}{\rho \mathfrak{a}},
\end{equation}
where $\mathfrak{a}$ is the effective radius of the central mass. We are now in a position to solve the extended Euler-Lagrange equation
\begin{equation}
\frac{d}{dt}\frac{\partial L}{\partial \dot{R}}- \frac{d^{2}}{dt^{2}}\frac{\partial L}{\partial\ddot{R}}=\frac{\partial L}{\partial R}
\end{equation}
which now includes, the gravitational effects of the bubble and the central mass. This Lagrangian leads to the following equation for $R(t)$.
\begin{eqnarray}\label{eq}
&&12G\ddot{R}(t)\left(M_{i}^{2}-M_{t}^{2}\right)-3R(t)\ddot{R}(t)c^{2}\left(M_{i}-M_{t}\right)\nonumber\\
&&-8R(t) \left[7 \dot{R}^{2}(t) +6 R(t)\ddot{R}(t) \right]    \beta Gc^{2}\alpha^{2}\pi\left(M_{i}+M_{t}\right)\nonumber\\
&&+6 R(t)\ddot{R}(t)\beta G\left(M_{i}-M_{t}\right)^{2}
+32 R^{2}(t)\beta\alpha^{2}c^{6}\pi\nonumber\\
&&+16 R^{2}(t) \left[ \dot{R}^{2}(t)  + R(t)\ddot{R}(t) \right] \beta c^{4}\alpha^{2}\pi \nonumber\\
%
&&=136 R(t)\beta G c^{4}\alpha^{2}\pi\left(M_{i}+M_{t}\right).
\end{eqnarray}

\begin{figure}
\epsfxsize=9.90cm
\centerline{\epsfbox{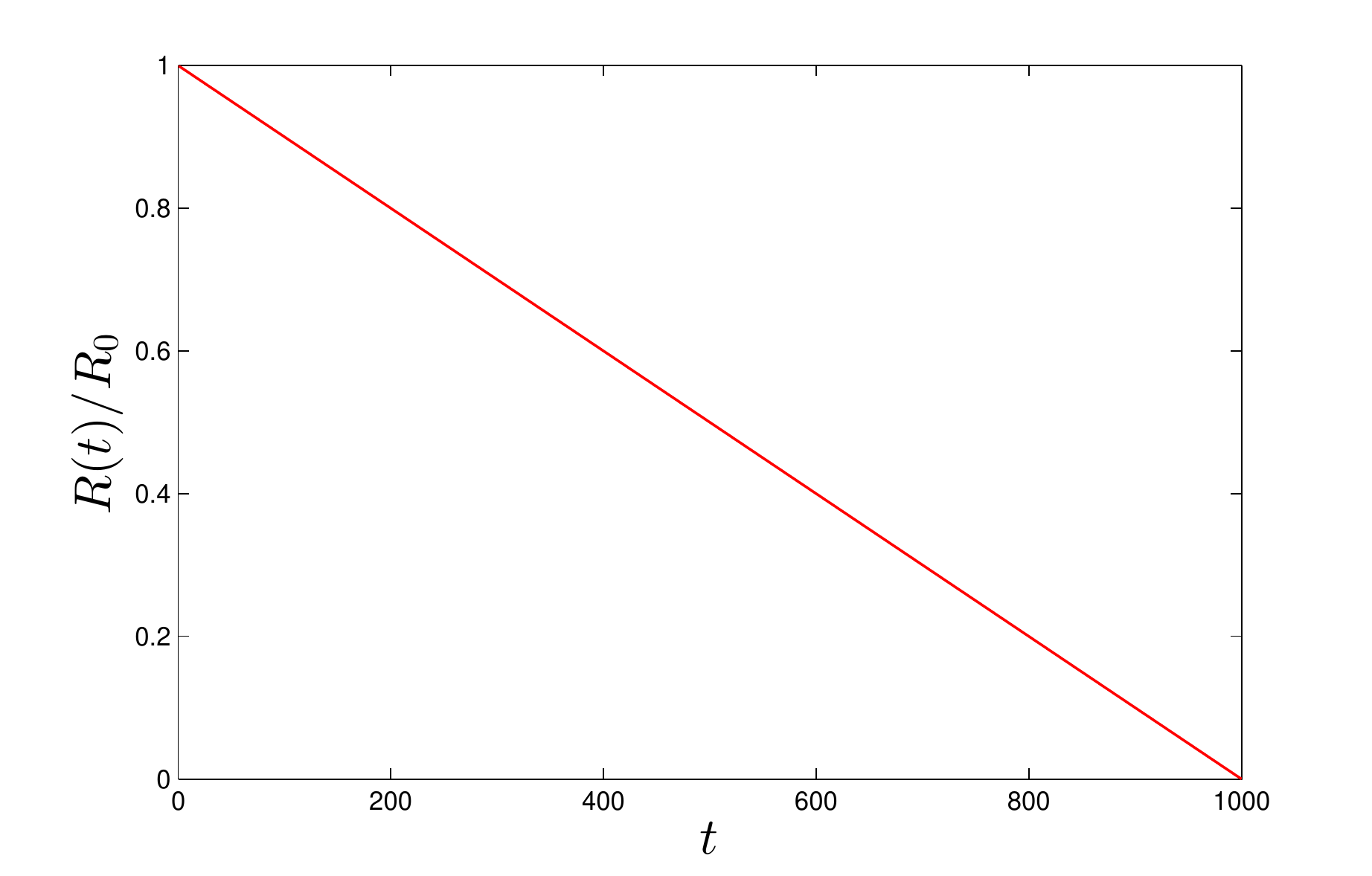}}\caption{The collapse of the domain wall bubble from rest until it collides with the central mass, in which the symmetron field is highly screened. The parameter chosen are, $M_{i}=10^{48}$kg, $M_{b}=1.5\times10^{43}$kg,$\alpha=3.33\times10^{26}\hbar/c^{3}$ and $\beta=2.36\times10^{-22}\sqrt{\hbar}/c^{\frac{3}{2}}$.}\label{compare}
\end{figure}

In order to obtain this equation, we have assumed that $R(t)$ is always much larger than the Schwarzschild radii $2GM_{i}/c^{2}$ and $2GM_{b}/c^{2}$. Furthermore, let us assume that the bubble mass $M_{b}$ is much smaller than the central mass (i.e. $\varsigma\equiv\frac{M_{t}}{M_{i}}=o(1)$). Using these approximations, we have numerically calculated $R(t)$
for a typical value of $M_{t}/M_{b}$ in Fig. \ref{compare}, which depicts that the collapse of the domain wall bubble from rest until it collides with the central mass, in which the symmetron field is highly screened.

\section{Summary and Conclusion}\label{Conclusion}

The four popular nonlinear scalar field systems SG, DSG, $\varphi^{4}$ and $\varphi^{6}$, extensively analysed in the literature, possess exact kink solutions in $1+1$ dimensions. Motivated by the symmetron analysis we employed these exact solutions to construct initial conditions for the collapse of a large (compared to the thickness of the wall) spherical domain wall. We first presented a simple analytical model, based on the conservation of the total energy of the wall and a reasonable assumption for the curvature effects. We then solved the corresponding equations numerically and compared the results with our simple analytical model. We showed that the analytical model fits the more accurate numerical results very well, until the full collapse, after which oscillations and scalar radiation take place. 

We then explored the effect of a central matter lump on the evolution of a spherical domain wall. We reached the conclusion that a central matter lump can prevent the full collapse and annihilation of the domain wall bubble, due to the repulsion between the domain wall and matter over-density within the symmetron model.
Furthermore, we have investigated the dynamics of the bubble with a central mass, in the presence of gravity for the specific $\varphi^{4}$ system.  Our results show that the collapse is almost linear $a(t)\approx-\sqrt{\Sigma}t$ until the wall collapses into a black hole, if direct interaction with the central matter is not taken into account (Figures \ref{vr}, \ref{tau} and \ref{compare}). The collapse halts as soon as the bubble reaches the central mass, due to the screening effect of matter, as seen from calculations which include direct interaction of $\varphi$ with matter (Figures \ref{fig:DSGp} and \ref{fig:al1r010pr25s02t1060}--\ref{fig:al1r010pr25s02t9060}).
In concluding, we mention that doing the calculations with the  simultaneous dynamical effects of gravitation, scalar field and scalar field-matter coupling proved to be too difficult to end up with a reliable solution. Thus, we considered them separately in this paper. An investigation of the combined effect is left for a future work.

\acknowledgments{We thank Pedro Avelino for helpful suggestions and comments.
MP acknowledges the support of Ferdowsi University of Mashhad via the proposal No. 32361.
NR acknowledges the support of Shahid Beheshti University Research Council.
FSNL acknowledges financial  support of the Funda\c{c}\~{a}o para a Ci\^{e}ncia e Tecnologia through an Investigador FCT Research contract, with reference IF/00859/2012, funded by FCT/MCTES (Portugal).}


\end{document}